\documentclass[
rmp,aps,
preprintnumbers,nofootinbib,amsmath,amssymb]{revtex4}
\usepackage{bm}
\usepackage{graphicx}
\usepackage{graphics}
\usepackage{epsfig}

\usepackage[pdftex,bookmarks,colorlinks,breaklinks]{hyperref}
\hypersetup{linkcolor=red,citecolor=blue,filecolor=dullmagenta}

\begin{document}

\title{Disclinations, dislocations and continuous defects: a reappraisal}
\author{M. Kleman}\email{maurice.kleman@mines.org}
\affiliation{Institut de Min\'{e}ralogie et de Physique des
Milieux Condens\'{e}s\\
\noindent(UMR CNRS 7590), \\
\noindent Universit\'{e}
Pierre-et-Marie-Curie, Campus Boucicaut, 140 rue de Lourmel, 75015
Paris, France}
\author {J. Friedel}
\affiliation{Laboratoire de Physique des Solides, \\
\noindent(UMR CNRS 8502), \\
Universit\'{e}
de Paris-Sud, B\^at. 510, 91405 Orsay c\'edex, France}

\begin{abstract}
Disclinations were first observed in mesomorphic phases. They were later found relevant to a number of ill-ordered condensed matter media, involving continuous symmetries or frustrated order. Disclinations also appear in polycrystals at the edges of grain boundaries; but they are of limited interest in solid single crystals, where they can only move by diffusion climb and, owing to their large elastic stresses, mostly appear in close pairs of opposite signs.

The \emph{relaxation} mechanisms associated with a disclination in its creation, motion, change of shape, involve an interplay with continuous or quantized dislocations and/or continuous disclinations. These are attached to the disclinations or are akin to Nye's dislocation densities, which are particularly well suited here. We introduce the notion of \emph{extended Volterra process} which takes these relaxation processes into account and covers different situations where this interplay takes place. These concepts are illustrated by some of a variety of applications in amorphous solids, mesomorphic phases and frustrated media in their curved habit space. These often involve disclination networks with specific node conditions.

The powerful topological theory of line defects only considers defects stable against any change of boundary conditions or of relaxation processes compatible with the structure considered. It can be seen as a simplified case of the approach considered here, particularly suited for media of high plasticity or/and complex structures. It cannot analyze the dynamical properties of defects nor the elastic constants involved in their static properties; topological stability cannot guarantee energetic stability and sometimes cannot distinguish finer details of structure of defects.

\end{abstract}

\date{\today}

\maketitle

 \tableofcontents

\section{Introduction}
\label{Introduction}
\subsection{General considerations}

Defects in mesomorphic phases (or liquid crystals) have been the subject of numerous investigations in the last thirty years. This research has emphasized the importance of \emph{disclinations}, line defects first defined by \textcite{volterra07} and the main specific defects of liquid crystals \cite{friedel22}. As a consequence, the role of defects of a similar nature has also been recognized in other media, most of them with non solid crystalline symmetries, but not only. This essay is devoted
to the geometrical and topological concepts relevant to this field of research, in various media where disclinations are acting in interplay with other defects, mostly other types of line defects (dislocations).

We are led to compare two different theoretical approaches to the classification of line defects, the Volterra process and the topological stability theory. They are \textit{not} equivalent, but rather complementary. The topological theory has attracted attention by its property of classifying, not only line defects, but also defects of any dimensionality, in a very general manner, on the basis of the topological properties of the order parameter. The Volterra process only applies to line defects, which it classifies by the elements of the symmetry group of the ordered medium. This approach allows to deal with the (static and dynamical) interplay between disclinations and other line defects. Continuous defects (which relate to continuous symmetries) thus control the shape of quantized disclinations, to which they are associated, either attached to them, or accompanying them at a small distance. These specific defects of the Volterra process have been little studied in recent years. Yet the sustained interest in mesomorphic and frustrated phases, and in other media whose structural properties are remote from 3D crystalline media, justifies a reappraisal of the subject.

As we shall see, the Volterra process yields the same main conclusions as the topological stability theory, but at a finer level, by properly handling boundary conditions and all the plastic relaxations, including those related to line-attached defects. This approach can be particularly useful when dealing with nanostructures or with dynamical aspects, when the viscosity is large.

We do not
approach the subject of mesomorphic phases straightaway.  The final situation depends of course on the
\emph{symmetries} of the media under consideration, a topic
that is not visited in the first part (section \ref{CONTINUOUSDEFECTS}) of our essay,
and on the physical nature of the relaxation processes that bring
the defects towards their final, stable or metastable, state. We shall consider several structures characteristic of ill-condensed media. But we shall first make a detour through isotropic uniform solid media, with the
sole purpose of understanding the generic geometrical relations between disclinations on one hand, dislocations on the other, and {of}
their interplay
A number of new results are inserted, whose weight in the general balance of the paper has forced us to limit the review of some topics (the foundations of the topological theory, geometrical frustration), especially when excellent reviews already exist. On the other hand, the basic ingredients of the theory of continuous defects, in the perspective we want to replace it, are rather scattered in the literature, and deserved anyhow some deeper analysis; the distinction between \textit{constitutive} and \textit{relaxation} line-attached dislocations is new and structures this topic.

As a major application, the concept of disclination in smectic A (SmA)
{and other mesomorphic }phases will be considered at some length. Other topics of some
importance will also be tackled:

\noindent  {- the role of disclinations in polycrystal structure and
deformation, of importance for polynanocrystals,}

\noindent - the structure of an undercooled liquid, which features a curved space crystal mapped into flat physical space. This is a typical example, along with liquid crystalline Blue Phases (BPs), of \emph{geometrical frustration}. Our analysis of defects in a 3D 'amorphous' curved space S${^{3}}$ is new and generalizes the results for the usual E${^{3}}$ amorphous phase.

Before listing the contents of this paper, we find it useful to recall the main feaures of the description of defects in ordered madia in terms of the Volterra process and its extension, the topological classification and the theory of continuous defects, all involved in our description of disclinations.
\begin{figure}
\includegraphics{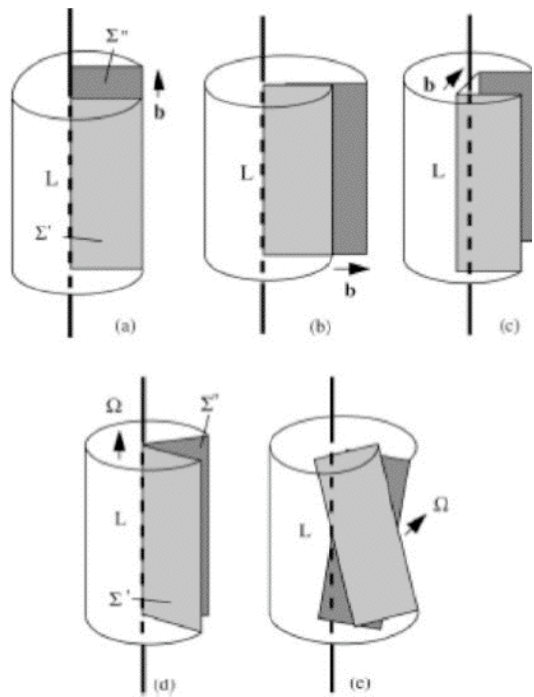}
\caption{ \label{vp} The Volterra process for elementary types of
\textit{dislocations}: (a) screw dislocation, (b) and (c) two
constructions of the same edge dislocation; and
\textit{disclinations}: (d) wedge disclination, (e) twist
disclination. The original cut surface $\Sigma$ is a vertical half
plane bounded by the axis of a cylinder of matter. View before the
void is filled ((c) and (d)) and the medium relaxed (all). From
Kleman and Lavrentovich, \emph{Soft Matter Physics, an
Introduction}, Springer, 2003, with permission}
\end{figure}

We recall the main characteristics of the Volterra process
for defect lines in a solid body \cite{volterra07}, cf. J. \textcite{friedeldisloc}.  Cut the matter along a surface $\Sigma$ (the
\emph{cut surface}) bound by a line L (a loop,~or an infinite
line), displace the two lips $\Sigma'$ and $\Sigma''$ of the cut
surface by a relative \emph{rigid} displacement that can be
analyzed as the sum of a translation $\mathbf{b}$ and a
rotation $\mathbf{\Omega}$, introduce matter in its
perfect state (\textit{i.e} elastically undeformed) in order to fill the void
left by the rigid displacement, or remove matter in the regions of
double covering, glue back together new matter and primitive matter along
the lips $\Sigma'$ and $\Sigma''$, let the medium relax
elastically;
Fig. \ref{vp} represents the Volterra process for different orientations of $\mathbf{b}$ and
*$\mathbf{\Omega}$ with respect to a straight dislocation line.
This process is certainly badly defined along the
line L itself (an elastic singularity remains along L,
the \emph{defect}) and generates a singularity of the order parameter on the cut surface; however this latter difficulty disappears if $\mathbf{b}$ and $\mathbf{\Omega}$ are translational and rotational symmetries of the medium (\emph{quantized}, \emph{perfect} \emph{defects}).  The strain field is small if\textbf{}
the rigid displacement is restricted to a translation of small amplitude, comparable to the atomic distance say; L is  then a
\emph{dislocation}, $\mathbf{b}$ its \emph{Burgers
vector}.
There always are large amplitude displacements if the {Volterra process} applies
to a rotation $\mathbf{\Omega}=\Omega \,\mathbf{t}$,
 except on the axis of
rotation $\mathbf{t}$ itself; L is then a \emph{disclination}.  This is the reason why
research in the field of defects presents such a
complexity and, thereby, such an accumulation of riches.

When the {Volterra process}
 involves at the same time a translation and a
rotation, one has a \emph{dispiration} \cite{harris70a,harris77}\footnote{Dispirations have been studied, but to a limited extent, only in some smectic phases where some symmetry elements include simultaneously a translation and a rotation. \textcite{tana92} have observed dispirations in a antiferroelectric smectic phase (SmC$_{A}$): the layer thickness $d_{0}$ is half the repeat distance of the polarization $\mathbf{P}$, which changes sign from one layer to the next. Hence a $d_{0}$ translation and a $\pi$ rotation of $\mathbf{P}$ constitute together a helical symmetry. See also \textcite{kucz99} for observations in a chiral antiferroelectric smectic phase (SmC$^{*}_{A}$), and \textcite{lejcek02}for theoretical considerations, as well as references therein.}.\\

\noindent \textsl{a. Quantized perfect disclinations.}

 \indent The situation is particularly simple if the line is straight, along the rotation vector (\emph{wedge} disclination), more complex in the {Volterra process} sense when the rotation vector is perpendicular to the line (\emph{twist} disclination), in which case it generally involves the simultaneous presence of \emph{perfect dislocations} attached to the twist line. Research in this domain is very lively:  the approach in the Volterra process terms is here supplemented by the theory
 of topological stability, see \ref{The TC} below.

 Stemming from the Volterra process picture of a defect,  line and surface geometry theory is specifically
employed for columnar and lamellar media \cite{kln} and, more
generally, for liquid media with quantized translations
\cite{achard05}.
Riemannian geometry is the main mathematical tool to treat perfect quantized disclinations in geometrically frustrated media \cite{kleman89}.\\

\noindent \textsl{b. Three important concepts in the development of {the Volterra process}.}

(i) \textit{Imperfect line singularities}. If the relative displacement ($\bm{\Omega}$, $\mathbf{b}$) is not a symmetry of the medium, the defect line is bordering a \emph{surface of misfit} along the cut surface of the {Volterra process}.  One will talk of imperfect dislocation, disclination, or dispiration.This is for instance the case, in a crystal, of partial dislocations bordering a stacking fault or of disclinations bordering grain boundaries.

(ii) \textit{Continuous distributions of defect lines of infinitesimal strength \textrm{(}$\mathbf{\Omega}$, $\mathbf{b}$}), already considered by \textcite{volterra07}.  Such planar distributions were introduced by \textcite{frank50b} for describing pure flexion and rotation grain boundaries, which can be attached to straight wedge and twist disclinations respectively ($\mathbf{\Omega}$ respectively parallel and perpendicular to the line).  Three dimensional distributions were introduced by \textcite{nye53} and others to describe plastic distortions of minimal energy.

(iii) \textit{Plastic relaxation}, leading to extended {Volterra process}. In a medium such as liquid crystals where some of the stresses can be released plastically in a \emph{viscous} way, this relaxation plays a large role in reducing the energy and increasing the mobility and flexibility of disclinations. The classical {Volterra process}, which refers to a solid (frozen) medium, has then to be completed by a stress relaxation that can be analyzed in terms of a three dimensional continuous distribution of dislocations of infinitesimal strength. This is usually obtained by replacing the elastic stress field of a solid medium by the elastic stresses referring to the liquid crystal considered (e.g. Frank and Oseen equations \cite{frank58a} replacing Hooke's elasticity).  We will in this case talk of an \emph{extended {Volterra process}}.

Somewhat similar plastic stress relaxations are possible in solids at the tip of slip lines or cracks, considered as dislocations or disclinations; the end of a sub-boundary produced by slip can also relax plastically by developing a localized crack.  One can also talk in those cases of extended {Volterra process}, but remember that the plastic relaxation can have now a finite elastic limit or at least a strongly viscous friction \cite{friedel59a}.

The motion and bending of disclinations involve the production of two dimensional continuous distributions of infinitesimal dislocations that can often disperse in a liquid crystal, or be plastically relaxed in a solid, or again be absorbed by another defect, as we shall show on some examples.

\subsection{The topological classification} \label{The TC}
\noindent \textsl{a. Why a topological classification}

We outline the general principles of this theory very briefly, and illustrate them on some simple
examples of liquid crystal phases, in order to bring out the concept of topological
stability (the essential contribution of this theory to the physics of defects), and
to comment on the effects of non-commutativity of the symmetries, which
are taken into account in a systematic way in this theory.

The topological theory started with the articles of \textcite{toulouse76, rogula76, volovik76}; for recent reviews, with emphasis on mesomorphic phases, see \textcite {michel, mermin, trebin}.
The topological classification has cured a certain number of difficulties raised by the Volterra process, when applied to ill-condensed matter, mostly liquid crystals. Liquid crystal defect features have no equivalent in usual solid crystals: disclinations whose core singularity vanishes away, point singularities, 3D knotted non-singular configurations.

- the Volterra process correctly describes straight wedge
 disclinations of strength $|k|=1/2$ (i.e., rotational symmetries of angle $\omega=\pm \pi$), but cannot be extended to $|k|>1/2$, i.e., angles $\omega=(2n\pm1)\pi, n\neq 0, n \in Z$, in its 'naive' version,

- twist disclinations cannot be constructed, except locally (for an illustration, see \cite{harris77}),

- the 'escape in the 3rd dimension' \cite{meyer73} obviously does not come out of the Volterra process.

However these difficulties can also be dealt with successfully by introducing the concept of \textit{extended Volterra process}.\\

\noindent \textsl{b. The order parameter space}

The topological classification of defects relies on the application to ordered media of the methods and concepts of algebraic topology; standard references are \textcite{stee57} or  \textcite{mas67}.  The manifold V of internal states, also called the vacuum or the order parameter space is the space of all the possible different positions and orientations of the perfect crystal in (flat) Euclidean space.  Let H be the group of symmetry of the ordered structure, G=E$^{3}$=R$^{3} \Box \textrm{O}(3)$ the group of Euclidean isometries of E$^{3}$.  The symbol $\Box $ is for the semi-direct product of groups.  R$^{3}$ is the 3D group of continuous translations, O(3)=SO(3)$\times \textrm{Z}^{2}$ the full group of rotations, with center of symmetry included.  The symbol $\times$ is for the direct product of groups, H is a subgroup of G. V is then the quotient space G/H.  Observe that V is not a group, generically, except if H is a normal subgroup of G.

Examples of order parameter spaces are given in \textcite{kleman00};  the order parameter space of a uniaxial nematic $\mathrm{V(N)=P}^{2}$, the projective plane; the order parameter space of a 3D crystal, regarding only the translations, is V(Xal)=T$^{3}$, the 3D torus.

These concepts extend in a natural way when G is the group of a space M of constant but $\neq 0$ curvature, and H a subgroup of G, i.e. the group of symmetry of a ordered structure of habit space M. We shall make use of this extension in the investigation of defects in frustrated media, see section \ref{TemplatesintheThree-Sphere}.\\

\noindent \textsl{c. The first homotopy group (the Fundamental Group) of the  order parameter space}

Let us now consider how the order parameter space V enters into the issue of the topological classification of defects.  Start from a distorted ordered medium, in which the order parameter is broken along a line L.  In order to test the topological nature of the breaking, surround L by a closed loop $ \gamma$ entirely located in the good ordered structure, good in the sense of the usual theory of dislocations in solids.  It is possible to attach to each point $\mathbf{r}$ belonging to $\gamma$ a 'tangent' perfect ordered structure, which maps in a unique way onto some point R $\in V$.  When $\mathbf{r}$ traverses the closed loop $\gamma$, R traverses a closed loop $\Gamma$ on V.  Call $\phi$ this well-defined continuous mapping
$$\phi :\gamma \rightarrow \Gamma.$$
The function $\phi$ can be extended continuously to the whole continuous domain D of the ordered medium in which the order parameter is well defined, since the order parameter is expected to vary continuously in D.  Therefore any continuous displacement of $\gamma$ in D maps on a continuous displacement of $ \Gamma$.  There is consequently a relation of equivalence between the different images $\Gamma$s; it is this equivalence that is described by the notion of homotopy, \cite{stee57}.  All the $\Gamma$s belonging to the same class of equivalence are represented by an element $\left[\Gamma \right]$ of a group, the so-called fundamental group (or first homotopy group) $\Pi_{1}(V)$.  More precisely, $\Pi_{1}(V)$ is the group of classes of oriented loops belonging to V, equivalent under homotopy, and all having the same base point. This latter technicality has no incidence on the classification of defects. More important is the remark that, in most cases, the fundamental group is not commutative, a property that is related to the fact that the \textit{topological charge} of a defect, i.e. the corresponding element in $\Pi_{1}(V)$, is modified when the defect $\left[\Gamma \right]$ has circumnavigated about a defect $\left[\Gamma' \right]$; it is changed to $\left[ \Gamma'\right]\left[\Gamma \right] \left[\Gamma'\right]^{-1}$, an element of the same conjugacy class.  It is therefore usual to consider that all the elements of a given class of conjugacy represent the same defect.  Examples are provided later on in this article.

\subsection{The theory of continuous defects}
This theory flourished in the early sixties, see \textcite{nye53}; \textcite{kondo,bilby,kroner};
it was at that time applied to solid crystals only.  It
concentrates mostly {on} the study of sets of line defects, whether
these defects, of the quantized type, are considered at such a
scale that the concept of defect density makes sense, or whether
the characteristic invariants carried by the defects (in the sense
of Volterra, i.e. translations, rotations) are
continuous, with the result that the notion of infinitesimal
defects $-$ with vanishing Burgers vector dislocation or vanishing
rotation vector disclination $-$ is significant. And indeed most of the
results of the theory of continuous defects relates to continuous
distributions of dislocations of infinitesimal strength, to
intrinsic point defects (interstitials, vacancies, etc), and, to a
lesser extent, to disclinations of infinitesimal strength.  The
main geometrical ingredients of the continuous theory of
dislocation line defects and intrinsic point defects are the
concepts of torsion and curvature on a manifold; the points of
contact with the Einstein's theory of generalized relativity are
therefore numerous, and have been stated a number of times
\cite{kroner,hehl}: {cosmic strings are spacetime topological defects that may be described in delta-function valued torsion and curvature components, carried respectively by translation and rotation symmetry breaking defects in a Minkowskian manifold} \cite{vilen94}; for {recent spacetime defect calculations} see {e.g.}
\textcite{letelier95}, \textcite{puntigam97}, {and references therein}. Defects such as disclinations whose
characteristic invariants belong to non-commutative groups cannot
so easily be turned into density sets.  Methods borrowed from the
theory of fields in high-energy physics have also been applied;
this is the so-called \emph{gauge field theory of defects}, cf. \textcite{julia79}, \textcite{dzyalo80}, \textcite{dzyalo81}.

It was expected that these continuous approaches would
open a way to a certain number of problems difficult to attack by
the physics of quantized defects when these defects are plenty. Thus the dynamical theory of dislocations in solids, the defect
melting theory, the defect content of deformed frustrated phases
(e.g. Frank and Kasper phases, quasicrystals, even
amorphous media and glasses, and among mesomorphic blue
phases or {twist grain boundary (TGB)} phases), also would provide a
suitable definition of other defect densities: of disclinations,
of point defects, etc. The rich and interesting courses given at
the Les Houches Summer School on Defects held in 1980
\cite{kroner,dzyalo81} take stock of the various advances in
continuous and gauge field
theory made at the time.

However, although the continuous theory of defects is of
a rare mathematical elegance, the applications have been scarce,
one of the most convincing being perhaps the analysis of
magnetostrictive effects in ferromagnets \cite{kleman67}.  At the Discussion
Meeting he organized in Stuttgart  \cite{discussion},
Kr\"{o}ner made the following remark:\\
\textsf{"Although the field theory of defects }[\textit{by field theory he
obviously meant the traditional theory of continuous defects as
well as its gauge field extension}] \textsf{has found many applications,
the early hope that it could become the basis of a general theory
 of plasticity has not been fulfilled. Among various reasons we mention first of all,
 that the defects namely the dislocations that above all are responsible for plastic flow,
do not form smooth line densities that can well be described by a
dislocation density tensor field.  Direct observation of
dislocations in crystals, for instance by means of electron
microscopy, shows that dislocations rather
 form \underline{three dimensional networks}} [\textit{our sub-lining}] \textsf{that are
 interconnected in practically immobile nodes and other often complex local
 arrangements... These networks have a strong statistical component,
 a fact that shows that a real physical understanding of plasticity
 requires also considerations in the frame of statistical physics.
 However, a statistical theory of interacting deformable lines that
 can be created, annihilated and change their length, has never been worked
 out".}

 To these remarks can be added that the material science
physicists, who certainly know best the problems and the traps of
the physics of defects, are in general not fully aware of the
mathematical tools (non-Riemannian manifolds, exterior calculus,
Grassmann algebra, differentiation on manifolds and fibre bundles)
that are at the very basis of continuous gauge theory.  This
language problem has little chance to be solved in the near
future, inasmuch as the gap between material scientists and field
theorists keeps widening.  By all means, we here avoid as much as
possible complex mathematical tools.

In any case, we are led to conclude that field theory is not a
panacea. On the other hand, as a matter of fact, the fundamentals
of plastic deformation and fracture of crystalline materials have
recently undergone a revival through the development of new
experimental methods that now explore nanometric scales, and
through the improvement of computer power and computing methods.

Dislocations (whose set of characteristic invariants
is isomorphic to an infinite abelian group) and intrinsic point
defects are the dominant defects in solid crystals; there
thereby is no real necessity to introduce in the theory the
group-theoretic description of other types of defects that would
require finite abelian groups or non-abelian groups, making the theory unmanageable. Notwithstanding this simplification, Kr\"{o}ner's criticism stands valid; even if one restricts to
static situations, the continuous theory neglects real
dislocation networks $-$ in the sense of \textcite{
frank50b} $-$, and the frictional effects due to
them. This is probably the true reason why the theory of
continuous dislocations has found so little use yet.

\subsection{Disclinations}
\noindent \textsl{a. Disclinations and continuous dislocations}

 There is
however a situation where dislocation densities retrieve their
true importance and where continuous theory plays a role, it is
with disclinations considered as \emph{singularities of
dislocation and disclination densities}; this is the point of
view which is taken here, and which has some relevance to
mesomorphic phases and perhaps also to quasicrystals, Frank and
Kasper phases, undercooled liquids, polynanocrystals, and more
generally
frustrated phases.

Disclinations can exist in solid crystals, whose
building blocks are atomic, point like; but, as already stated,
the continuous theory has been but little applied to
disclinations in solid crystals, where such objects have a huge
energy. On the other hand, disclinations are the rule in
mesomorphic phases, whose building blocks are anisometric
molecules (rod like, disk like, etc). Again, these disclinations
quite often appear as {the singularity set of a dislocation density}.
Hence the interest in revisiting the continuous theory of
defects, although it appears at once that new concepts have to
emerge. The case of mesomorphic phases requires an
extension of the theory of continuous defects for solids to the
case when there is locally only one physical direction (the
director), i.e. no local trihedron of directions, as in
the uniaxial nematic N, the SmA and the columnar D cases\footnote{The continuous theory of defects makes use of lattice
manifolds, whose points carry local trihedra.}; see also a
related remark in \ref{Nstarphase}. The role of stress relaxation
is especially
important and complex in such mesomorphic phases.\\

\noindent \textsl{b. Three dimensional networks}

The question also arises whether disclinations form \textit{3D networks} in amorphous systems, liquid crystals such as cholesteric Blue Phases, undercooled liquids $-$ as already stated, it has been hypothesized that glass disorder can be described in terms of disclinations in a icosahedral curved crystal \cite{kleman79} $-$ or in nanocrystals, clusters \cite{friedelprague} and quasicrystals, where they are indeed documented \cite{frank58b,frank59, nelson83a}.  In such 3D networks, the disclinations have to be somewhat flexible, which is possible, whether these disclinations are quantized or not, only if other defects, dislocations or disclinations, continuous or not, attach to them.  Thus the question of the interplay between dislocations and disclinations goes beyond mesomorphic phases.  This question will be presented and discussed in this essay.

\subsection{Outline of the article }
 To summarize, the continuous theory of defects, in its
primitive form, only considers dislocation densities, which are singularities of continuous
fields.  It does not consider finite
defects like disclinations, neither grain boundaries, nor Frank
networks. This essay, contrariwise, assumes the coexistence of
finite and infinitesimal defects. Grain boundaries are
introduced.  The field theory instruments are not employed; but
this is possible (see \textcite{kleman82b}, in a
similar context, but in reason of the advances presented here, the results
have to be revisited).  An important aspect of disclinations in
mesomorphic phases is how their flexibility and mobility concur to
the relaxation of stresses imposed by the boundary conditions
(static or dynamic). Both dislocation and disclination densities,
we shall advance, play a leading role in such relaxation processes.

 Section \ref{CONTINUOUSDEFECTS} is about the description, in geometrical terms,
of the defect structure of disclinations, without taking into
account the constraints due to the symmetries of the medium.  Therefore
this section applies to amorphous media and isotropic liquids, but
at this stage it does not provide more than the geometrical tools
to study the role of stress relaxation mediated by continuous
defects in various media, and how it affects disclination line
flexibility and mobility, and the geometrical rules to building
disclination networks. It is therefore directly applicable to solid media only.

The results of section \ref{CONTINUOUSDEFECTS} are employed in section \ref{crystallinesolidsandnanocrystals} to shed a new light on the properties of nanocrystals, and are
extended in section \ref{quantizedDisclinationsinMesomorphicPhases} to
quantized disclinations, examples being taken in mesomorphic phases. This is the place where we discuss the
relationship between topological stability and the kind of
stability that stems from the Volterra process.  Section \ref{quantizedDisclinationsinMesomorphicPhases} and section \ref{FocalConicsinSmA'sasquantizedDisclinations} (more especially devoted to focal conics in SmA's as quantized disclinations), are thereby about the nature of disclinations in partly ordered materials, and their interplay with continuous and$/$or quantized dislocations.

Defects in frustrated phases are discussed in section \ref{DefectsinCurvedMaterials}. taking into account the frustrated local order, it extends to three dimensional \emph{spherical} amorphous media the results obtained in section \ref{CONTINUOUSDEFECTS} for three dimensional \emph{euclidean} amorphous media, making full use of the quaternion representation of the geometry of S$^{3}$.  Important results are (i) that the 'dislocations', like the translational symmetries they break, are non-commutative $-$ we call them \emph{disvections} $-$, (ii) that infinitesimal disclinations (rather than dislocations) are attached to 'twist' finite disclinations. This result emphasizes the role of \emph{disclination networks} in frustrated media. A part of \ref{DefectsinCurvedMaterials} is devoted to the classification of defects in the spherical \{3,3,5\} polytope, which has been used as a template for amorphous media with local icosahedral order. Section \ref{3,3,5decurved:geometry,topology,anddefects} discusses some characteristics of the decurving process of the just alluded to curved media. \\
\indent The discussion in section \ref{Discussion} bears on a comparison between the extended Volterra process and the topological theory $-$ a question that runs as a red thread through the entire essay $-$ and expatiates on the question of plastic relaxation, i.e. the role in various media of continuous $-$ mostly $-$ and quantized defects in stress relaxation.\\

\section{Continuous Defects in Isotropic Uniform Media. Geometrical Interplay between Disclinations and Dislocations}
\label{CONTINUOUSDEFECTS}

An amorphous metal, considered at a scale larger than the atomic
size, is  an example of an isotropic uniform solid medium. The Volterra process allows the consideration of continuous,
non-quantized, dislocations and disclinations that carry
stresses. On the other hand a result of the topological
theory of defects is that these are not topologically stable. The
objects to which this section is devoted are therefore, to the
best, metastable.

 In
subsections \ref{Dislocationcontent} to \ref{Omegaconstant.Genericdisclinationline} we investigate disclinations of finite
strength whose rotation vector $\bm{\Omega}$ is constant in
modulus and direction. Such objects are attended by two types of attached
infinitesimal dislocations (constitutive and relaxation dislocations), {from which the concept of}
\emph{extended Volterra process}. We comment on the equivalence between
these infinitesimal dislocation sets and grain boundaries in
subsection \ref{Disclinationsandgrainboundaries}. In subsections \ref{Polygonaldisclinationlines.Attacheddisclinations} and \ref{Genericdisclinationlines.Disclinationdensities} we investigate the case when
the rotation vector is varying along the line, in direction \emph{and} in
modulus. The ensuing considerations directly yield an expression
for the fundamental invariant of a disclination, which we call the
\emph{Frank vector}. The Frank vector is for disclinations
what the Burgers vector is for dislocations; in particular it
satisfies a Kirchhoff relation at disclination nodes. The detailed study of quantized
disclinations is postponed to sections \ref{quantizedDisclinationsinMesomorphicPhases} and \ref{FocalConicsinSmA'sasquantizedDisclinations}.

This case of anisotropic uniform medium is chosen for the simplicity of the relaxations in the extended Volterra process. As stressed above, it is somewhat artificial to distinguish the relaxations of stresses involved in the motions of a given (non quantized) disclination from that for the dispersion of the disclination itself: the same plastic properties are involved in both and should occur in similar lengths of time in non viscous liquids. It is only for very viscous liquids or better for amorphous solids with slow atomic diffusion that one can reasonably assume the short range relaxation of stresses due to the motion of the dislocations to be more rapid than their dispersion life time, which we shall consider here as infinite. The dynamical competition of the two processes in somewhat viscous liquids has so far not been much studied.

The models developed and applied in sections \ref{CONTINUOUSDEFECTS} and \ref{crystallinesolidsandnanocrystals} will be extended in sections \ref{quantizedDisclinationsinMesomorphicPhases} and \ref{FocalConicsinSmA'sasquantizedDisclinations}, with no such difficulty, to quantized and topologically stable disclinations.

\subsection{Dislocation content of a straight wedge disclination}
\label{Dislocationcontent} 

\begin{figure}
\includegraphics{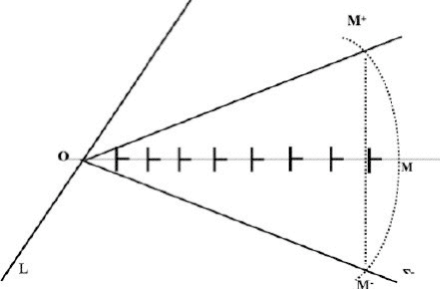}
   \caption{ \label{fig1} {Wedge disclination L and its edge dislocation
content. $\textrm{OM}=\textrm{OM}^{-}=\textrm{OM}^{+}$;
$\textrm{M}^{+}\textrm{M}^{-}=2\textrm{OM}
\sin\dfrac{\Omega}{2}$}}
\end{figure}

Let us consider an infinitely long wedge disclination line L, of
rotation angle $\mathbf{\Omega}=\Omega \, \mathbf{t}$, the
rotation axis \textbf{t} being along L, Fig.~\ref{fig1}.

The Volterra process consists in opening a dihedral void of matter
(for the sake of simplicity we consider a disclination of negative
strength). The relative displacement of the lips of the cut
surface $\Sigma$, which we take to be a half-plane, at a point
\textbf{M} of $\Sigma$, is written as equal to $\,\sin
\dfrac{\Omega}{2} \, \,\mathbf{t} \times  \mathbf{OM}$, \textbf{O}
being any origin on L; this displacement can also be the result of
a set of edge dislocations $-$ \cite{friedeldisloc}, chapter 1
$-$, located uniformly in $\Sigma$, and whose total Burgers vector
$\mathbf{b_{\mathbf{M}}}$ is precisely $2 \, \sin
\dfrac{\Omega}{2} \,\, \mathbf{t} \times \mathbf{OM}$, for those
dislocations lying between the {edge of} the dihedron and
\textbf{M}, i.e. with a density
 \begin{equation} \label{e1}
\emph{d} \mathbf{b}_{\mathbf{M}} = 2\,
\sin\dfrac{\Omega}{2} \enskip\mathbf{t} \times \emph{d}
\mathbf{M}
\end{equation}
thus producing a tilt boundary of rotation $\bm {\Omega}$ along $\Sigma$.

Let us give a few examples:

(i) \textit{In an \underline{amorphous solid} or a glass}
the angle $\Omega$ can take any value; a continuous distribution
of dislocations thereby yields a continuous wedge disclination.
This tilt boundary introduces a mismatch of short range atomic order, which can be suppressed if some local atomic diffusion is allowed at short range.

(ii) \textit{In a \underline{ferromagnetic solid}}, the meeting
line of several magnetic walls is a continuous wedge disclination
whose angle $\Omega$ relates directly to the magnetoelastic
constants \cite{kleman74}. Again, this disclination can be
analyzed in terms of continuous dislocations.

(iii)  \textit{\underline{Non-quantized} wedge
disclinations in a \underline{crystalline} solid} are the limits of \emph{tilt boundaries}, which are split
in the usual way into finite dislocations parallel to that limit; this is the type considered up to now.  $\Omega$ can take any
value; it is tuned by the density of edge dislocations.  For small values of $\Omega$, the continuous distribution of infinitesimal dislocations can also regroup into parallel dislocations of finite strength allowed by the crystal structure (cf \ref{Nanocrystals} for a more detailed discussion). In the
general case, this is an imperfect disclination with a stacking
fault that is a tilt boundary.  Such disclinations have very large energies, as soon as $\Omega$ is finite, in the absence of any
plastic relaxation. They can nevertheless be produced, for
instance when a slip line crosses a low angle grain boundary,
during plastic deformation of polygonized crystals; they can also
be associated in parallel pairs of equal rotation strength and of
opposite signs. The stress concentrations produced at the cores of
the disclinations are often relaxed by the development of cracks \cite{friedeldisloc}.\\

\textit{Remark}: The Volterra process is properly defined for
$|\Omega|\leq\pi$. Observe indeed that Eq.~\ref{e1} does not distinguish
 between (i)- $\Omega=2\pi$ and $\Omega=0$, (ii)-
between the angles  $\Omega=2\pi-\alpha$ and $\Omega=\alpha$. This
equation {puts limits} to the application of the Volterra process and to the use of Eq. 1.  Notice also that it is inconsistent to consider
a unique Volterra process with an angle $|\Omega| \geq 2\pi$, since this requires
removing matter (for $\Omega>0$) or adding matter (for $\Omega<0$) in at
least a full space. Hence, for any angle
$\Omega=(2n+1)\pi+\alpha$, $|\alpha|<\pi$, one has to consider
$(2n+1)$ successive applications of the Volterra process, followed by a Volterra process of angle $\alpha$.\\

\indent We now deepen the relationship of a wedge disclination with its accompanying  dislocations by first considering the displacement of the entire line parallel to itself (
\ref{Emitted/absorbeddislocations} ), secondly displacing a part of the line only (\ref{Twistcomponent}).

\subsection{Emitted/absorbed dislocations. Constitutive and relaxation dislocations}\label{Emitted/absorbeddislocations}

It is clear that the energy of the disclination becomes
prohibitive if the rotation axis \textbf{t} does stay in place when the
disclination is displaced; if some plastic relaxation is allowed, \textbf{t}
moves to a new position by the \emph{emission} or
\emph{absorption} of a certain number of dislocations.  We
have to specify the direction of this motion, as the stress
about L depends on the position of the cut surface $\Sigma$ in
space, even if the medium is isotropic.

\subsubsection{Motion \emph{in} the cut surface.}
\label{thecutsurface}
\begin{figure}
\includegraphics{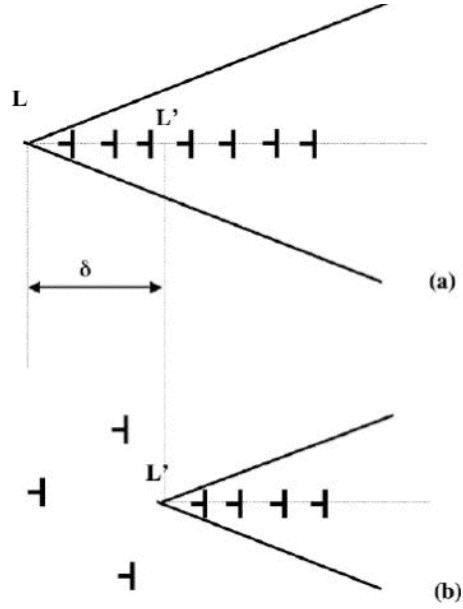}
  \caption{ \label{fig3} {Displacement of a wedge disclination from L (a) to
L' (b) by emission of dislocations that disperse away}}
\end{figure}

If L is moved in the plane of $\Sigma$ by a displacement vector
$\bm{\delta}$, the total Burgers vector
$\mathbf{b}_{\bm \delta}$ of the emitted/absorbed dislocations is
equal to $\pm 2 \, \sin  \tfrac{\Omega}{2} \enskip
\mathbf{t}\times \bm{\delta}$: these are edge
dislocations; $\mathbf{b}_{\bm{\delta}}$ is the sum total
$\mathbf{b}_{\bm{\delta}}=\displaystyle \sum \mathbf{b}_e$ of
elementary dislocations $\mathbf{b}_e$ allowed by the
symmetries of the medium, Fig.~\ref{fig3}.

In an amorphous solid or a glass the
$\mathbf{b}_e$s may have any modulus. In an ordered
solid, the $\mathbf{b}_e$'s have to be equal to
translation symmetries of the medium.

 Notice that in both cases the emitted/absorbed dislocations
contribute to the relaxation of the sample that has suffered the
displacement of the disclination.  In order to illustrate this
point in the amorphous, solid, case, observe that it would make no
sense if the emitted dislocations stay in place in the
continuation of the cut surface, because this would {not modify} the strains and stresses previously carried by the medium,
before the line has moved.  In other words, one is led to
recognize the existence of two types of dislocation densities:
those belonging to the actual cut surface of the disclination,
which we call \emph{constitutive dislocations}, and those
left away in the wake of the moving disclination, which we call
\emph{relaxation dislocations}. In principle, since the
stress field attached to a wedge line L, measured in a frame of
reference attached to L, is independent of the position of L, the
relaxation dislocations should carry no stress at all at complete
relaxation; they are dispersed in the entire space with vanishing
Burgers vectors, if they are continuous, or vanish at the
boundaries of the sample, if they are quantized.

 {Of course, a
part of the stresses carried by the disclination can also be
relaxed by defects that are not attached to the line, e.g.
infinitesimal dislocations nucleated in the bulk; these are Nye's
dislocations, which we study in some detail later when we come to
layered media (smectics) (\ref{Nye'srelaxationdislocations}), and  in Appendix \ref{appnye}. But we shall not consider non-attached defect densities in the present section.  The final result depends on the material properties of the medium, whether it has solid elasticity (amorphous) or viscous behaviour (liquid). In this latter case the relaxation can be complete.}
\subsubsection {Motion \emph{off} the cut surface}
\label{offthecutsurface}
 If L is translated off the plane of $\Sigma$ by ${\bm{\delta}}$, the total
Burgers vector of the absorbed dislocations is also equal to $\pm 2 \, \sin  \tfrac{\Omega}{2} \enskip
\bm{\nu}\times \bm{\delta}$,
where $\bm{\nu}$ is the new axis of rotation.  A new piece of cut
surface, parallel to the $\{\bm{\nu},\bm{\delta}\}$ plane, is created.

\subsection{Twist component of a disclination}
\label{Twistcomponent}
\begin{figure}
\includegraphics{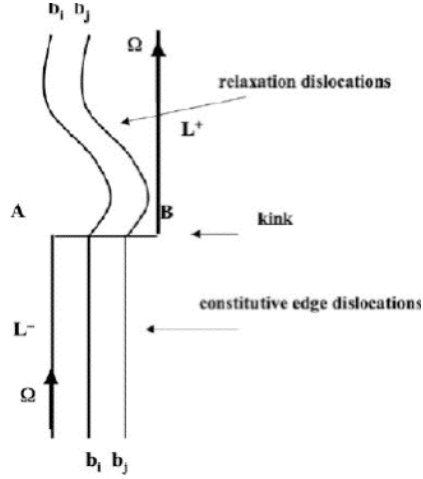}
 \caption{
   \label{fig4}
   {Kink AB linking the two wedge half-lines $\textrm{L}^{-} \,\textrm{and} \,\textrm{L}^{+}$. The cut surface $\Sigma$ is supposed to be on the right of the disclination $\textrm{L}^{-}, \, \textrm{AB}, \,\textrm{L}^{+}$.
The upper segments of the dislocations that traverse AB tend to disperse away (plastic relaxation), while keeping attached to the constitutive segments on the kink. }}
\end{figure}

We now {turn} our attention to a line L made of three segments, namely
two parallel semi-infinite wedge segments $\textrm{L}^{-}$ {and} $ \textrm{L}^{+}$,
joined by a third {perpendicular} segment $\mathbf{AB}$ of small length,
called a \emph{kink}, Fig.~\ref{fig4}. We assume that the cut surface $\Sigma$ is a
plane that contains the three segments.

The $\textrm{L}^{+}$ segment is the result of a displacement of a part of
the entire line parallel to itself by a translation $\bm{\delta
}= \mathbf{AB}$, by emitting or absorbing
dislocations.  In this process, $\mathbf{\Omega}$ stays parallel to itself.
According to the results above, we have, on the $\Sigma$ side of
\textbf{AB}, \emph{constitutive dislocations} of total Burgers vector
 \begin{equation} \label{e2}
\mathbf{b}_{e}(\mathbf{AB}) =
\pm 2 \thinspace \sin  \dfrac {\Omega}{2}\enskip \mathbf{t} \times \mathbf{AB}
\end{equation}

\noindent on the other side of $\mathbf{AB}$, \emph{relaxation
dislocations} of total Burgers vector
\begin{equation}  \label{e3}
\mathbf{b}_{{d}}(\mathbf{AB}) =
\pm 2 \sin  \dfrac {\Omega}{2} \enskip \mathbf{t}\times
\bm{\delta}
\end{equation}

These two quantities being equal, we see that the dislocations
cross the segment $\mathbf{AB}$, but behave quite
differently on either side.

A classical
Volterra process, acting once for all on the cut surface of the
disclination line, is not relevant to the present geometry,
because such a Volterra process can be performed only if the rotation vector
$\bm{\Omega}$ is fixed in space.

\subsection{$\mathbf \Omega$ constant.  Generic disclination line}
\label{Omegaconstant.Genericdisclinationline}

Again, we restrict {our attention} to the isotropic case.  Consider a curved
disclination line: Fig.~\ref{fig5} shows a disclination with $\bm{\Omega}$
constant in length and in direction from point to point
along L.  The angle between L and $\bm{\Omega}$ varies from point to point. Let $\mathbf{P}$ and $\mathbf{Q} =
\mathbf{P} + \dfrac {d\mathbf{P}}{ds}
ds$ be two points infinitesimally close on \textrm{L}.  The Burgers vector of the
infinitesimal dislocation introduced by the variation of position
of $\mathbf{\Omega}$ from $\mathbf{P}$ to
$\mathbf{Q}$ is, by reasoning on the cut surface
$\Sigma$ as above, equal to $\mathrm{d}\mathbf{b}_{\mathrm{PQ}}=\mathbf{d}_{\mathrm{Q}}(\mathbf{M})-\mathbf{d}_{\mathrm{P}}(\mathbf{M})$, where $$\mathbf{d}_{\mathrm{P}}(\mathbf{M})=2 \sin \dfrac {\Omega}{2}\, \mathbf{t} \times \mathbf{PM},$$ $$ \mathbf{d}_{\mathrm{Q}}(\mathbf{M})=2 \sin \dfrac {\Omega}{2}\, \mathbf{t} \times \mathbf{QM}$$
are the displacements of the cut surface at \textbf{M}, any point on $\Sigma$, seen respectively from \textbf{P} and \textbf{Q}. Hence:
\begin{equation} \label{e4}
\mathbf{d}_{\mathrm{Q}}(\mathbf{M})-\mathbf{d}_{\mathrm{P}}(\mathbf{M})
=-2 \sin  \dfrac {\Omega}{2} \enskip\mathbf{t} \times  \frac{d\mathbf{P}}{ds} ds
\end{equation}

\noindent This dislocation, which we denote $d\ell_{\mathrm{PQ}}$, can be
thought of as attached to the line at the infinitesimal arc \textbf{PQ}. Of course
$d\mathbf{b}_{\mathrm{PQ}}$ has to be a translation allowed by the symmetry of the phase.  The shape taken by $d\ell_{\mathrm{PQ}}$ resulting from plastic relaxation optimizes the energy carried by the disclination.

\subsubsection{Two types of dislocation continuous
distributions} \label {Two types of dislocation}
The question arises whether any infinitesimal relaxation
dislocation $d\ell_{\mathrm{PQ}}$, attached to L at the arc
\textbf{PQ}, crosses the line L, and transforms
\begin{figure}
\includegraphics{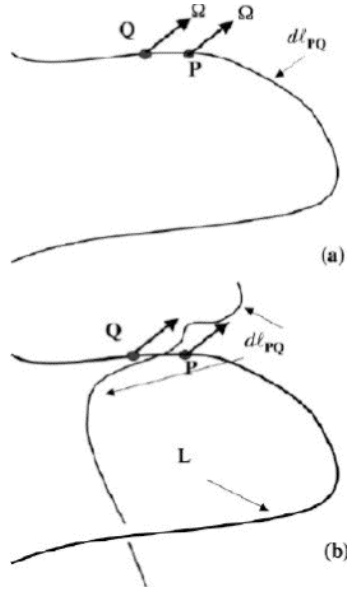}
   \caption{ \label{fig5}{(a)- The infinitesimal dislocations
$d\ell_{\mathrm{PQ}}$ are constructed assuming first that the cut surface
bound by \textrm{L} is \textit{common} to all the $d\ell$s attached along L.  (b)- Then each  $d\ell_{\mathrm{PQ}}$ is
deformed at fixed P, Q; the two parts of $d\ell_{\mathrm{PQ}}$ on both
sides of the infinitesimal arc \textrm{PQ} do not have the same elastic distribution.  L is the
limit between the two distribution types}}
\end{figure}

on the other side of
L into a constitutive infinitesimal dislocation, with the
same Burgers vector, as it does in the simple case investigated in
II.C.  The answer is positive and the demonstration is as follows.
\textbf{M} being on the cut surface $\Sigma$ of L, Eq.~\ref{e4} is as a result valid on the full area of $\Sigma$.  In other words, the
infinitesimal dislocation line $d\ell_{\mathrm{PQ}}$ with Burgers
vector $db_{\mathrm{PQ}}$ has the same cut surface as L, and consequently meets
L, is fixed between P and Q, and is closed in the manner of L, Fig.~\ref{fig5}(a).  In fact,  with the local rotation
vector defined as above, the disclination L is the result of creating
a density of infinitesimal $d\ell_{\mathrm{PQ}}$ by infinitesimal Volterra processes on the same cut surface.

Now let us deform these $d\ell_{\mathrm{PQ}}$ dislocation lines
(this is an allowed operation), opening them into infinite lines
(or line segments ending on the boundaries of the sample) in such
a way that now they cross the imaginary line L, which divides each
of them into two semi infinite arcs.  This process traces out the bounds of the cut surface $\Sigma$ along L if one
imposes different types of distribution for the line arcs on both
sides of L, yielding different elastic distributions.  In the spirit of Fig. \ref{fig4}, one can imagine on one side a 2D surface tiled with constitutive dislocations, a kind of generalized misorientation boundary, on the other
side, relaxation dislocation segments dispersed in space.
See Fig.~\ref{fig5}(b).

\subsubsection{Line tension of twist \textit{vs.} wedge segments}  \label{Linetensionoftwist}
Notice that we have realized a dislocation geometry that {display}
\textit{two} metastable configurations of a set of infinitesimal
dislocations, namely those of the misorientation boundary and
those fully dispersed. Therefore the core region of a twist
disclination line, where these two configurations merge, has a
contribution to the total energy that scales as its length.  The
other contribution is the energy of the dislocation lines,
essentially that of the constitutive dislocations, which
scales as the area of the cut surface, i.e., the square of the
length of the line.  One expects that this contribution is larger
than the first one.  Therefore the line energy per unit length of
line is proportional to the length of the line.  The latter
contribution is worth comparing with the energy per unit length of
a wedge line, which scales as the square of the transverse size: the line tension of a disclination is thus a linear function of their length which reduces to a proportionality for wedge ones, whereas, for twist ones, one must add a small constant term due to their relaxed dislocations.

\subsection{Disclinations and grain boundaries}
\label{Disclinationsandgrainboundaries} 

As in crystalline solids (see \ref{Dislocationcontent}), wedge lines in
solid amorphous materials carry a large energy, except in the same
special circumstances as indicated above. The existence of twist
lines or partially twist lines is even less probable, and their
mobility and change of curvature {is }certainly negligible, since it
would require the climb of the attached dislocations, which
requires plastic deformation.  Hence a \textit{caveat}: except in the later case of polycrystals, the discussion that
follows assumes implicitly that there is no restriction to
reaching low energy states by plastic relaxation; it therefore
applies to some sort of amorphous material endowed with a finite
viscosity, which operates through the existence and mobility of infinitesimal relaxation dislocation densities. One does not expect that such processes are possible in a solid crystal. But the comparison between crystals and amorphous media is worth carrying out, especially through the parallel concepts of grain boundary and cut surface.

 \subsubsection{Frank's grain boundary and Friedel's
disclination compared}   \label{Frank'sgrainboundaryandFriedel'sdisclinationcompared} Equation~\ref{e1}, integrated
along a segment \textbf{MN} of the disclination line L, gives
the total Burgers vector of the dislocations that are \emph{attached} to any segment
\textbf{MN} of L (and lie along its cut surface $\Sigma_{L}$).
\begin{equation} \label{e6}
\Delta \mathbf{b}_{\mathbf{MN}}=2 \sin
\dfrac {\Omega}{2} \enskip\mathbf{t} \times
\mathbf{MN}
\end{equation}

This
expression is similar to Frank's formula \cite{frank50b},
for a crystal grain boundary
$\Sigma_{GB}$ of angle of misorientation $\bm{\omega} = \omega \, \mathbf{s}$, $|\mathbf{s}|=1$; Frank's
formula yields the total Burgers vector of the (quantized) dislocations that
cross any segment \textbf{PQ} belonging to $\Sigma_{GB}$.
\begin{equation} \label{e7}
\Delta \bm{\beta}_{\mathbf{PQ}}=2 \sin
\dfrac {\omega}{2} \enskip\mathbf{s} \times
\mathbf{PQ}
\end{equation}

This similarity does not come as a surprise, after the foregoing
discussion.  In particular, one can deduce from Frank's approach
that any closed line in a grain boundary can be chosen as a
disclination line, provided the 'exterior' dislocation segments
are let to relax. Identifying Eq.~\ref{e6} and Eq.~\ref{e7} and assuming
$\mathbf{M}=\mathbf{P}$, $\mathbf{N}=\mathbf{Q}$, we have
\begin{equation} \label{e8}
\sin  \dfrac {\Omega}{2} \enskip \mathbf{t} \times
\mathbf{MN} = \sin  \tfrac
{\omega}{2} \enskip \mathbf{s} \times \mathbf{MN}
\end{equation}

\noindent which yields either
\begin{equation} \label{e9}
\sin \dfrac {\Omega}{2} \enskip \mathbf{t}-\sin
 \dfrac {\omega}{2} \enskip \mathbf{s}=0
\end{equation}

\noindent or
\begin{equation} \label{e10}
\sin \dfrac {\Omega}{2} \enskip \mathbf{t}-\sin
 \dfrac {\omega}{2} \enskip
\mathbf{s}\propto\mathbf{{MN}}
\end{equation}

Assume that $\bm{\omega}$ is a constant vector, i.e.
that the cut surface of the disclination line belongs to a unique
grain boundary.  Equation \ref{e9}, which yields $\Omega =
\omega$, $\mathbf{t} =\mathbf{s}$, applies when $\bm{\Omega}$ is a
constant vector, which is what we have assumed up to now.  Equation \ref{e10}
on the other hand expresses that other possibilities exist, with
$\Omega \neq \omega$, $\mathbf{t} \neq \mathbf{s}$, with the tangent
$\mathbf{t}$ to the disclination ($\mathbf{MN} =
\bm{\tau} ds$) belonging to the plane \{$\mathbf{t}$, $\mathbf{s}$\}.
Although $\bm{\omega}$ is a constant vector, Eq.~\ref{e10}  applies to a
variable $\bm{\Omega}$, a situation more fully discussed in
\ref{Polygonaldisclinationlines.Attacheddisclinations} and \ref{Genericdisclinationlines.Disclinationdensities}. We give later on an important example obeying
Eq.~\ref{e10} (see Appendix \ref{appell}).\\

\textit{Remark}: The $2\,\sin\dfrac{\Omega}{2}$ (or $2\,\sin\dfrac{\omega}{2}$) factor in Eq. \ref{e1}, \ref{e6}, or \ref{e7}  deserves some comments.

\begin{figure}
\includegraphics{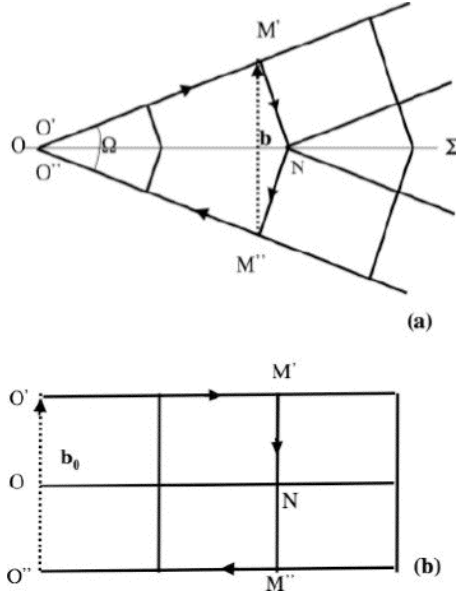}
 \caption{ \label{fri}{Burgers vector. A crystal structure has been superimposed to show more clearly the
 Burgers circuit O'M'NM"O"OO' $-$ in the final state (a) and in the initial state (b) $-$ and the rotation $\bm \Omega$
 of the Volterra process
}}
\end{figure}

The integral Burgers vector $\mathbf{b}$ of the cut $\Sigma$ is counted as a lack of closure of a Burgers circuit in the final state (more properly in the final state of the Volterra process before elastic relaxation of the sector M'OM" here introduced, Fig.\ref{fri}a).

It is usual for dislocations to count the Burgers vector $\mathbf{b}_{0}$ as a lack of closure in the initial state; one would then have (Fig.\ref{fri})
\begin{center}
$|\mathbf{b}|\,=\,${M}'{M"}\,=\,2\,OM'\,$\sin\dfrac{\Omega}{2}$
\end{center}
\noindent and\begin{center}
$|\mathbf{b}_{0}|\,=\,${OO}'\,=\,2\,OM' \,$\tan\dfrac{\Omega}{2}.$
\end{center}

Such a difference originates from the non-commutativity of the Burgers circuit and the
rotation of the disclination. Obviously, the formula in $2\,\sin\dfrac{\Omega}{2}$ is to be preferred as it correctly describes the final state, the only one of interest here. The difference is only noticeable for large $\Omega$s, where the Burgers vector $\mathbf{b}$ of the constitutive dislocation is smaller by a factor  $\cos \frac{\Omega}{2}$ than that $\mathbf{b}_{0}$ of a    crystal dislocation.
 In the sequel we use the Burgers vector formula $\mathbf{b}$; {it appears as follows in the Frank vector introduced in \ref{Generalizedpolygonalloops.Kirchhoff'srelation.}}:
 \begin{equation} \label{efri}
\mathbf{f}(\bm \Omega)=2\,\sin \dfrac {\Omega}{2} \enskip \mathbf{t}.
\end{equation}

 \subsubsection{Isolated twist segments}  \label{Isolatedtwistsegments}  Equation~\ref{e7}  has been
derived by Frank for a finite segment
$\mathbf{PQ}$, independently of any loop to which
${\mathbf{PQ}}$ could pertain; the dislocations
which cross this segment (the crossing set) in the grain boundary
have two parts (relaxation on one side of
${\mathbf{PQ}}$, constitutive on the other), with
no violation of any conservation law on Burgers vectors.  Such a
segment and its set of attached dislocations constitute a sort of
\emph{stripe} in the grain boundary, geometrically and
topologically independent of it (certainly not energetically, but
we shall pay no attention to this question).

\begin{figure}
\includegraphics{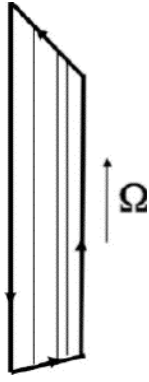}
   \caption{\label{fig6}{An elementary disclination of mixed character. The
constitutive dislocations (inside) are drawn, but not the
relaxation dislocations (outside)}}
\end{figure}

Therefore, proceeding with our analysis, we conclude that a
continuous disclination line can be made of a sequence of
independent segments $\mathbf{M}_{i}\mathbf{M}_{i+1}$, possibly
infinitesimal, each of them carrying a different rotation vector
$\bm{\Omega}_{i,i+1}$. Each stripe
$\mathbf{M}_{i}\mathbf{M}_{i+1}$ defines a grain boundary of
finite width.  Therefore, any isolated, finite, segment MN can
divide each of the dislocations (continuous or quantized)
belonging to the crossing set into constitutive (on one side) and
relaxation (on the other side) dislocation segments. Each stripe
has two edges parallel to $\bm{\Omega}$ and two edges of mixed
character, see Fig.~\ref{fig6}.

A stripe is an elementary disclination, with a twist (or mixed, but
not pure wedge) segment transversal to the stripe. The
longitudinal boundaries of the stripe, along the dislocations that
construct the stripe, can be considered as wedge segments.  Thus,
a stripe, considered as a disclination loop, is necessarily of
mixed character.

\subsection{Polygonal disclination lines. Attached disclinations}
\label{Polygonaldisclinationlines.Attacheddisclinations} 
Disclinations with a rotation vector varying in length and
direction are possible; they require attached disclinations (or
attached disclination densities). For the sake of clarity we do
not introduce disclination densities straightaway and develop the
theory for attached disclinations of finite strength. A
disclination can be thought of as the sum of infinitesimal stripes
that tile its cut surface and partition it into long stripes,
elongated along the constitutive dislocations; the situation is
reminiscent of the tiling of the cut surface of a dislocation into
elementary dislocation loops.
 \subsubsection{Wedge polygonal loops and bisecting
disclination lines} \label{Wedgepolygonalloops}

Consider for instance Fig.~\ref{fig7}(a),
which represents a disclination
line made of two semi-infinite \textit{wedge} segments L and L'.
The stripes divide into two sets \textit{parallel} to the
disclination segments; the continuity of the constitutive
dislocations is insured, when $\mathbf{\Omega}'=\mathbf{\Omega}$, if the twist edges of the stripes are
along the line that bisects L and L'; it is indeed easy to show
that the Burgers vector of the constitutive dislocation segments
parallel to L and L' are continuous across this line.

A \emph{wedge loop} is nothing else than a continuous
generalization of Fig.~\ref{fig7}(a), applied to a closed polygonal
disclination; $\mathbf{\Omega}$ is constant in length and everywhere tangent to the loop,
the constitutive dislocations close into loops entirely located in
the cut surface.

The \emph{bisecting line} just introduced has a special stability because it has no relaxation dislocations attached to it. It is indeed a \emph{wedge} disclination, as established by the analysis that follows.

 \subsubsection{Disclinations meeting at a node. Kirchhoff relation. Frank vector}  \label{Generalizedpolygonalloops.Kirchhoff'srelation.}
\begin{figure}
\includegraphics{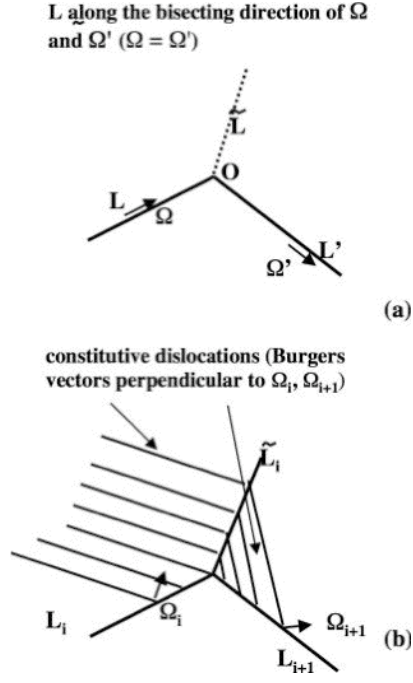}
   \caption{\label{fig7}{A polygonal disclination; (a) two semi-infinite
wedge lines L and L' meeting at \textbf{O}; (b) polygonal disclination made
of segments $\textrm{L}_{i }$ of mixed character (twist-wedge);
the Burgers vectors are continuous across the line parallel to
$\bm{\tilde{ \tau}}_{i}$}}
\end{figure}

 The foregoing considerations generalize to a polygonal
disclination made of segments $\textrm{L}_{i}$ of mixed character
(twist-wedge), with $\mathbf{\Omega}_{i}$  varying in direction but also
in modulus ($\mathbf{\Omega}_{i} \neq \mathbf{\Omega}_{j}$), Fig.~\ref{fig7}(b).
The considerations that follow apply to a situation where there
are no restrictions on $\bm \Omega_i$, $\bm \Omega_{j}$, and the directions
of the segments ${\textrm{L}}_{i}$, ${\textrm{L}}_{j}$; they are not
necessarily in the same plane.  Let
${\textrm{L}}_{i}(\mathbf{\Omega}_{i})$,
${\textrm{L}}_{i+1}(\mathbf{\Omega}_{i+1})$ be two consecutive
disclination segments. The constitutive dislocation segments meet
without discontinuity of the Burgers vector on the half line
parallel to the direction
\begin{equation} \label{e11}
\bm{\tilde{ \tau}}_{i}=2\,\sin \frac{\Omega_{i}}{2}\enskip \mathbf{t}_{i}-2\,\sin \frac{\Omega_{i+1}}{2}\enskip \mathbf{t}_{i+1},
\end{equation}
as, following Eq. \ref{efri}, the density of dislocations must be counted along the common edge $\widetilde {\textrm{L}}_{i}$ of the two boundaries.

In effect, if $\mathbf{O}_{i}\mathbf{P}$ is parallel to
$\bm{\tilde{ \tau}}_{i}$ , we have
\begin{equation} \label{12}
2\,\sin \frac{\Omega_{i}}{2}\enskip \mathbf{t}_{i} \times
\mathbf{O}_{i}\mathbf{P}=2\,\sin \frac{\Omega_{i+1}}{2}\enskip \mathbf{t}_{i+1}\times
\mathbf{O}_{i}\mathbf{P}
\end{equation}

Equation~\ref{12} means, according to Frank's formula, that the
Burgers vector is continuous across any segment parallel to
$\bm{\tilde{ \tau}}_{i}$.  The half line
$\widetilde{\textrm{L}}_{i}$ is obviously a generalization of the
bisecting line introduced just above. Notice that in general the planes
$\{\widetilde{\textrm{L}}_{i},\bm{\tilde{ \tau}}_{i} \}$ and
$\{\widetilde{\textrm{L}}_{i+1},\bm{\tilde{ \tau}}_{i+1} \}$
are not tilt planes for the constitutive dislocations;
this happens only if $\mathbf{\Omega}_{i}$, $\mathbf{\Omega}_{i+1}$, and
the directions of the segments $\textrm{L}_{i}$,
$\textrm{L}_{i+1}$ are all four coplanar directions.\\

\noindent \textsl{a. {Kirchhoff relation.}}  \label{Kirchhoffrelation}Let us introduce, instead of the rotation vector
$\widetilde{\bm \Omega}_i=\widetilde{\Omega}_i\,\tilde{\mathbf{t}}_i$, the expression
\begin{equation} \label{13}
\bm{\tilde{\tau}}_{i}=2\,\sin
\frac{\widetilde{\Omega}_{i}}{2}\enskip\mathbf{ \tilde{t}}_{i}
\end{equation}

 Equation \ref{e11} then takes the form
\begin{equation} \label{14}
2\,\sin \frac{\Omega_{i}}{2} \, \mathbf{{t}}_{i}=2\,\sin
\frac{\widetilde{\Omega}_{i}}{2}\,
\tilde{\mathbf{t}}_{i}+2\,\sin \frac{\Omega_{i+1}}{2}\, \mathbf{t}_{i+1}
\end{equation}

\noindent which can be written
\begin{equation} \label{15}
\mathbf{\Omega}_{i}=\widetilde{\mathbf{\Omega}}_{i}+\mathbf{\Omega}_{i+1}
\end{equation}

\noindent if the rotation angles are small.  In these expressions, the signs
are such that the line $\textrm{L}_{i}$ is oriented inward (toward
$\mathbf{O}_{i}$), whereas $\textrm{L}_{i+1}$ and
$\widetilde{\textrm{L}}_{i}$ are oriented outward.  If the
orientations are so chosen that they all are outward or all
inward, one gets
\begin{equation} \label{e16}
\sum_{\mathrm{P}} 2\,\sin \frac {\Omega_\mathrm{P}} {2}\enskip
\mathbf{t_{\mathrm{P}}}=0.
\end{equation}

\noindent This equation is valid for any number of
disclination segments meeting at the same point
\textbf{O} (see below).

Equation \ref{e16} does not contain any reference to the directions of
the disclinations segments meeting in \textbf{O}; it is akin to a
\emph{Kirchhoff relation} for the vectors  $2\,\sin \frac {\Omega_\mathrm{P}} {2}\,
\mathbf{t_{\mathrm{P}}}$. Notice
in particular that Eq.~\ref{15}, which is valid for small angles,
obtains straightforwardly by considering Frank circuits about the
lines $\textrm{L}_{i}$, $\textrm{L}_{i+1}$ and $\widetilde{\textrm{L}}_i$. We
shall christen the vector $\mathbf{f}=2\, \sin
\frac {\Omega}{2}\,\mathbf{t}$ that plays for a disclination line the same role
as the Burgers vector \textbf{b} plays for a dislocation line, the \emph{Frank
vector}. It is oriented in the same direction as
the \emph{rotation vector} $\Omega \,\mathbf{t}$.

By insuring that line $\widetilde{\textrm{L}}_i$ has no attached relaxed dislocations, condition \ref{12} expresses the fact that $\widetilde{\textrm{L}}$ has a wedge character. Such a choice of $\widetilde{\textrm{L}}$ assumes that the twist lines have a larger line tension than the wedge ones because of their core energy, as discussed in \ref{Disclinationsandgrainboundaries}.\\

\noindent \textsl{b. {Lines meeting at a node.}} \label{Linesmeetingatanode}
\begin{figure}
\includegraphics{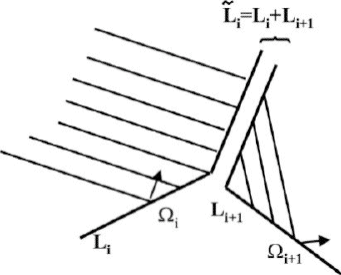}
   \caption{\label{fig7bis}{Polygonal disclination as in Fig. \ref{fig7}: splitting of
$\widetilde{\textrm{L}}_{i}$}}
\end{figure}

We now deepen the disclination
nature of $\widetilde{\textrm{L}}_i$ in the case considered above when three
disclinations meet in $\mathbf{O}_{i}$. As defined above, it is
obviously a \textit{wedge} line $-$ the Frank vector
$\tilde{\mathbf{f}}_i= 2\,\sin
\dfrac{\widetilde{\Omega}_{i}}{2}\, \tilde{\mathbf{t}}_i$ is along the line $-$
that is split into \textit{two mixed} lines, with
$\tilde{\mathbf{f}}_i=2\,\sin
\dfrac{\widetilde{\Omega}_{i}}{2}\, \tilde{\mathbf{t}}_i$ Frank vectors and
$\tilde{\mathbf{f}}_{i+1}=2\,\sin
\dfrac{\widetilde{\Omega}_{i+1}}{2}\,\tilde{\mathbf{t}}_{i+1}$, Fig.~\ref{fig7bis}.
Again, there are no relaxation dislocations along $\widetilde{\textrm{L}}_i$. It is therefore a very
special wedge line.

Now, line fluctuations $\delta \widetilde{\textrm{L}}_i$ would
break the continuity of the Burgers vector on $\widetilde{\textrm{L}}_i$,
and generate relaxation dislocations with Burgers vector $$\delta
\mathbf{b}= 2\,\sin
\frac{\widetilde{\Omega}_{i}}{2}\,
\tilde{\mathbf{t}}_i \times \delta \widetilde{\textrm{L}}_i -  2\,\sin
\frac{\widetilde{\Omega}_{i+1}}{2}\,
\tilde{\mathbf{t}}_{i+1} \times \delta \widetilde{\textrm{L}}_i,$$ according to
Eq.~\ref{e1} ($\delta \mathbf{b}= \delta \mathbf{b}_i-
\delta \mathbf{b}_{i+1}$), by adding the effects of the
two lines involved in the splitting),
i.e., from Eq.~\ref{e11},
\begin{equation} \label{e17}
\delta
\mathbf{b}= 2\,\sin
\frac{\widetilde{\Omega}_{i}}{2}\,
\tilde{\mathbf{t}}_i \times \delta \widetilde{\textrm{L}}_i -  2\,\sin
\frac{\widetilde{\Omega}_{i+1}}{2}\,
\tilde{\mathbf{t}}_{i+1} \times \delta \widetilde{\textrm{L}}_i,\ \\ =\tilde{\bm \tau}_i \times \delta \widetilde{\textrm{L}}_i
\end{equation}

\noindent from which equation it emerges that $\tilde{\bm \tau}_i$
is the Frank vector of the $\widetilde{\textrm{L}}_i$ disclination. These
considerations also confirm that Eq.~\ref{e16} is a Kirchhoff relation
at a node where three disclinations meet.

\begin{figure}
\includegraphics{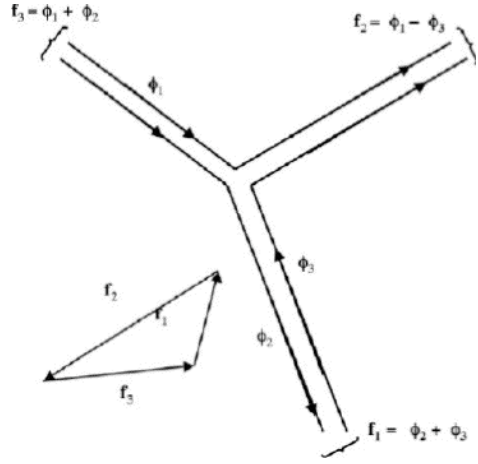}
    \caption{\label{fig8}{Three broken disclinations of Frank vectors
$\bm{\phi}_{1}$, $\bm{\phi}_{2}$, $\bm{\phi}_{3}$, meeting at a
point and composing three disclinations obeying Kirchhoff
relation $\mathbf{f}_{1} + \mathbf{f}_{2} + \mathbf{f}_{3} = 0$ (insert)}}
 \end{figure}

Notice that our reasoning has given a specific role to
one of the three disclinations, $\widetilde{\textrm{L}}$, but this
restriction can be easily removed.  Consider three Frank vectors
$\mathbf{\phi}_{1}$, $\mathbf{\phi}_{2}$, $\mathbf{\phi}_{3}$, and
construct three disclination segments meeting at a common point \textbf{O},
such that:
\begin{equation} \label{e18}
\mathbf{f}_{1} = \mathbf{\phi}_{2} + \mathbf{\phi}_{3}, \quad
\mathbf{f}_{2} =-\mathbf{\phi}_{3} + \mathbf{\phi}_{1}, \quad\mathbf{f}_{3} = -
\mathbf{\phi}_{1} -\mathbf{\phi}_{2}.
\end{equation}
This can easily be done by giving a sharp corner to the
three disclinations ${\bm{\phi}}_{i}$ at \textbf{O} and joining their
straight segments two by two, as in Fig.~\ref{fig8}.  If all the segments
$\textrm{L}_{i}$ are along $\mathbf{f}_{i}$, the $\textrm{L}_{i}$s
are all of wedge character, and there are no relaxation
dislocations attached to them. The appearance of
relaxation dislocations attached to the $\textrm{L}_{i}$s
would make them change direction.  The final geometry depends on the
energy balance between the grain boundaries (the constitutive
dislocations), the relaxation dislocations, and the core energy of
the disclinations. Notice that the three
$\textrm{L}_{i}$s are coplanar if they are of wedge character.

The extension to any number of disclinations is obvious, which
justifies our claim concerning Eq.~\ref{e16}. Notice however that, whereas Fig. \ref{fig8} represents three grain boundaries merging two by two along three disclinations, the case of \textit{four} disclinations (say) meeting at a node requires, in the most general case, four grain boundaries merging three by three along the four disclinations;  each of them is then split into three subdisclinations. Such a geometry occurs by nature in ideal polynanocrystals, see \ref{Structure of an ideal nc}.

\textit{Remark:} In accordance with the remark at the
end of \ref{Dislocationcontent}, Eq. \ref{e16} does not apply properly if one of
the angles $|\Omega_{p}|>\pi$.

 \subsubsection{Disclinations merging along a line}  \label{Disclinationsmergingalongaline}
The situation
where three lines $\textrm{L}_{1}$, $\textrm{L}_{2}$,
$\textrm{L}_{3}$ merge along a unique line L is also worth
considering.  One expects that the Kirchhoff relation
\begin{equation} \label{Kirmerg} \mathbf{f}_{1} + \mathbf{f}_{2} +
\mathbf{f}_{3} = 0 \end{equation} is satisfied. This case is physically
represented in a ferromagnet by three Bloch walls merging along a Bloch line, the
dislocations being the sources of the magnetoelastic stresses
\cite{kleman74}. More generally, one expects that, in an \emph{amorphous medium}, $n$ disclinations $\cdots  \mathbf{f}_{i} \cdots $ merging along a line yield a unique disclination of Frank vector $\mathbf{f} = \sum_{i}\mathbf{f}_{i}$.

\subsection{Generic disclination lines.  Disclination densities}
\label{Genericdisclinationlines.Disclinationdensities} 
We now come to the generic case when a disclination line L is
smoothly curved and its Frank vector varies smoothly.  Consider
two infinitesimally close points \textbf{P} and \textbf{Q} on L, with Frank vectors
$\mathbf{f}_\mathrm{P}$ and $\mathbf{f}_\mathrm{Q}$; we write, with
obvious
notations,
\begin{equation} \mathrm{Q}=\mathrm{P}+\mathbf{s} \,
\delta s, \qquad \mathbf{f}_{\mathrm{Q}}- \mathbf{f}_{\mathrm{P}}  = 2\,\sin
\frac {\Omega_\mathrm{Q}}{2} \, \mathbf{t}_\mathrm{Q} - 2\,\sin
\frac {\Omega_\mathrm{P}}{2} \,\mathbf{t}_\mathrm{P}, \\\quad \mathbf{s}_{\mathrm{Q}}-\mathbf{s}_{\mathrm{P}}
= \frac{d\mathbf{s}}{ds} \delta s =\frac{\mathbf{n}}{R} \delta s.\end{equation}
The
variation between \textbf{P} and \textbf{Q} of the displacement on the cut surface
of L in \textbf{M}, can be written:
\begin{equation} \label{e19}
\delta\mathbf{b}=\mathbf{f}_{\mathrm{Q}}\times \mathbf{QM}- \mathbf{f}_{\mathrm{P}}\times
\mathbf{PM}  \\ =- \mathbf{f}_{P}\times\mathbf{s} \, \delta s+(\mathbf{f}_{\mathrm{Q}}-
\mathbf{f}_{\mathrm{P}}) \times \mathbf{PM}
\end{equation}

The first term ($-\mathbf{f}_{\mathrm{P}}\times\mathbf{s}\,\delta s$) measures
the relaxation dislocation densities attached to the line between
\textbf{P} and \textbf{Q}.  We specialize to the second one, which measures the
relaxation disclination densities. According to Kirchhoff's
relation, we have $-\delta \mathbf{f}=\mathbf{f}_{\mathrm{Q}}- \mathbf{f}_{\mathrm{P}}$, the
sign being chosen so that the attached disclination densities
$\displaystyle \dfrac{d\mathbf{f}}{ds}$ are oriented outward (as
$\mathbf{f}_{\mathrm{Q}}$), and $\mathbf{f}_{\mathrm{P}}$ inward.

If we assume that L is a wedge disclination, i.e.
$\mathbf{t}_{{\mathrm{P}}} =\mathbf{s}_{{\mathrm{P}}}$,
$\mathbf{t}_{{\mathrm{Q}}} =\mathbf{s}_{{\mathrm{Q}}}$, we have
\begin{equation} \label{e20}
\delta \mathbf{f}=\delta \Omega_{\mathrm{P}}\,\mathbf{t}_{\mathrm{P}}+
\frac{2}{R}\,\sin
\frac {\Omega_\mathrm{P}}{2}\,\mathbf{n}\, \delta s
\end{equation}

The first term of the right member
($\delta \Omega_{{\mathrm{P}}}\,\mathbf{t}_{{\mathrm{P}}}$) measures the effect
of the variation in modulus of the rotation vector. In its
absence, the rotation vector of the attached disclinations is
along the principal normal. This is obviously reminiscent of the
bisecting disclination line. If the attached disclinations have a
wedge character, i.e. are along the principal normal, again as
above, there are no supplementary dislocations accompanying them,
apart those constituting L, and we might expect that the energy is
minimized. Another result, not visible in the previous analysis
(\ref{Generalizedpolygonalloops.Kirchhoff'srelation.}), is that the curvature of the disclination line L is
directly related to the presence of the attached disclinations.

In the generic case ($\mathbf{t}_{{\mathrm{P}}} \neq
\mathbf{s}_{{\mathrm{P}}}$, $\mathbf{t}_{{\mathrm{Q}}} \neq \mathbf{
s}_{{\mathrm{Q}}}$, $\delta \Omega_{{\mathrm{P}}} \neq 0$), the Frank vectors
of the attached disclinations are no longer along $\mathbf{n}$, but, as shown
in the previous analysis, there is still a choice for the
direction of the attached disclinations for which the
supplementary dislocations are cancelled. And the conclusion on
the relation between curvature and attached disclinations is still
valid.

\section {Coarse-Grained Crystalline Solids, Grain Boundaries, Polynanocrystals}
\label{crystallinesolidsandnanocrystals}

\subsection {Coarse-grained crystalline solids}
 \label{Coarse-grainedcrystallinesolids}
\emph{Non-quantized} wedge disclinations in a crystalline solid have been mentioned previously; they are akin to limited tilt boundaries discussed in \ref{CONTINUOUSDEFECTS}.

Grain boundaries of \emph{small misorientation} angle (subgrain boundaries) are well documented; they are limited by wedge, twist or mixed disclinations, according to the geometrical interactions between Burgers vectors and Frank vectors discussed previously. These interactions are restricted to those that imply Burgers vectors equal to translation symmetries of the medium.

\emph{Large misorientation} angle grain boundaries are especially important in polynanocrystals.

\begin{figure}
\includegraphics{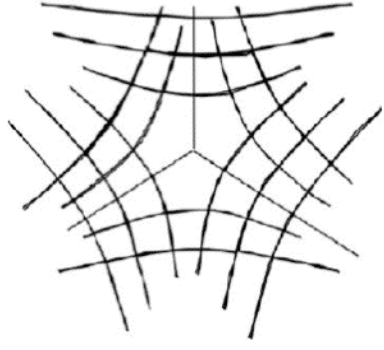}
  \caption{   \label{fig2}{Quantized wedge disclination in a crystal,
$\Omega=-\pi$}}
\end{figure}
\emph{A quantized} disclination in a crystalline solid carries an
angle $\mathbf{\Omega}$ of rotational symmetry of the crystal; in
the wedge case, the line is itself the axis of this symmetry, see
Fig.~\ref{fig2}. As already emphasized, the related energy is
extremely large and thereby their existence improbable. A related
imperfect disclination occurs when the grain boundary is a plane
of geometry of large atomic density for the two grains. If this
imperfect disclination is rejected outside the sample, one has a
low energy \emph{twin}.

Quantized disclinations in mesomorphic media
are discussed in  \ref{quantizedDisclinationsinMesomorphicPhases} and \ref{FocalConicsinSmA'sasquantizedDisclinations}.

\subsection{Grain Boundaries}
\label{GrainBoundaries} 
\subsubsection{Classification of grain boundaries and of
continuous disclinations} \label{GBand
disclinationclassifications}

 Grain boundaries in solids are classified according to the orientation of the rotation vector $\bm{\omega}$ with respect to the plane of the boundary  $\Sigma_{GB}$: \emph{tilt grain boundary} when $\bm{\omega}$ is in $\Sigma_{GB}$, \emph{twist grain boundary} when $\bm{\omega}$ is perpendicular to $\Sigma_{GB}$. This classification makes sense: a tilt grain boundary can be split into a set of parallel identical edge dislocations whose Burgers vectors are perpendicular to the boundary; a twist grain boundary can be split into two sets of parallel identical screw dislocations whose Burgers vectors belong to the boundary. Such splittings are currently observed in small misorientation grain boundaries (also called sub-boundaries). Each set of screw lines of a twist grain boundary carries non-vanishing stresses, but the two sets cancel their stresses at long distance\footnote {It is believed that there is only one set of screw lines in the twist grain boundaries of the TGBA liquid crystalline phase, see \textcite{renn88}, \textcite{kamien99}; in that case the long distance cancellation of the stresses carried by each grain boundary is due to the presence of a family of parallel grain boundaries at periodic distances.}.
\begin{figure}
\includegraphics{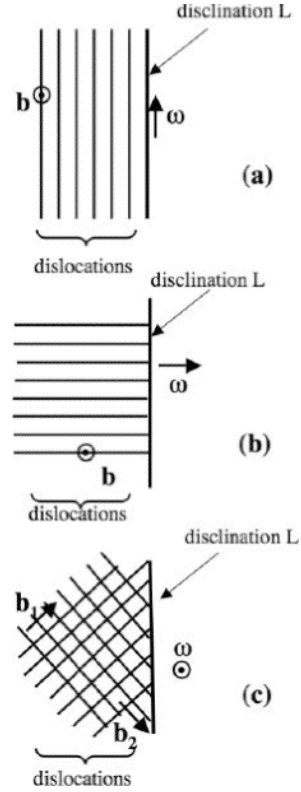}
   \caption{\label{fig3GB}{Classification of continuous disclinations in a solid: (a) wedge disclination; (b) normal tilt disclination; (c) pure twist disclination}}
\end{figure}

We have classified disclination lines according to the orientation of the rotation vector $\bm{\omega}$ with respect to the line direction L: \emph{wedge line} when $\bm{\omega}$ is along L, \emph{twist line} when $\bm{\omega}$ is perpendicular to L. This classification is perfectly adequate when no account is taken of the presence of a grain boundary attached to the line, e.g. quantized disclinations (no grain boundaries), but is not consistent with the grain boundary classification. For instance a tilt grain boundary can be limited either by a wedge disclination or a twist disclination.

A finer classification of continuous disclinations (i.e. carrying a grain boundary) seems therefore appropriate:

(i) \emph{wedge} disclination line, Fig.~\ref{fig3GB}(a): L is parallel to the constitutive dislocations of a tilt boundary; $\bm \omega$ is parallel to L.

(ii) \emph{normal tilt} disclination line, Fig. \ref{fig3GB}(b): L is perpendicular to the constitutive dislocations of a \underline{tilt boundary}; $\bm \omega$ is perpendicular to L.

(iii) \emph{pure twist} disclination line, Fig.~\ref{fig3GB}(c): L belongs to a \underline{twist boundary}; $\bm \omega$ is perpendicular to the boundary, thus to L. We have taken $|\mathbf{b_{1}}|=|\mathbf{b_{2}}|$, hence $\mathbf{b}=\mathbf{b_{1}}+\mathbf{b_{2}}$ perpendicular to L, as required.

 \subsubsection{Polycrystals as compact assembly of polyhedral crystals} \label{Polycrystalsascompactassembly}
 Nearly perfect polycrystals, as possibly created by annealing, can be viewed as compact assemblies of polyhedral grains. Their common facets are commonly triangular and fairly flat grain boundaries $\Sigma$, each surrounded by a disclination line L of strength equal to the rotation $\bm \omega$ of the grain boundary.

 \begin{figure}
\includegraphics{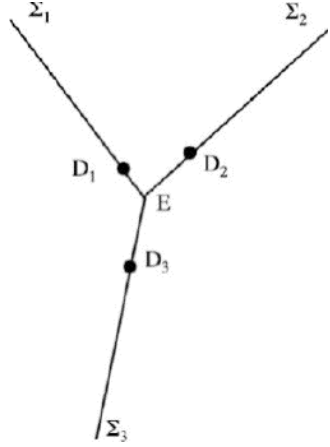}
   \caption{\label{fri2}{The three parallel disclinations D$_{i}$, with rotations $\bm \omega _{i}$, along an edge E between three grain boundaries $\Sigma_{,i}$}}
\end{figure}

Three grains meet along a fairly straight edge E bordering such a facet, where the three parallel disclinations combine along the edge, Fig.~\ref{fri2}. In such stress free annealed polycrystal, each such triplet of parallel disclinations must compensate their long range stresses.\\

 \noindent \textsl{a. {Kirchhoff relations.}}  \label{Kirchhoffrelationsaagg}
 To analyze the stresses due to the three disclinations D$_{i}$ of an edge E, we have again to distinguish the contributions of the wedge components  $\bm\omega_{i, \|}$ (parallel to E, cf Fig.~\ref{fig3GB}(a)) to those of the normal tilt and pure twist components $\bm \omega_{i, \bot}$, with attached dislocations, Fig.~\ref{fig3GB}(b) and~(c).

 For this second part $\bm \omega_{i, \bot}$, the compensation of the three families of relaxation dislocations lead to the same condition as above for a node:
 \begin{equation}   \label{kir1}
\sum_{i} \mathbf{f}(\bm \omega_{i, \bot})\,=\,0,
    \end{equation}
    with $
\mathbf{f}(\bm \omega)=2\,\sin \dfrac {\omega}{2} \enskip \mathbf{t}
$.

   \begin{figure}
\includegraphics{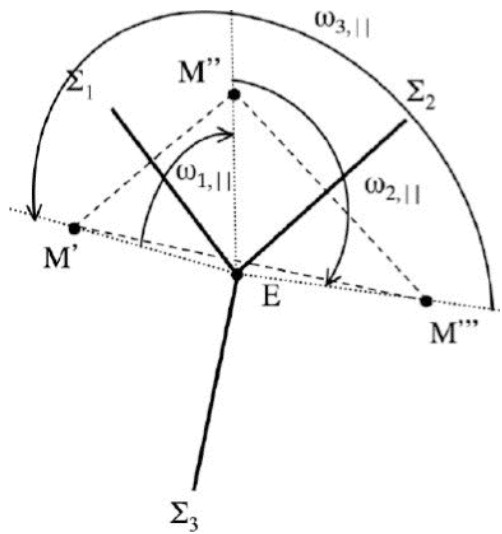}
   \caption{\label{fri3}{Composition of the three edge disclinations of parallel rotations $\bm\omega_{i, \|}$}}
\end{figure}

To have a completely stress free edge, the wedge components $\bm\omega_{i, \|}$ must also compose to zero:
 \begin{equation}   \label{efri3}
\sum_{i} \bm\omega_{i, \|}=\,0,
    \end{equation}

        It is clear from Fig.~\ref{fri3} that the Volterra process which produces $\bm\omega_{1, \|}$ and $\bm\omega_{2, \|}$ by moving M' to M'' and M" to M'" sum up to an effect opposite to that which $\bm\omega_{3, \|}$ produces by moving M'" to M'. Thus, using Eq.~\ref{e1} and the fact that M', M'' and M'" are on a circle centred at E, one gets
    \begin{equation*}
    \mathrm{M' M''} \cos \angle (\mathrm{M"M'M'''})+\\
    \mathrm{M'' M'''} \cos \angle (\mathrm{M''M'''M'})-
    \mathrm{M' M''} =0
    \end{equation*}
    because
    \begin{equation*}
    \sin\frac{\omega_{1, \|}}{2}\,\cos \frac{\omega_{2, \|}}{2}+\sin\frac{\omega_{2, \|}}{2}\,\cos \frac{\omega_{1, \|}}{2}=\\
    \sin\frac{\omega_{1, \|}+\omega_{2, \|}}{2}=-\sin\frac{\omega_{3, \|}}{2}
    \end{equation*}

  \noindent \textsl{b. {Subboundaries.}}  \label{Subboundaries}
  For small misorientation boundaries, and also for boundaries whose orientation does not differ much from a small energy twin, all or part of the continuous dislocations cluster into a periodic distribution of quantized dislocations \cite{burgers39,read50}. This \emph{polygonization} was  first observed by X rays and described in these physical terms by \textcite {crussard44}, after annealing of FCC single crystals strained in multislips (stages II and III). Later observations after etching of low angle boundaries \cite{lacombe48} provided the first experimental proof of the decomposition of these boundaries into rows of dislocations, and various techniques such as electron microscopy for metals and semiconductors and pinning of dislocations by precipitates in transparent ionic solids analyzed the details of the dislocation networks on the subboundaries and the way these dislocations connect at the edges of the grains, as assumed above, cf. J. \textcite{friedeldisloc,friedel87}. These dislocations can slip under stress, especially after annealing of crystals strained in single slip (stage I), which produces especially simple networks such as pictured Fig.~\ref{fri3} \cite{washburn52}; in the more general case, the bowing under stress of the dislocations of the various subboundaries decreases by a large fraction the effective elastic moduli \cite{friedel55}.\\

  \noindent \textsl{c. {Large misorientation boundaries.}}   \label{Largemisorientationboundaries.}
 Following G.\textcite{friedel26} who calls them 'macles par m\'eri\'edrie'. \textcite{bollmann} has established similar dislocation arrangements for large misorientation boundaries, in terms of a crystallographic network  common  to both grains in contact along the boundary. A common crystallographic network has also been put forward by G. \textcite{friedel26} for what he calls 'macles g\'en\'erales'.{ But} generally it is believed that the boundary is  an amorphous contact on an atomic thickness, with possible ledges along which one of the grains can overlap into the other.
 These configurations, as a whole, obey the same conditions of stability as small angle boundaries, but allow more stressed states than the former ones, like roughness, glide, lateral motions of the grains, $\cdots$\\

  \noindent \textsl{d. {Specific complications that arise from the crystal structure.}}  \label{Specificcomplicationsthatarisefromthecrystalstructure.}
 The coalescence of continuous distributions of infinitesimal dislocations, as mostly considered in this paper, into quantized crystal dislocations can introduce some complications that should be stressed, as they have no equivalent in liquid crystals or magnetic structures. Some of the following are presented in \textcite{friedeldisloc}.

 (i) In the simplest cases, the infinitesimal dislocations coalesce into quantized dislocations all with the same directions of line and Burgers vector. This condition optimizes the energy of contact between the grains at the expense of an elastic distortion of the crystal, over a distance from the grain boundary of the order of the distance $l=$ ON between dislocations, Fig.~\ref{fri}. This is the case {for} three {grain boundaries} meeting along wedge disclinations, as pictured Fig.~\ref{fig3GB}a and~\ref{fri3}, when one set of edge dislocations rearrange after straining in stage I of a single slip system; another example is given by two (or three) systems of screw dislocations building a network of increasing density into a twist boundary, such as in Fig.~\ref{fig3GB}c: this can be obtained by torsion of an hexagonal lattice along the hexagonal axis of symmetry, e.g. in graphite \cite{degennes07} and in HCP metals (M. Fivel, private communication).

 (ii) In most cases, however, where dislocations of a number of slip systems have been developed by straining, the subboundaries af the polygonal structures obtained by recovery are composed each of a distribution of two or more systems of dislocations, so as to produce subboundaries of mixed natures and more or less random orientations. Such subboundaries are somewhat less stable, as their elastic distortions average out at a larger distance from the subboundaries for a given rotation, owing to the mixing of dislocations of several slip systems.

 (iii) Even in the simpler case first considered, Fig.~\ref{fri} shows that the possible positions of the crystal dislocations can only occur at specific positions such as O
and N along the subboundary. The periodic distribution of such dislocations must be coherent with the crystal structure along the boundary; it must correspond to a discrete series of angles $\omega_{i}$, with intervals increasing with the angle at least for small angles $\omega_{i}$. A subgrain boundary with an angle between  $\omega_{i}$ and  $\omega_{i+1}$ can only be built of successive segments corresponding to $\omega_{i}$ and  $\omega_{i+1}$, and the distance of elastic relaxation away from the boundary is of the order of two such successive domains.

 (iv) Fig.~\ref{fri}  shows schematically the structure of such a symmetric \emph{tilt boundary}, in its initial state with a continuous distribution of infinitesimal dislocations. Their regrouping into quantized crystal dislocations can only occur at distances $l_{i}$, where the $l_{i}$ are integer multiples of the lattice period along the boundary (n = 2 in Fig.~\ref{fri}); $l_{i}$ is related to $\omega_{i}$ by
  \begin{equation}  \label{efri8}
\textrm{f}_{i}=\frac {b}{l_{i}}=\sin\omega_{i}.
    \end{equation}
    For increasing $\omega_{i}$, this special 'Frank vector' is increasingly smaller than Eq.~\ref{efri} for the general case.

 (v) These $\omega_{i}$s correspond to especially coherent structures, of lower energy. In fact, in crystal structures with a center of symmetry, they are \emph{twins}; and Eq.~\ref{efri8} leads then for $\omega_{i} =\pi$ to a perfect coherence of the two crystals, with vanishing boundary tension. In the absence of a center of symmetry, $\omega_{i} =\pi$ leads to a twin with especially low boundary tension and again no crystal dislocations. In both cases, the increase of $\textrm{f}_{i}$ when $\omega_{i}$ decreases from $\pi$ can be described in terms of an increasing density of crystal dislocations.
More generally, perfect matching without crystal dislocations and vanishing grain boundary tension occurs when $\omega_{i}$ is an allowed rotational symmetry of the crystal, such as $\omega_{i} ={\pi}/{2}$ for the case of Fig.~\ref{fri} with a cubic crystal symmetry. The same occurs of course in all cases for $\omega_{i} =\pm 2\,\textrm{n}\,{\pi}$.

  (vi) \begin{figure}
\includegraphics{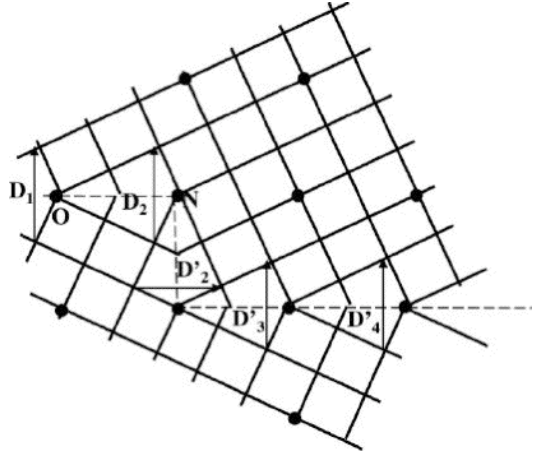}
   \caption{\label{fri8}{Ledge in a merihedric twin, cf \textcite{friedel26}, \textcite{bollmann}. The superlattice is underlined by black dots}}
\end{figure}

In cubic structures, such twins are of the kind considered by G.\textcite{friedel26} and \textcite{bollmann}, in that they can be built on a common superlattice of the two crystals. Such twins, with not too large $l_{i}$s so $\omega_{i}$ {is }not too far from $\pi/2$, can present ledges as sketched Fig.~\ref{fri8}, which introduce supplementary dislocations such as D'$_{2}$.
\subsection{Polynanocrystals}
\label{Nanocrystals} 
   Polynanocrystalline materials are assemblies of small polyhedral grains. The grain size ranges from a few nm to 1$\mu$m. Their plastic properties, especially investigated in pure metals, pure semiconductors, and ceramics, show several features. {One example is} the presence of huge internal stresses, that distinguish them from the classic picture of dislocation driven phenomena in usual coarse-grained crystalline materials $-$ see e.g., \textcite{weert}; \textcite{kumar2003, ovid05, wolf05}; \textcite{van06b}. Polynanocrystals provide a remarkable physical example to discuss some of the notions just introduced. We are interested in the relation between the polynanocrystal plastic properties and the disclination and grain boundary structure.

\subsubsection{Data on the plastic deformation of polynanocrystalline materials} \label{PlasticDeformation}

A dominant mechanism of coarse-grained crystal plastic deformation (work-hardening) at low temperature is the slip of dislocation pile-ups; this mechanism obeys the well-known Hall$-$Petch relation $\sigma_{Y}=\sigma_{Y_{0}}+k\,l^{-1/2}$ $-$ where $\sigma_{Y_{0}}$ is the single crystal yield stress (the friction contribution), $k$ a material dependent constant and $l$ the grain size, see e.g. J. \textcite{friedel59b}.

But the yield stress does not increase without limit when $l$ approaches atomic sizes. After reaching considerable values in the nanoscopic range, it decreases somewhat below some cross-over size $l_{c}$ $-$often referred as the \emph{strongest} size \cite{yip98}; the material becomes ductile and even 'superplastic'. This latter property is not met in all samples, and depends crucially on the absence of porosity (high density) and of nanocracks. It has been observed e.g. in an iron alloy, \cite{branag03}, in nickel, \cite{macfa99,schuh02}, in copper, \cite{lu00,wang02, koch03,champion03,zhu04}. Nanocrystalline nickel, for instance, exhibits a Hall-Petch strengthening as the grain size decreases down to $l_{c}\approx14 \,\mathrm{nm}$, thus reaching internal stresses {of order at least} ten times those observed in usual coarse-grained samples.

High ductility requires suppression of plastic flow localization, i.e., strain hardening stabilizing the tensile deformation, see \textcite{ovid05}. Several mechanisms at the origin of this strain hardening are currently discussed in the literature.

(i) \emph{Partial dislocations emitted by grain boundaries. }The size range just below $l_{c}$ is characterized in FCC metals by the appearance of Shockley partials with $1/6<1 1 2>$ Burgers vectors bordering $\{ 1 1 \bar{1}\} $ stacking faults and the (correlated) formation of \emph{twin lamellae}, during which process the yield stress is still increasing; see e.g., for Al, \textcite{chen03}. In effect, one can imagine that the formation of partials on some glide system impedes the motion of partials on another one.
A {large number} of atomistic calculations, critically reviewed and developed in \textcite{van06c}, supports a few experimental results pointing in this direction.

We shall keep in mind that this type of dislocation-biased plastic deformation must put into play glides and probably lateral displacements or growth/shrinking processes of grain boundaries; the published data do not clearly assert whether these glides are analogous to low temperature glide or employ diffusion mechanisms, at the emission/absorption of dislocations.

(ii) \emph{Hardening by annealing and softening by deformation}. \textcite{huang06} have recently reported on the necessity of extremely high stresses in order to nucleate partials in well annealed, equilibrated, ultrafine nc grains of Cu with no intragranular isolated dislocations nor twins; on the other hand, the appearance of these grain boundary or/and disclination (probably) nucleated dislocations softens the material. This is reminiscent of the plastic behavior of whiskers \cite{ma06}, and suggests that grain boundary perfection is an important factor. Indeed, the computer simulations alluded to above employ grain boundaries that are not totally relaxed and  show up ledges that act as sources of partials.

(iii) \emph{grain boundary sliding and grain rotation.} It seems that this is the dominant mechanism at high temperatures and/or grain size below $l_{c}$, with \textcite{coble63} diffusion inside the grain boundaries. See \textcite{schiotz98}; \textcite{ovid02,van02}; \textcite{shan04}; \textcite{ma04}.

We distinguish in the sequel
 an ideal polynanocrystal configuration, with continuous disclinations and their constitutive dislocations,
 from actual configurations, with their defects and unusual large internal stresses.
  \subsubsection{Structure of the ideal polynanocrystal}     \label{Structure of an ideal nc}
  The picture developed above for annealed coarse-grained polycrystals should apply to polynanocrystals. Thus the computer simulations of \textcite{van00} indicate that there is no difference between boundaries in polynanocrystals and in coarse-grained materials, and that the degree of organization is often rather high.

  Two main differences arise from the conditions of preparation:

  - sintering can produce large internal stresses or intergranular cavities. These large stresses should displace possible low angle boundaries, which are fairly common in coarse-grained polycrystals \cite{friedel53}: such subboundaries should be mobile enough to glide towards large angle boundaries with which they should be integrated.

  - the usual size of the mosaic structure of crystals, whether in the form of a simple Frank network or of a finely polygonized structure, should not be present in polynanocrystals near their maximum elastic limit, both because of their destruction by the high internal stresses just referred to, but more generally because annealing conditions should destroy mosaic structures of sizes less than typically 10 microns (J.\textcite{friedeldisloc} p. 240).
  With these provisoes, the conditions of stability of a polynanocrystal should follow the same Kirchhoff relations developed above for coarse-grained materials.

The surface tension $\mathrm{E}$ of a subboundary depends on the misorientation $\omega$; $\mathrm{E}\approx \mathrm{E_{0}}\,\omega$, where $\mathrm{E_{0}}=\dfrac{\mu \mathrm{b}}{4\pi (1-\nu)}$; on the other hand, for large misorientations, above some value $\omega>\omega_{max}\approx 20\, \textrm{degrees}$, $\mathrm{E}$ can be taken independent of the direction of  the grain boundary with respect to the two grain orientations ($\mathrm{E}\approx \mathrm{E_{0}}\,\omega_{max}$), $-$ see J. \textcite{friedel53}, J. \textcite{friedeldisloc}, chapt.~10 $-$ as long as the grain boundary is not along or close to a lattice direction common to the two grains.

 \begin{figure}
\includegraphics{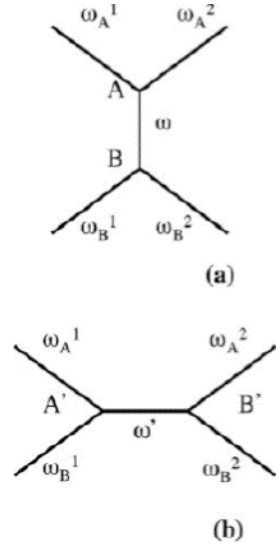}
   \caption{ \label{figfoam1}{A modification process frequent in foams and possible in polynanocrystalline media}}
\end{figure}
Most of the grain boundaries in polynanocrystalline materials are large angle ones; as a consequence they have an energy fairly independent of the angle of misorientation. If submitted to this sole constant surface tension, the grain boundaries should form angles of $\approx$ 120 degrees at triple junctions. Other forces are originating from the wedge disclination segments. The equilibrium can be reached, in principle, by processes of the form described Fig.~\ref{figfoam1}, proposed by J. \textcite{friedel87} for solids
and well known in foams \cite{weaire00}, which consists, first, in the vanishing of the boundary segment AB, and  subsequently, after a passage through an unstable quadruple junction, the appearance of the boundary segment A'B'. In this figure A, B, A', B' are triple junctions seen end on, and $\omega_{A}^{(1)},\,\omega_{A}^{(2},\,\omega_{B}^{(1)},\,\omega_{B}^{(2)},\,\omega_{A'}^{(1)},\,\omega, \,\omega'$ are the \underline{wedge component} misorientations of the GBs whose sections with the plane are drawn; these rotation vectors are all directed along the normal to the figure.

 The triple junction Kirchhoff relations can be written:

\noindent along A: \,\,\, \qquad $\omega\,+\,\omega_{A}^{(1)}\,+\,\omega_{A}^{(2)}\,=\,\delta \omega_{A},$

\noindent along B: \qquad $-\omega\,+\,\omega_{B}^{(1)}\,+\,\omega_{B}^{(2)}\,=\, \delta \omega_{B},$

\noindent along A': \,\,\, \qquad $\omega' \,+\,\omega_{A}^{(1)}\,+\,\omega_{B}^{(1)}\,=\,\delta \omega_{A'},$

\noindent along B': \qquad $-\omega' \,+\,\omega_{A}^{(2)}\,+\,\omega_{B}^{(2)}\,=\, \delta \omega_{B'}.$

They yield \\
 \indent  \,\,\, $ \delta \omega_{A}+\delta \omega_{B}= \delta \omega_{A'}+\delta \omega_{B'} \,(= 2\delta \omega).$ \\
Hence we can write \\
\indent  \,\,\, $\delta \omega_{A}=\delta \omega + \delta \varpi,\quad\delta \omega_{B}=\delta \omega - \delta\varpi,$\\
\indent  \,\,\, $\delta \omega_{A'}=\delta \omega + \delta \varpi',\quad\delta \omega_{B'}=\delta \omega - \delta \varpi',$ \\where\\
\indent  \,\,\, $ 2\delta \varpi= \delta \omega_{A}+\delta \omega_{B},\,2\delta \varpi'=\delta \omega_{A'}+\delta \omega_{B'}.$

$\delta \omega$ measures the repulsive terms between A and B, A' and B'; they are equal. The attractive terms $\delta \varpi,\,\delta \varpi'$ are different; thus the modification is favored if $|\delta \varpi' |\,<\, |\delta\varpi|$. In the case when $\delta \omega=\delta \varpi=\delta \varpi'=0$, the modification is ruled by the surface tension.

There is probably no chance, although it is in principle possible, that a polynanocrystal reaches the same structural configuration as foreseen for a foam, and even less that this structure coarsens, as 2D foams do \cite{neu52}, a result extended recently to 3D foams \cite{hilgen01}.

 \subsubsection{Plastic deformation of a polynanocrystal} \label{deformationactualpolynanocrystal}

The lack of a mosaic structure in fine polynanocrystal prevents the presence of internal Frank Read sources in the grains. This is well recognized by most authors. We shall consider two possible processes:  production of partials and grain boundary sliding, grain rotation and more generally grain boundaries as sources and sinks of dislocations.\\

 \noindent \textsl{a. {Production of partials.}} There probably are several mechanisms possible for the production of partials, some relating to the grain boundaries, other ones to the disclination lines (triple junctions).
    \begin{figure}
\includegraphics{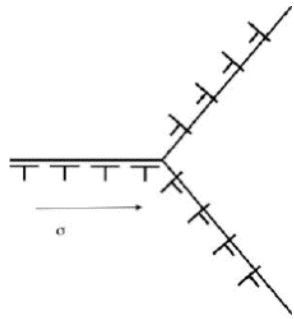}
    \caption{\label{figpileup}{Slip of a pile up, left half-plane, against a normal tilt disclination}}
\end{figure}

 (i) \emph{Constitutive edge dislocations}:  the role of the applied stress in a possible bowing of the dislocations of the boundary has already been put forward. This is probably  a low temperature mechanism, by glide, but a number of computer simulations (see \textcite{van06c}) as well as experimental works indicate that the process is thermally activated \cite{jlm06}, and displays a small activation volume .
   The same dislocations are free, given suitable stresses, to climb in the plane of the boundary and thus to displace the bordering disclination, thereby displacing the two other boundaries that merge along the common edge along which the disclination is located.

(ii) \emph{Slip of dislocation in the boundary}, whose Burgers vectors are \emph{in} the plane, Fig. \ref{figpileup}.

Such dislocations can pile up along the edges and be at the origin of stress concentrations large enough to nucleate new dislocations and/or move the other grains, as proposed by \textcite{ovid05}. Notice that the dislocations considered here are not constitutive or relaxation dislocations of the boundary to which they belong (they do not obey Eq.~\ref{e1}). In fact they form a classical piling up.

  (iii) \emph{grain boundaries and triple junctions as sources and sinks for dislocations}. The emitted/absorbed dislocations are not constitutive dislocations of the grain boundaries, neither pairs of dislocations of opposite signs (dipoles). Assume the presence of a ledge on a grain boundary bordered by a triple junction. If this ledge affects only one (possibly two) of the three grain boundaries merging along the triple junction, there is necessarily some kind of splitting of the triple junction into its constitutive disclinations, in the region of the ledge.
\begin{figure}
\includegraphics{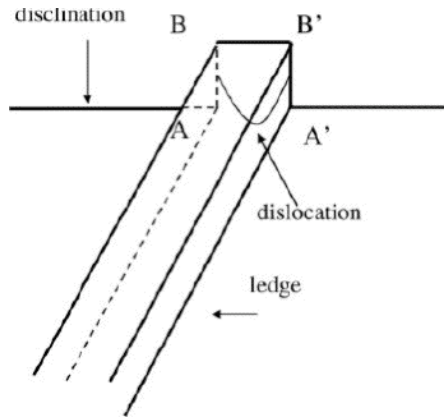}
    \caption{\label{fig2kinks}{Double kink on a disclination. The dislocation Burgers vector is in the plane of the grain boundary or close to it; see text. }}
\end{figure}
Consider one of these 'stripped' disclination segments; the ledge
determines on it a double kink AB, A'B', Fig.~\ref{fig2kinks}. According to Eq.~\ref{e2}, the Burgers vector of the dislocation which necessarily joins the two kinks is perpendicular to the kinks and to the rotation vector $\bm \omega$ of the kinked disclination; $\bm \omega$ is certainly close in direction, if not parallel, to the disclination line, according to the analysis of \ref{Structure of an ideal nc}. Thus a large component of the Burgers vector of the dislocation is in the plane of the grain boundary; this might favor slip in this plane. Also, the smaller the kink lengths AB, A'B', Fig.~\ref{fig2kinks}, the smaller the Burgers vector of the dislocation attached to them, the smaller the energy to nucleate the double kink. One can therefore speculate that this is the origin of partials (rather than perfect dislocations) and of the related small activation energy and volume (by two orders of magnitude as compared
to the values observed in coarse-grained metals).\\

  \noindent \textsl{b. {Grain Boundary Sliding and Grain Rotation.}} \label{GrainBoundarySlidingandGrainRotation}
These mechanisms of deformation have been mentioned, \ref{PlasticDeformation}. Notice that a relative rotation $\delta \omega$ of two grains with a common grain boundary requires that the constitutive dislocation densities be modified accordingly, i.e. that an exchange of grain boundary dislocations with the intergranular medium takes place. \textcite{huang06} have observed that the number of subboundaries decreases under high T annealing, so that the grains necessarily 'roll' \cite{shan04}, the trend being seemingly toward a polynanocrystal with large misorientation grain boundaries only.

\section {Quantized Disclinations in Mesomorphic Phases}
\label{quantizedDisclinationsinMesomorphicPhases}

There is no conceptual difficulty to construct in the manner of
the Volterra process a quantized \emph{wedge} disclination
in a medium endowed with finite rotational symmetries;
\textit{e.g.}, a nematic phase $\Omega=\pm \pi$, a smectic phase
($\Omega=\pm \pi$), liquid crystals in general, a solid crystal
($\Omega=\pm \tfrac{2 \pi}{n}$, $n = 1, 2, 3, 4,$ or
$6$), a quasicrystal \cite{bohsung87} ($n = 5, 8, 10, 12,$ etc) or a
3D spherical or hyperbolic curved crystal \cite{kleman89}. In 3D solid crystals, the line energies are so large
that disclinations are observed in {special} conditions only. See
\ref{Dislocationcontent} (3D solid
crystal wedge lines) and \ref{Disclinationsandgrainboundaries} (continuous twist
lines) for comments.

{The} disclinations that are observed in various liquid
crystals usually differ widely from what is expected to result
from a pure (not extended) Volterra process. These
differences originate in various liquid crystal symmetries, thereby various
types of relaxation defects.  For, as in the
previous section, one can consider the interplay of the quantized
wedge disclination, either with attached dislocations, which
transform it into a twist or mixed disclination with the same
rotation vector $\mathbf{\Omega}$, or with attached disclinations, which
yield a $\mathbf{\Omega}$ variable in direction that illustrates the large
rotation deformations that a liquid crystal can suffer, or again with unattached dislocations that result from their motion. {Quantized disclinations constructed by such
extended Volterra processes can be described in terms of twist, wedge, or mixed
segments,} in addition to their physical property of topological stability.

Two questions therefore arise; a)- how do the Volterra
characteristics of wedge, twist, or mixed character (i.e.
different types of extended Volterra processes) reflect in the topological
classification, b)-whether any quantized disclination,
empirically given, can be constructed in a systematic way by an
extended Volterra process.  To point a) we have a partial answer, namely that
disclinations, when differing only by constitutive dislocations, belong
to the same conjugacy class of $\Pi_1(\textrm{V})$.  Point b) is reported to the final discussion (\ref{Discussion}).

\subsection{Quantized wedge disclinations and their transformations}
\label{quantizedwedgedisclinationsandtheirtransformations}

This subsection is devoted to the molecular
configurations carried by quantized disclinations, when the
\emph{limitations} that are imposed in the Volterra process by the specific
symmetries of the medium are taken into account. Examples given in
this subsection relate to nematics (N) and cholesterics (N*), and
in the next subsection to SmAs. It appears that important
disclination properties (shape, flexibility, interplay
between them and with other defects, etc.) escape an analysis
based on the sole topological classification.
\subsubsection{N phase} \label{Nphase}The order parameter space of the nematic
phase is the projective plane $\textrm{P}^2$, whose first homotopy group $\Pi_1(\textrm{P}^2)$ is
$Z_{2}$, the group with two elements $\{e, a\}$, $a^2=e$, $e$ being the
identity. All topologically classified defects belong to a unique class of
homotopy, namely $a$; all Volterra defects of strength $|k| = n +
\tfrac{1}{2},\, |\Omega|=(2n+1) \pi,\, n \in Z^{+}
\bigcup \{0\}$ can be mapped on $a$. Most experimental
observations yield $n = 0$, i.e. two different Volterra disclination
types, $k = +\tfrac{1}{2}$ and $k = -
\tfrac{1}{2}$, which the topological theory classifies under the same heading; indeed, the topological theory predicts that it is possible to transform smoothly
a $k = +\tfrac{1}{2}$ into a $k = -
\tfrac{1}{2}$ \cite{8}. In fact, since the Volterra process predicts also $k = n$ (these lines, again, are not topologically stable), the cases $k = +\tfrac{1}{2}$ and $k = -
\tfrac{1}{2}$ differ by the not topologically stable but Voterra line $k = 1
$. The Volterra classification thereby proves useful
when analyzing experimental results.

Consider a wedge line. Any axis orthogonal to the director
is a twofold symmetry axis. The wedge line has to be along such an axis. As a
consequence, there is
a director that is orthogonal to the line in its close vicinity.

The deformation of a wedge straight line implies a
translation and/or a rotation of $\bm{\Omega}$ along
the line. Let us apply the considerations of \ref{Genericdisclinationlines.Disclinationdensities}, Eq. \ref{e19} and
 \ref{e20}.  The translation and rotation of $\bm{\Omega}$
along the line bring

 (i) \emph{a non-vanishing density of \underline{attached} \underline{dislocations}}
\begin{equation} \label{e21}
\delta \mathbf{b}_{Tr}=-2\thinspace \mathbf{t}_\mathrm{P} \times \mathbf{s}_\mathrm{P}\, \delta s
\end{equation}

\noindent that vanishes if the Frank vector is along the tangent $\mathbf{s}_\mathrm{P}$  in \textbf{P},
i.e. if the deformed disclination is still of wedge character,

 (ii) \emph{a non-vanishing density of \underline{attached} \underline{disclinations}}
\begin{equation} \label{e22}
\delta \mathbf{f}_\mathrm{P}= \tfrac{2}{R} \mathbf{n}_\mathrm{P} \,\delta s,
\end{equation}

\noindent (this expression being valid in the pure wedge case
only) whose infinitesimal Frank vectors $\delta
\mathbf{f}_\mathrm{P}$ are along the principal normal $\mathbf{n}_\mathrm{P}$ to the
line in \textbf{P}.  Such a defect configuration is licit if the direction
of $\mathbf{n}_\mathrm{P}$ is along an actual director, because any director is an
axis of continuous rotation symmetry for the N phase.  This is in
agreement with our remark above, according to which there is a
director that is orthogonal to a wedge line in its close vicinity.
The density of attached disclinations is vanishing if $\bm{\Omega}$ suffers a
pure translation along the disclination line.

Continuous defects belong to the identity element of the
first homotopy group; this is why their distribution, which is a
function of the shape of the line and of the $\mathbf{\Omega}$  field, does not
influence the topological invariant, always \textit{a} for any $|k| =
\tfrac{1}{2}$ disclination line.

A $|k| = 1$ disclination line does not require special
configuration rules for the director in the near vicinity of the
line, because any axis is a $2\pi$ rotational symmetry axis.  It
is then possible to align the director along the line, at the
expense of special densities of defects \cite{kleman73}.\\

\textit{Remark:} The above description of the curvature of a disclination line in a N phase in terms of defect densities might look overdone; but, at least in what concerns dislocation densities,  it is no more so than the description of the strains and stresses in an amorphous medium in terms of defects. There is no difference of nature between the non-vanishing density of dislocations in a nematic and that one of an amorphous medium, except that in the first case they originate on disclinations and that the rules of elastic relaxation differ. In both cases the 3D continuous translational symmetries render the dislocation densities trivial (this is not so in most liquid crystalline phases, the biaxial nematicN$_{B}$ being a exception, see below). The disclination densities (continuous rotational symmetries about the directors) are superimposed on the dislocation densities but are independent; it would be interesting, knowing the field of distortions of a N phase, to separate what is due to dislocation densities from what is due to disclination densities.

\subsubsection{N* phase} \label{Nstarphase}   The interplay between continuous
dislocations and quantized disclinations in cholesterics has been
discussed by the present authors \cite{friedel69} in the very
paper where this concept was first introduced.
 \begin{figure}
\includegraphics{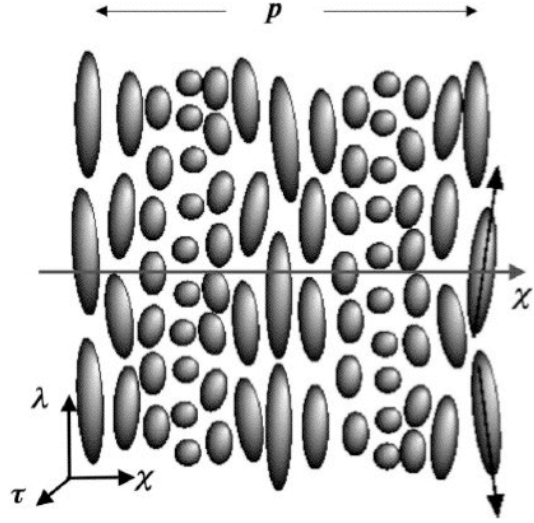}
   \caption{ \label{Chol}{Cholesteric phase N*. The $\bm \lambda$ and $\bm \tau$  directors indicated on the left side belong to the extreme left molecules. The $\bm \chi$  director is constant all over the figure.}{From Kleman and Lavrentovich, \emph{Soft Matter Physics, an Introduction}, Springer, 2003, with permission} }
\end{figure}
There are 3 types
of quantized rotational symmetries $\mathbf{\Omega}$ in a N* phase, all
multiple of the angle $\pi$: 1)- along the molecular axis, which
we denote $\bm \lambda$; 2)- along the helicity axis, $\bm \chi$; 3)-
along the transverse axis, $\bm \tau$. See Fig. \ref{Chol}. The related Volterra processes yield similar
results (similar limitations) to those above for the N phase,
since  $\bm \lambda$, $\bm \chi$ and $\bm \tau$ are directors; being a twofold
axis, any direction orthogonal to $\bm \lambda$, e.g. $\bm \chi$
and $\bm \tau$, is necessarily singular on the core of a
Volterra-constructed related disclination, except if the rotation vector
is an integer multiple of $4\pi$, as shown by \textcite{anderson}.
Hence Volterra processes are perfectly defined for $k_{\lambda}=2n+
\tfrac{m}{2}$, $n$, $m \in Z$ (and by an obvious extension for
$k_{\tau}=2n+ \tfrac{m}{2}$, and
$k_{\chi}=2n+ \tfrac{m}{2}$). Topologically stable disclinations are
classified by the elements of the non-Abelian quaternion group
$Q_8$, whose elements are usually denoted $\{\pm 1, \pm i, \pm j, \pm k\}$
in quaternion notation, or $\{ \pm e, \pm i \sigma_1, \pm i \sigma_2, \pm i
\sigma_3 \}$ in a 2$\times$2 matrix representation where the
$\sigma_i$'s are the Pauli matrices and $e$ the unit 2$\times$2 matrix.
We employ the quaternion notation, which is more appropriate to
crystals in curved spaces of constant positive curvature \cite{coxeter91}, see
Section \ref{DefectsinCurvedMaterials}.  $\{+ 1\}$ corresponds to $k_{\lambda}$, $k_{\chi}$,
$k_{\tau}=2n$, $\{- 1\}$ to $k_{\lambda}$, $k_{\chi}$,
$k_{\tau}=n$ odd, $\{\pm i\}, \{\pm j\}, \{\pm k\}$ to $k_{\lambda}$,
$k_{\chi}$, $k_{\tau}$ half integers, respectively
\cite{mermin, kln}.

Employing the Volterra
method, we now look for the possible attached continuous defects
and their role in the flexibility of quantized disclination
lines.\\

\noindent \textsl{a. {Attached defects: \underline{continuous dislocations;}} }\label{Attacheddefects:continuousdislocations} $k_{\lambda}$, $k_{\chi}$, and $k_{\tau}=\pm \tfrac{1}{2}$ lines.

 The only
possible continuous dislocation Burgers vectors are parallel to
the cholesteric planes (orthogonal to the helicity axis $\bm \chi$),
which are invariant under any in-plane translation. Consider then
a disclination line L, with tangent vector ${\mathbf{t}}$ at
some point \textbf{P} of L. The associated attached Burgers vectors are
along the direction $\bm{\Omega} \times
\mathbf{t}$, according to Eq. 1. Therefore there is no
topological obstruction to the flexibility of a line L in a plane
perpendicular to $\bm \tau$ if L is a $\lambda$ disclination, and in a
plane perpendicular to $\bm \lambda$ if L is a $ \tau$ disclination, but no
other types of flexibility are allowed for these lines. On the
other hand, a $ \chi$ line could curve in any plane; another way to
state this latter result is to notice that a $ \chi$ disclination line of
strength $k$ is also a dislocation of Burgers vector
$\mathbf{b}=-k\,p\, \bm \chi$, because of the equivalence of a
$\pi$ rotation along the $\bm \chi$ axis with a $
\tfrac{1}{2}p$ translation along the same axis ($p$ is the pitch), cf.
\textcite{friedel69} and \textcite{bouligand2}.\\

\noindent \textsl{b. Attached defects: \underline{continuous dispirations;}} \label{Attached defects:continuousdispirations}  $k_{\chi}$ and $k_{\tau}= \pm  \tfrac{1}{2}$ twist lines.

The other
continuous symmetries in a N* phase are helical rotations
$\{\delta \bm{\Omega}, -p \, \tfrac{\delta
\bm{\Omega}}{2\pi}\}$, $\delta
\bm{\Omega}=\delta \Omega \,\bm{\chi}$ along
the $\bm \chi$ axis, i.e. the combination of a translation
and of a rotation. The corresponding Volterra defect is a
continuous \emph{dispiration} that combines a dislocation and
a disclination. Applying Eq.~\ref{e1} to the translational part $-p \,
 \dfrac{\delta \bm {\Omega}}{2 \pi}$, the tangent $\mathbf{t}_p$ to
the line at \textbf{P} has to be in the $\{\bm \lambda, \bm \tau\}$ plane
(perpendicular to $\bm \chi$). The rotation vector direction
$\mathbf{\Omega}_\mathrm{P}$, which we write $\mathbf{\Omega}_\mathrm{P}=\Omega \, \bm{\varpi}_\mathrm{P}$,
$|\varpi_\mathrm{P}|=1$, is along $\bm \lambda$ or $\bm \tau$. The vector $\delta
\bm\varpi=\bm \varpi_\mathrm{Q} - \bm \varpi_\mathrm{P}$ (it appears in Eq.~\ref{e19}, where it is
denoted $\delta \mathbf{t}$) has to be along $\bm \chi$. We
have, after Eq.~\ref{e19}, and assuming that the disclination line
strength is $k=\tfrac{1}{2}$:
$$\, \, \, \,
2\, \mathbf{\varpi} \times \mathbf{t} \, \delta s=-p  \frac{\delta \mathbf{\Omega}}{2\pi} \ \, \small{\textrm{(dislocation component)}}$$
\begin{equation}  \label{e23} 2\, \delta \mathbf{\varpi}=\delta \mathbf{\Omega} \quad
\small{\textrm{(disclination component,)}}
\end{equation}

\noindent and by elimination of $\delta \mathbf{\Omega}$,\\
\begin{equation} \label{e24}
 \frac{d \bm{\varpi}}{ds}= \frac{2 \pi}{p} \, \mathbf{t} \times \bm{\varpi}
\end{equation}

This equation means that the rotation rate of $\mathbf{\Omega}_\mathrm{P}$
is $\dfrac{2 \pi}{p} \mathbf{t}_\mathrm{P}$. Therefore the rotation
rate of the Frenet trihedron attached to the disclination line at
\textbf{P} is
\begin{equation} \label{e25}
\bm{\omega}_\mathrm{P}= \frac{2\pi}{p} \, \mathbf{t}_\mathrm{P}+ \frac{1}{\rho(s)} \, \bm{\varpi}_\mathrm{P}
\end{equation}

Because, according to Frenet's formulae, we have
$\bm \omega= \tau^{-1} \, \mathbf{t}_\mathrm{P}+\rho^{-1} \, \mathbf{b}_\mathrm{P}$ (with obvious notations), we
infer that the disclination line has a \emph{constant
torsion} ($\tau =  \dfrac{p}{2\pi}$), the same
for all lines of this type, and that
$\bm{\varpi}_\mathrm{P}=\mathbf{b}_\mathrm{P}$ is along the
\emph{binormal} of the disclination line. The rotation vector
$\mathbf{\Omega}_\mathrm{P}$ is therefore along the binormal, which is either a
$\bm \lambda$ or a $\bm \tau$ direction. Because of Eq.~\ref{e23}, and because
$\delta \bm{\Omega_\mathrm{P}} \propto \bm{\chi}$ ,
it follows that the local $\bm \chi$ axis is along the principal
normal $\mathbf{n}_\mathrm{P}$, that the rotation vector ($\bm \lambda$
or $\bm \tau$) is along the binormal, and that the tangent to the line
is a  $\bm \tau$ or $\bm \lambda$ direction. The disclination is of pure
twist character. The search for constant torsion curves starts
with \textcite{darboux} and has been the subject of recent
investigations, relating them to the celebrated B\"{a}cklund
transformation and the classification of surfaces of constant
negative curvature; see e.g. \textcite{calini}. Among the solutions, the
simplest ones are those for which $\rho(s)$ is a constant; the
curve is then a circular helix with pitch $P= \displaystyle\frac{4
\pi^2}{p} \Big(\displaystyle \frac{1}{\rho^2}+\displaystyle
\frac{4\pi^2}{p^2} \Big)^{-1}$ and radius $R=\displaystyle
\frac{1}{\rho} \Big( \displaystyle \frac{1}{\rho^2}+\displaystyle
\frac{4\pi^2}{p^2} \Big)^{-1}.$ The $\rho=\infty$ limit case is
simple and interesting. The disclination line is straight and
orthogonal to the $\bm \chi$ axis. Such an object is highly energetic,
but can be stabilized by the presence of another straight
disclination of opposite sign, parallel to the first one and at a
short distance. Continuous dispirations link the two lines in the
ribbon in between.  Now, because the set of two disclination lines
of opposite signs is equivalent to a dislocation, the ribbon can
take any shape, which is comparable to the case discussed by
J. \textcite{friedel69}.

As {reviewed} by \textcite{calini}, there is a
considerable variety of closed constant torsion curves with
different knot classes. The search for disclinations affecting
such shapes in N* phases remains to be done.

The foregoing analysis of the flexibility of
disclinations does not leave place for curved lines in the
$(\bm \lambda,\bm \tau)$ plane orthogonal to the $\bm \chi$ axis.  But such
lines are known to exist, e.g. ellipses belonging to
focal conic domains (FCD), much akin to focal conic domains in SmA's
\cite{bouligand3, 6a, bouligand7}. In the limit where the pitch is small
compared to the size of the sample, and the size of the FCD so
large compared to the pitch that the inner helicity of the N*
layers is a negligible phenomenon, the analysis of these ellipses
(and of their conjugate hyperbolae) can be conducted similarly
to FCDs in SmAs (see below).  But in most experimental cases
the situation is more delicate and a thorough investigation is
all
wanting.

\indent \textit{Remark 1}. The above discussion relates to
disclinations of strength $|k|=\tfrac{1}{2}$, of
homotopy classes $\{\pm i\}, \{\pm j\}$, or $\{\pm k\}$. The case of $|k|=1$ defects
(homotopy class $\{- 1\}$) requires a different approach, since
Eq.~\ref{e1} is no longer valid, and we have to treat the $|k|=1$ defect
as a sum of two  $|k|=\tfrac{1}{2}$ defects.  There is
no objection in principle to uniting two disclinations of the same
strength $ \tfrac{1}{2}$ along the same line, making
then a disclination of unit strength. For instance, in the
constant torsion case, such a process of adding two
$|k|= \tfrac{1}{2}$ is a way of canceling the
singularity on the core, if the rotation vector is along the
$\bm \tau$ director (the $\bm \lambda$ director is then along the line);
however the singularity of the order parameter itself, which
consists of the three director trihedron, is not canceled. One can infer that, in the set of defects just investigated (\ref{Attached defects:continuousdispirations}), the twist $|k|=1$ line is favored.

A situation where a non-singular defect of seemingly
$|k|=1$ strength cannot be split into two $|k|=
\tfrac{1}{2}$ defects is discussed by  \textcite{bouligand4}, but in
fact it relates to non-singular \emph{topological configurations}, in the
sense of \textcite{michel}, classified by the Hopf index for the
director field, not \emph{line defects}.

 \textit{Remark 2}. The biaxial nematic phase $(\textrm{N}_\mathbf{B})$ has
the same topological classification of disclinations as the N* phase \cite{volovik,toulouse}.  But
the symmetry group is different. $\textrm{N}_\mathbf{B}$ is invariant under any
translation (i.e. belonging to $\textrm{E}^3$) but there are no
continuous rotation symmetries. Hence, the only possibility to
curve for a $\textrm{N}_\mathbf{B}$ line is by attaching continuous dislocations, in
strong contrast with N.  This is an example where the Volterra process appears to
give a more detailed view of the defect conformation than the topological theory.

\subsection{SmA phase}
\label{SmAphase}

 \begin{figure}
\includegraphics{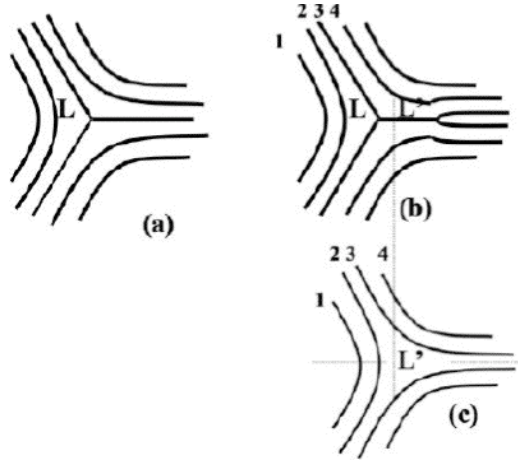}
    \caption{\label{fig9}{The disclination located at L (a, b)) is
displaced to L' (c) by the absorption of a dislocation of Burgers
vector unity ($b = 1$, $\textrm{LL}' = \tfrac{1}{2}$). The
dislocation visible in (b) has disappeared in the process (c)}}
\end{figure}
\subsubsection{Wedge disclinations} \label{Wedgedisclinations}

Figure~\ref{fig9} shows a $k =
-\tfrac{1}{2}$ wedge line in a Sm phase\footnote{The
same picture is valid for a 3D crystal, see the 2D cut along a lattice
plane Fig. \ref{fig2}.},
displaced from L to L' by the absorption of
an edge dislocation $|{\mathbf{b}}|=d_0$ whose Burgers
vector is twice as large as this displacement $
\tfrac{1}{2}d_0$.

The nature of the core has changed.  The absorption of a second
dislocation equal to the first one, along the same route, would
displace the line by the same amount, the total effect of the two
displacements being equal to the repeat distance ${d}_{0}$ of the layers,
L being moved to a position L" (not drawn Fig.~\ref{fig9}), where the original core is retrieved. The
analogy with the displacement of a wedge continuous line described in \ref{Emitted/absorbeddislocations} is
striking; the configuration of L", compared to L, displays a full
new layer equivalent to a dislocation of Burgers vector
$|\mathbf{b}|=2 d_0$ for a displacement $\mathbf{d}_0$ of the
disclination line, as obtained by applying Eq.~\ref{e1}. The inverse
displacement requires the appearance of a dislocation line on the
core, which relaxes and eventually disappears far away from the
line\footnote{It is often taken for granted that dislocations
always nucleate by pairs of opposite signs. But we have here a situation where a dislocation line is nucleated with no partner of the opposite sign. We believe that this possibility is relevant in some important cases. For example, one might this way nucleate screw dislocations all of the same sign at the SmA $\to$ TGBA transition.}.

F. C. Frank was the first to point out, in an oral
communication at the 1969 Montpellier Symposium on Liquid
Crystals, that the displacement of a disclination in a liquid
crystal (he used a cholesteric phase N* to illustrate his
comment) involves the emission/absorption of dislocations, which are quantized in the case he considered.

\subsubsection{Nye's relaxation dislocations} \label {Nye'srelaxationdislocations} 
 \begin{figure}
\includegraphics{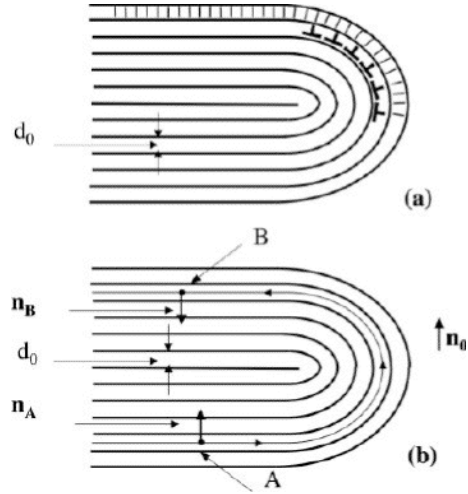}
   \caption{\label{fig10}{The $k= \tfrac{1}{2}$ wedge SmA
disclination; (a)- Nye's infinitesimal dislocations as agents of
layer curvature; (b)- Calculating the sum total of the
infinitesimal dislocations belonging to one layer}}
\end{figure}
There is a feature not apparent in the analysis of constitutive
and relaxation dislocations carried out in \ref{CONTINUOUSDEFECTS}, namely the
possibility of continuous relaxation dislocations that are
directly related to the curvature of the layers, and not attached
to the line.  Figure~\ref{fig10}(a) is illustrative of the case
$k= \tfrac{1}{2}$. Let us turn our attention to one of
the layers inside the disclination wedge. As a liquid layer, its
'inner' group of symmetry is $\textrm{E}^2$; it thereby admits
continuous dislocations whose Burgers vectors are parallel to the
layer. Those dislocations determine the curvature $\dfrac{1}{\rho}$ of the layers, the relation between the
dislocation density and the curvature being  $\dfrac{db}{ds}= \dfrac{d_0}{\rho}$, where $d_0$ is the
layer thickness. This relation was first established by
\textcite{nye53} for solid crystals (the curvature of the lattice planes is a function
of the dislocation densities), see Appendix \ref{appnye}.

Let $\mathbf{t}$ be a unit vector along the tangent to the
layer in a section perpendicular to the wedge line, and let us
traverse a path AB in this section, everywhere tangent to the
layer along $\mathbf{t}$, in the part of the layer which
is curved, see Fig.~\ref{fig10}(b). The total Burgers vector measured
along the path from \textbf{A} to \textbf{B} ( \textbf{A}, \textbf{B} any points on the upper and lower horizontal parts of the path) is:
\begin{equation} \label{e26}
\mathbf{b}_{\mathbf{AB}}=\displaystyle \int_\mathbf{A}^\mathbf{B} \mathbf{t} \,
\dfrac{db}{ds}\,ds=d_0 \displaystyle \int_\mathbf{A}^\mathbf{B }
\dfrac{\mathbf{t}}{\rho}\,ds \\ =-d_0 \displaystyle \int_\mathbf{A}^\mathbf{B}
\dfrac{d\mathbf{n}}{ds}
\,ds=d_0\,(\mathbf{n}_\mathbf{A}-\mathbf{n}_\mathbf{B}) \end{equation}

i.e. $\mathbf{b}_{\mathbf{AB}} =2d_0\,
\mathbf{n}_0
$, where $\mathbf{n}_0$ is the normal to the layers
far from the disclination (we have employed one of Frenet's
formulae to transform the second integral into the third, namely
$ \dfrac{d\mathbf{n}}{ds}=\dfrac
{\mathbf{b}}{\tau}-\dfrac{\mathbf{t}}{\rho}$, where
${\tau}(s)$ is the torsion and
$\mathbf{b}(s)$ the binormal at a point $s$ on the path).

Observe that $\mathbf{b}_{\mathbf{AB}}$ is of a sign opposite to the Burgers vector
of the constitutive dislocations of the disclination line; observe
further that the result does not depend on the precise shape of
the layer, whose possible strain at finite distance does not
invalidate the result of Eq.~\ref{e26}, as long as the
layers are parallel and planar far away from the disclination.

\begin{figure}
\includegraphics{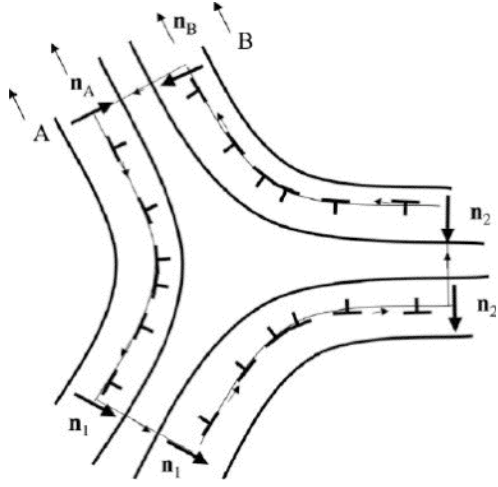}
   \caption{\label{nye}{The $k= - \tfrac{1}{2}$ wedge SmA
disclination;  A, $\mathbf{n_{A}}$ and  B, $\mathbf{n_{B}}$ are at infinity along the asymptotic directions of the disclination configuration. The edge dislocation density vanishes at infinity}}
\end{figure}

Equation~\ref{e26} establishes that the set of infinitesimal dislocations
attending a $k= \tfrac{1}{2}$ line relaxes
the elastic stresses due to the constitutive dislocations, in the geometry we are considering. This
is also true for a $k =- \tfrac{1}{2}$ line; the application of the method of Eq.~\ref{e26} to Fig.~\ref{nye} shows that the equivalent Burgers
vector is  $\mathbf{b}_{\mathbf{AB}}=2 d_0 \mathbf{n}_{0} \,\,(=2 d_0 \mathbf{n}_{\mathbf{A}})$.

{Any layer
curvature can be analyzed in terms of Nye's dislocations, see section
\ref{FocalConicsinSmA'sasquantizedDisclinations} for the
case of Focal Conic Domains in SmAs}.

The example that we have just developed evidences the main features of the relaxation processes relating to Nye's dislocations.

(i) the relaxation of the elastic stresses carried by a disclination which results from a pure (non extended) Volterra process is obtained by the glide of the Sm layers past each other, which accumulates infinitesimal dislocations;

(ii) the accumulation of these dislocations along the layers is equivalent to a set of infinitesimal disclinations attached to the master disclination, analogous to those discussed in \ref{Genericdisclinationlines.Disclinationdensities} (see a similar discussion relating to focal conics, which are special types of disclinations, in \ref{FocalConicsinSmA'sasquantizedDisclinations}). It is also equivalent to a piling of subboundaries;

(iii) the stresses resulting from the fact that the molecules inside the layers are compressed at one end and stretched at the other introduce an elastic constant at the origin of the Frank$-$Oseen splay constant for smectics;

(iv) finally, the Nye's dislocations geometry screens the long distance stresses carried by the master disclination, which considerably reduces its line tension.\\
\subsubsection{Topological stability and Volterra process compared in SmAs. Twist disclinations} \label{TSand VPcomparedinSmA's.Twistdisclinations}
How do {the topological stability approach} and the Volterra process approach compare in SmAs?
The elements of the first homotopy group $\Pi_1(\textrm{V})$ classify line
defects (dislocations and {disclinations}). For a SmA,
$\Pi_1(\textrm{V}_{SmA}) \sim Z\Box Z_2$. Let $(n,
\alpha)$ denote an element of $Z\Box Z_2$, with $n \in Z$ and
$\alpha \in Z_2  (= \{e, a\})$. With these notations, the identity (null)
defect is denoted $(0,e)$, a dislocation is denoted $(n,e)$, where $n$ stands for the Burgers
vector, and a $|\Omega|=\pi$ disclination is denoted $(0,a)$, irrespective of the sign of $\Omega$;
$a^2=e$.

Clearly enough, $(n,a)$ is a $|\Omega|=\pi$ disclination that has absorbed a dislocation $(n,e)$, i.e. that has suffered a translation  $nd_{0}$. Therefore {in Fig. \ref{fig9}}, if
$\textrm{L}$ is represented by the homotopy class $(0,a)$, (which we write
$\textrm{L} \mapsto (0,a)$), {then} $\textrm{L'} \mapsto (1, e)(0,a) = (1,a)$.  Notice that the nature of the core has changed. The absorption of another dislocation of the same sign yields a disclination $\textrm{L"}
\mapsto (2,e)(0,a) = (2,a)$ (not represented Fig. \ref{fig9}), with the same core as L.  The product of two disclinations yield the identity  (0,e). All these operations are summarized in the multiplication rules:
\begin{equation} \label{e27}
(n,\alpha)(m,\beta) = (n + \alpha(m),\alpha \beta), \\ e(m) = m, a(m) = - m.
\end{equation} Observe that any disclination, whatever the nature of its core, can be chosen arbitrarily as the origin disclination (0,a).

With the above analysis, $\textrm{L}$, $\textrm{L'}$ and $\textrm{L"}$ are given three different topological
invariants. But $\textrm{L}$ and $\textrm{L"}$ can also be gathered under the
same heading in the frame of the topological theory; they belong indeed to the same
conjugacy class of $\Pi_1(\textrm{V}_{SmA})$ $-$see Appendix~\ref{appconjug}$-$ and this is
enough to consider them as the same topologically stable defect, according to the
general topological theory \cite{michel, kleman89}.  Consider the
two equalities:
\begin{equation} \label{e28}
(2,a)=(0,a)(2,e),\\ (2,a)=(-1,e)(0,a)(1,e).
\end{equation}
The first equality means that $\textrm{L"}$ is obtained by
'adding' the dislocation $\textrm{L}_{2d} \mapsto (2,e)$ to $\textrm{L}$. In the Volterra
sense, it is an addition; in the topological stability theory sense, it is the product of
two homotopy classes $(0,a)$ and $(2,e)$.  The second equality,
which expresses that $(0,a)$ and $(2,a)$ are conjugate in
$\Pi_1(\textrm{V}_{SmA})$, has a simple physical image: it expresses the
effect of a complete circumnavigation of the disclination L
$\mapsto (0,a)$ about a dislocation $\textrm{L}_d \mapsto (1,e)$. Such an operation cannot change the nature of the
circumnavigating defect, although its homotopy class is modified.
We refer the reader to \textcite{mermin} for a nice pedagogical discussion and an illustration of
this property.

In terms of Volterra invariants, L, L" and all disclinations of the same conjugacy class carry rotations about twofold axes located in planes between layers; L' and all disclinations of the same conjugacy class carry rotations about twofold axes located in the middle planes of layers.

\subsection {Nature of the defects attached to a quantized disclination}
\label {Natureofdefectsattachedtoaquantizeddisclination}
Taking stock of the specific examples discussed above, we now derive some general properties relating to attached defects.
Various cases arise:
\subsubsection{{Continuous} attached defects (dislocations,
disclinations,
dispirations) and kinks} \label {Continuousdefects(dislocations,disclinations,dispirations)andkinks}

\noindent \textsl{a. {Topological stability.} } \label{Topologicalstability1} Continuous defect densities belong to the identity
homotopy class, therefore they do not modify the homotopy class of the
master disclination $\textrm{L}$ all along it. Or, said otherwise, they are
not visible when mapping a closed loop of the deformed medium into
the order parameter space V.\\

\noindent \textsl{b. {Volterra process.}} \label{Volterraprocess1} Because the existence of continuous defects has to
comply, in an ordered medium, with the existence of broken
symmetries, not all continuous defects are realizable, and thereby
limits are put on the possible realizations of master
disclinations, in particular their shapes (in dynamic terms, their
flexibility).  Equation~\ref{e19}, which derives from Eq.~\ref{e1} and has
been established for an isotropic uniform medium, is still valid,
these limits being taken into account.

\subsubsection{{Quantized} attached defects of the first type; full kinks} \label{quantizeddefectsofthefirsttype;fullkinks}

\noindent \textsl{a. {Topological stability.} } \label{Topologicalstability2} Inspired by the SmA example, we first refer to a
case where $\textrm{L} \mapsto \{a\}$ is transformed into a disclination
$\textrm{L"} \mapsto \{a\textrm{"}\}$ belonging to the \emph{same conjugacy
class} $\{a\textrm{"}\} = \{u\}\{a\}\{u^{-1}\}$ of the first homotopy group
$\Pi_1(V)$ as $\textrm{L}$, by the absorption or the emission of a
dislocation $\{v\}$. This dislocation hits the master disclination
at some 'node', where we have a relation of the Kirchhoff's type,
which is written in the language of the topological theory $\{a\textrm{"}\} = \{a\}\{v\}$; this relation
also reads:
\begin{equation} \label{e29}
\{v\} = \{a^{-1}\}\{a"\} = \{a^{-1}\}\{u\}\{a\}\{u^{-1}\},
\end{equation}

\noindent therefore the attached defect is a \emph{commutator} of
$\Pi_1(\textrm{V})$. The result is not restricted to attached dislocations;
it is valid for a defect $\{v\}$ of any type.  We now establish a
reciprocity theorem, which extends the as yet not understood concept of node.

Let us recall that the commutators of a group $\textrm{G}$
generate an invariant subgroup $\textrm{D[G]}$, also called the derived
group.  Not all elements of the derived group are commutators,
but all are products of commutators.  From the point of view of
the physics of defects, it is important to observe that $\textrm{D}$
contains entire conjugacy classes of $\Pi_1(\textrm{V})$, and that the
cosets of $\textrm{D}$ in $\Pi_1(\textrm{V})$ are also composed of entire
conjugacy classes \cite{kleman77, trebin84}. We take an
example. In a SmA, the dislocations having the same Burgers
vector parity all belong to the same coset of
$\Pi_1(\textrm{V}_{SmA})/\textrm{D}$; let us, for the sake of clarity, consider
only even dislocations: the two homotopy classes $(2r,e)$ and
$(-2r,e)$, $r \neq 0$, form an entire class of conjugacy and
they belong to the same coset as the homotopy class $(0,e)$ -
the null defect, which constitutes by itself a full conjugacy
class. This very fact means that the even dislocations are
\emph{equivalent} to the null dislocation $(0,e)$, in the
following sense. It is possible to split the even dislocation
$(2r,e)$ into two equal dislocations $(r,e)$; let one of them
circumnavigate about a $(m,a)$ {disclination}, which
circumnavigation brings it to the conjugate state
$(-m,a)(r,e)(m,a)=(-r,e)$, while the other one stays in place.
It is then possible for $(2r,e)$ to self-annihilate by letting
the fixed and the returned mutually annihilate, i.e. generate a
defect of homotopy class $(0,e)$. Similarly, any odd dislocation $(2r+1,e)$ is
\emph{equivalent} to  $(1,e)$; but
odd dislocations are not commutators.

One can therefore establish the following reciprocity theorem:

\noindent \textsf{if
the attached defect $\{v\}$ is a commutator, then $\{a\}$ and
$\{a"\}$ are either in the same conjugacy class, or belong to two
conjugacy classes belonging to the same coset of $\Pi_1(\textrm{V})/\textrm{D}.$
}

This is also true, by an easy extension, for any element of $\textrm{D}$,
not only commutators. We can thereby state in all generality that

\noindent  \textsf{any defect whose homotopy class belongs to the derived
group} $\textrm{D}[\Pi_1(\textrm{V})]$ \textsf{{is eligible as an attached defect,
and separates the master disclination into segments whose homotopy classes belong to
the same coset of}} $\Pi_1(\textrm{V})/\textrm{D}.$

We name such a defect an
\emph{attached defect of the first type.}\\

\noindent \textsl{b. {Volterra process.}} \label{Volterraprocess2} Insofar as defects of the first type are
commutators, they can terminate on a singular point, since they
are equivalent (in the sense just defined) to the identity
homotopy class, and this singular point can be the node where they
meet the master disclination. This is the main result one can
reach from the analysis of topological properties, but not more, because the topological theory
does not separate properly dislocations and disclinations.  In
other words, the topological analysis does not say anything about the shape
(the flexibility) of the master line, even though there is no
doubt that the attachments are the tools for its changes of shape
and the relaxation of the stresses.

\begin{figure}
\includegraphics{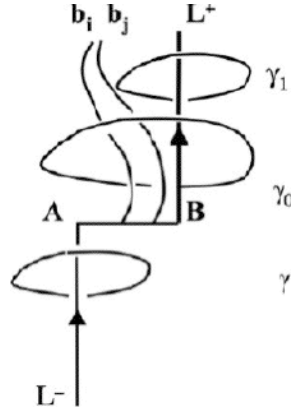}
   \caption{ \label{fig11}{$\textrm{L}$ is a quantized disclination, AB a kink.  The
relaxation dislocations no longer cross the master line}}
\end{figure}

Let us consider the geometry of a kink on a wedge disclination
(Fig.~\ref{fig11});
this geometry differs somewhat from that of Fig.~\ref{fig4} for
a continuous disclination. According to the above discussion, the
same dislocations that have been absorbed by the wedge segment
$\textrm{L}^-$, say, are still outside the wedge segment $\textrm{L}^+$; we still
have relaxation dislocations, terminating on the kink, an allowed
process insofar as these dislocations are of the first type.

The Friedel's relation (Eq.~\ref{e1}) is established for an \textit{a priori} Volterra
description of line defects. Let us assume that, in Fig.~\ref{fig11}, $\textrm{\textrm{L}}^+$
and $\textrm{L}^-$ are $k = \pm \tfrac{1}{2}$ wedge
disclination segments in a SmA, as in Fig.~\ref{fig9}. According to
Friedel's relation, which can be written, with $\mathbf{\Omega} = \pm \pi \mathbf{t}$\\
 \begin{equation} \label{e30}
\sum_i \mathbf{b}_i= \pm\thinspace 2\thinspace \mathbf{t} \times \mathbf{AB}
\end{equation}

\noindent the relaxation dislocation Burgers vectors are perpendicular to
the figure plane.  If we assume that we are in the conditions of
application of this relation in a SmA, the lines are drawn in a
medium that is not yet deformed by the Volterra process, the layers are parallel to
the $\textrm{L}^+$ and $\textrm{L}^-$ lines.  We do not {lose generality} assuming
further that the layers are perpendicular to the plane of the
figure; $\mathbf{AB}$ is along the layer normal. Therefore
the Burgers vectors $\mathbf{b}_i$ are parallel to the
layers; the related dislocations are \emph{continuous}. The
layers in the transition region between $\textrm{L}^+$ and $\textrm{L}^-$ suffer
extra curvatures, which represent these dislocations.  This is
certainly not a small
energy geometry.
\noindent But the geometry of Fig.~\ref{fig11} can be understood differently. If the
relaxation dislocations are quantized, there should be layers
parallel to the plane of the figure in the $\mathbf{AB}$
region (perpendicular to the Burgers vector), whereas
$\mathbf{AB}$ is along the normal to the layers. This is
possible only if we are considering the medium already deformed by
the disclinations.  The Friedel's relation still works for
$\mathbf{AB}$ joining a segment along $\textrm{L}$ to a segment
along $\textrm{L}'$, for example, in fig.~\ref{fig9}. It therefore works when
applied locally to the 'tangent' undistorted medium, \emph{on
either sides of the disclination}.  We give an example in a SmA
phase section \ref{FocalConicsinSmA'sasquantizedDisclinations}. We name \emph{full kinks} the kinks that
separate master disclination segments belonging to
the same coset of $\Pi_1(\textrm{V})/\textrm{D}$.

\subsubsection {{Quantized} attached defects of the second type;
partial kinks} \label {quantizeddefectsofthesecond type;partialkinks}
The discussion of the SmA case has shown that it is perfectly licit to
consider a master line made of two wedge segments $\textrm{L}$ and $\textrm{L}'$
that do not belong to the same coset of $\Pi_1(\textrm{V})/\textrm{D}$. The
dislocation $\{\nu\}$ that hits the (now \emph{partial}) kink
is not a commutator, and the relation $\{a'\} =\{a\}\{\nu\}$ is no
longer a trivial relation, in a way. In the above case it was
possible, at least in a thought experiment, to abolish the node by
smoothly turning the two segments to the same homotopy class -
abolishing simultaneously the full kink. This is now forbidden. On
the other hand, the attached defect exists only on one side of the
master line, because $\{a'\} =\{a\}\{\nu\}$ is now the TS
expression of a true Kirchoff's relation for three defects meeting
at a node. In the Volterra process language, $\{a'\} = \{a\}\{\nu\}$ is nothing
else than Eq.~\ref{e30}.

\section{Focal Conics in Smectic A phases as quantized Disclinations}
\label{FocalConicsinSmA'sasquantizedDisclinations}

The most original defects in SmA phases are focal conic domains
(FCD), whose geometrical properties were first investigated by
 \textcite{friedel10} and \textcite{friedel22}. A FCD
consists of a pair of confocal conics (an ellipse E and an
hyperbola H), which are the focal lines of the set of normals to a
family of parallel smectic layers folded into Dupin cyclides \cite{hilbert64}. For a recent account of the physics behind this geometry, see \textcite{K+L,kln};  an
essential property of E and H is that they are disclination
lines.

\begin{figure}
\includegraphics{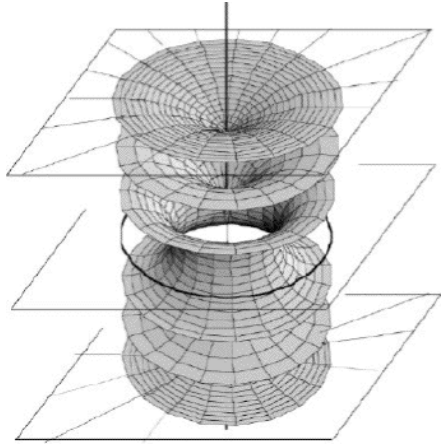}
   \caption{\label{fig12}{Toric FCD}}
\end{figure}

Consider the simple case where E is degenerate into a
circle C and thereby its conjugate conic C' is a straight line
orthogonal to the plane of C, passing through its centre. The
Dupin cyclides are then nested tori (Fig.~\ref{fig12}). They are restricted in this figure to the cyclides with a Gaussian curvature of negative sign, with planar continuations outside. This is the most frequent, if not the only one, empirical occurrence of toric domains.

 It is apparent that
C and C' are both wedge disclination lines, C of strength $k = \frac{1}{2}$, C' of strength $k = 1$. According to the analysis in \ref{Genericdisclinationlines.Disclinationdensities}, there
are no attached dislocations, only attached disclinations, whose
density can be written
 \begin{equation}  \label{(31)}
\delta \mathbf{f}= \tfrac{2}{R} \mathbf{n}\, \delta s=2 \mathbf{n}\, \delta \vartheta
\end{equation}

\noindent The Nye's edge dislocations that
follow the latitude lines of the tori  (call them of the first type) are of the type that attends
such disclinations; their (infinitesimal) Burgers vectors $\delta
\mathbf{b}_{rot}$ are along the meridian lines; they
obviously introduce extra matter that curves the C disclination.
On the other hand the Nye's edge dislocations (of the second type)
that follow the meridian lines of the tori, whose (infinitesimal)
Burgers vectors $\delta \mathbf{b}_{tr}$ are along the
latitude lines, relax the quantized dislocations that attend the
C wedge disclination, after the manner already discussed for a
straight $k = \tfrac{1}{2}$ line, Fig.~\ref{fig10}. Note that
the Burgers vector $\delta \mathbf{b}_{tr}$ is variable
along a dislocation of the second type, since the local frame of
reference is continuously rotated by the dislocations of the first type.

\begin{figure}
\includegraphics{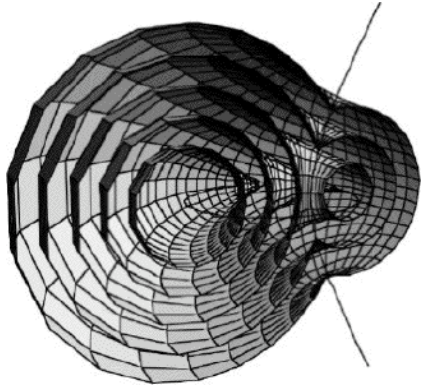}
   \caption{ \label{fig13}
   {FCD with positive and negative Gaussian curvature surface elements}
   \scriptsize{From Kleman and Lavrentovich, \emph{Soft Matter Physics, an Introduction}, Springer, 2003, with permission}
   }
\end{figure}

All these results extend to more general FCDs. Figure
\ref{fig13} represents such a FCD with positive and negative Gaussian curvature surface elements.

In
a local frame of reference where the 1- and 2-axes are along the
lines of curvature of the Dupin cyclide, the 3-axis  along the
normal, the dislocation density tensor (defined in Appendix \ref{appnye}) has components

\begin{center}
\begin{tabular}{|c|c|c|}
\hline $\alpha_{11}=0$ & $\alpha_{12}=\displaystyle \frac{1}{R_2}$
&
$\alpha_{13}=0$\\
\hline $\alpha_{21}=-\displaystyle \frac{1}{R_1}$ &
$\alpha_{22}=0$ &
$\alpha_{23}=0$\\
\hline $\alpha_{31}=\displaystyle \frac{1}{\rho_{g1}}$ &
$\alpha_{32}=\displaystyle \frac{1}{\rho_{g2}}$ &
$\alpha_{33}=0$\\
\hline
\end{tabular}
\end{center}
\bigskip
\noindent where $\dfrac{1}{\rho_{g1}}$ and
$\dfrac{1}{\rho_{g2}}$ are the geodesic curvatures of
the curvature lines.  The geodesic curvatures vanish in the toric
case, since then the lines of curvature are also geodesic lines.
We recall that $\alpha_{ij}$ measures a dislocation content; the
integral $\displaystyle \iint_{\gamma} \alpha_{ij} dS_i$ over
an area bound by a loop $\gamma$ is the $j\text{-component}$ of the
Burgers vector of the dislocations going through this area along
the $i\text{-direction}$. The dislocation tensor being a tensor is
invariant under a change of coordinates; the choice of the
curvature lines as coordinate lines has a simple physical
interpretation in terms of $\delta \mathbf{b}_{tr}$ and
$\delta \mathbf{b}_{rot}$, for the components
$\alpha_{12}$ and $\alpha_{21}$. The extra dislocation densities
$\alpha_{31}$ and $\alpha_{32}$ have also Burgers vectors parallel
to the smectic layers, as they must, for the symmetry reasons
already alluded to. They correspond to edge lines along the
normals to the layers, and contribute to the stability
of the shape of the conics and their relaxation.

 \begin{figure}
\includegraphics{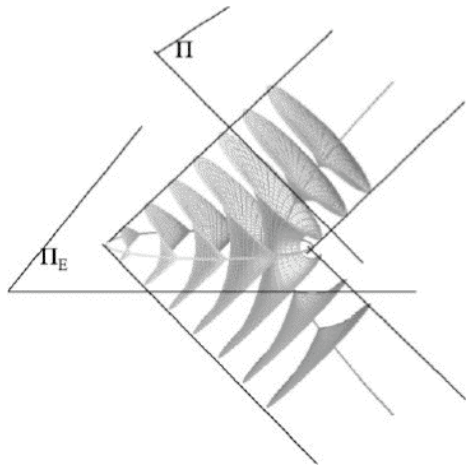}
   \caption{ \label{fig14}{Generic FCD with negative Gaussian curvature layers, see text}}
\end{figure}
As most generally observed \cite{kln}, FCDs are restricted to the negative Gaussian curvature parts $\sigma_{1}\, \sigma_{2}<0$ of the cyclides, Fig.~\ref{fig14}. The FCD is therefore confined inside a double cylinder lying on the ellipse, whose generatrices are parallel to the asymptotes of the hyperbola. Any layer inside this double cylinder is bordered by a curvature line of the cyclides, i.e. a circle, Fig.~\ref{fig14}. The plane $\Pi$ containing the circle is tangent to the cyclide, so that the outside extension of the layer can be along $\Pi$, without any cusp. $\Pi$ is perpendicular to one or the other of the asymptotic directions of the hyperbola, thus the set of all planes $\Pi$ form two families of parallel planes that eventually cross on the plane of the ellipse. By limiting the corresponding material layers to the half space above or below the ellipse plane $\Pi_{\textrm{E}}$, $\Pi_{\textrm{E}}$ appears as a tilt grain boundary of misorientation $\omega=2\,\sin^{-1} e$, where $e$ is the
ellipse eccentricity. E itself is a $k = \tfrac{1}{2}$
disclination line to which are attached the constitutive dislocations of the tilt grain boundary. See Appendix \ref{appell} for a detailed investigation of its characteristics.

\section  {Geometrical Frustration. Role of Disclinations}
\label{DefectsinCurvedMaterials}

\subsection{Geometrical frustration, a short overview}
\label{Geometricalfrustration,anoverview}
The concept of \emph{frustration} covers a number of {inhomogeneous} structures which are all describable in terms of defects, in fact \emph{disclinations}; see \textcite{kleman89} for a review. The word itself was introduced by \textcite{tou77b} in the frame of the theory of antiferromagnetic systems, where there exist closed paths of atoms with nearest neighbour exchange interactions that cannot be satisfied simultaneously.

 \subsubsection{Unfrustrated domains separated by defects}
  \label{Unfrustrateddomainsseparatedbydefects}

    By \emph{geometrical frustration} \cite{kleman85,kleman87}, we mean an extension of the concept of frustration such that: a) it connotes systems where the short range interactions are so much dominating that they completely determine the local configuration, and b) it is possible to describe  these interactions in geometrical terms $-$ in a sense geometrical frustration is an extension of the notion of steric hindrance. The concept is therefore of interest when the local configuration is incompatible with long range \emph{Euclidean} ordering, i.e. is non-cristallographic in the usual sense of this term. The frustrated medium is thereby divided into small, \emph{unfrustrated}, domains, of size $\xi$ say, separated by defect regions where  the short range order is broken.

         In three dimensions, it is fruitful to introduce a crystal template where the unfrustrated domains spread without obstruction, if such a description is feasible.  Such templates, where this local order extends homogeneously without distortion, are necessarily curved, non-Euclidean, with Riemannian habit spaces of \emph{constant curvature}. The decurving of the template into an Euclidean medium employs \emph{disclinations} of these curved crystals; they delimit the unfrustrated domains of the actual medium. Geometrical frustration connotes the existence of a particular type of incompatibility resulting from the different interactions in competition.
Disorder at scale $\xi$ does not prevent that at larger scales there exist correlations between the unfrustrated domains; the resulting frustrated medium can be either a crystal with broken translations $-$ if the weak long-range interactions take over at some scale $-$ or a medium truly disordered at all scales $>\xi$ $-$ if these long-range interactions are very small. Very similar but simpler approaches in one or two dimensions were developed earlier to describe approximate epitaxy and surface reconstruction of crystals, helical magnetic order, charge or spin density waves (cf \textcite{friedel77b} for an introduction). Chemists have also talked in the same spirit since before {World War II} of non alternate electronic structures of molecules and solids.

We first present a short overview of frustrated media with emphasis on the presence of disclinations. In most of the examples below, the 3-dimensional sphere S$^{3}$ is employed as a template habit space.

\subsubsection{Covalent glasses, disclinations}
\label{Covalentglasses}
For \emph{covalent glasses}, frustration originates in the constancy of the coordination number z, which can be incompatible with certain ring configurations.  For example 5- and 7- membered rings do not generate Euclidean order.  A large literature exists on the subject, starting with the celebrated model of \textcite{zach32} for continuous random networks; cf. \textcite{mosseri90} for covalent frustration models.

As pointed out by \textcite{rivier87}, the underlying  geometrical structure of a covalently bound material is a \emph{graph} with a constant coordination at each node, except for dangling bonds and possible double bonds. There is no reason why covalent bonds should form polyhedra. A stacking of polyhedra is a particular graph.  In that latter case, Rivier's theorem, which states that

 \textsf{odd-membered rings are not found in isolation, but are threaded through by uninterrupted lines which form closed loops or terminate on the boundaries of the specimen} \cite{rivier79},

 \noindent is quite useful.  But it applies also to non-polyhedral structures, as soon as rings are recognizable. Rivier's lines are obviously \emph{disclinations}.  See also \textcite{tou77b}.

\subsubsection{Double-twisted configurations of liquid crystal directors and polymers, disclinations} \label{Double-twistedconfigurationsofLC}
The most studied phenomenon of frustration in liquid crystals is the \emph{double-twist} molecular arrangement of short molecules met in Blue Phases, and whose presence has been advocated in cholesterics of biological origin (DNA, etc.), see  \textcite{livolant86}; \textcite{ livolant87}; \textcite{giraud88}. The chromosome of  dinoflagellate \cite {livolant80} displays also a double twisted geometry of DNA molecules of a type somewhat different from the type met in Blue Phases \cite {kleman85a, friedelprague}. It has also been advanced that double-twist is present in amorphous molten polymers \cite{kleman85}.

\begin{figure}
\includegraphics{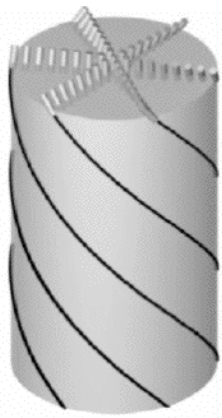}
   \caption{ \label{Figdt} {Double-twist configuration. \scriptsize{From Kleman, Lavrentovich and Nastishin, \emph{Dislocations in Solids}, vol. 12, Elsevier, 2005}, with permission
   }}
\end{figure}

           Consider a liquid crystal of chiral molecules. If the director is an axis of cylindrical symmetry, all the directions orthogonal to any director can act effectively as axes of helicity $-$ in a classical N* phase, this symmetry is broken and there is only one axis of helicity. The local unfrustrated arrangement  can be described as in Fig.~\ref{Figdt},
where the integral lines of the director are helices
of chirality opposite to the chirality of the rotation of the
director about the radii. We have $n_{r}=0,\,\,n_{\vartheta}=\sin
\psi(r),\,\,n_{z}=\cos \psi(r)$, with $\psi(0)=0$. The molecules
rotate with the inverse pitch  $$q(r) =\frac{2\pi}{p}=
-\mathbf{n}\cdot \bigtriangledown\times\mathbf{n} =
\dfrac{\mathrm{d}\psi }{\mathrm{d}r} + \dfrac{\sin 2\psi}{2r}$$
which is the same order of magnitude as the inverse pitch of the
cholesteric phase. Double-twist is entirely satisfied only for the
director along the $z$-axis ($r = 0$); at a distance $r = p/4$
from the $z$-axis, the director has rotated by $\pi/2$ along the
$r$-axis and is now perpendicular to the $z$-axis; its vicinity is
no longer double-twisted; frustration sets in.

Blue phases are made of elements of double-twisted cylinders of matter assembled in space. Three cylinders can stack along 3 orthogonal directions; the region of highest frustration, in-between, may show up a singularity of the director field, a $k=-\tfrac{1}{2}$ disclination. This local arrangement is found in several cubic symmetries showing a 3D disclination segment network, \cite{meiboom83,barbet84}. The blue fog is amorphous, and the disclination lines are seemingly at random; it has also been suggested that the blue fog is icosahedral, as in quasi-crystals \cite{rokhsar86, hornreich86}, achieving thereby another type of frustration.

    Double-twist can be thought of as resulting from a competition between a tendency to \emph{dense packing}, insuring parallel alignment of the integral lines, and a tendency to \emph{chirality}. But a geometry with \emph{equidistant helices}, which would result from such a competition \cite{kleman85}, is not homogeneous in Euclidean space.

The template proposed by \textcite{seth85} in a S$^{3}$  curved space insures homogeneous unfrustrated double-twist; we denote it \{dtw/{S}$^{3}$\}. The director is along a family of great circles of the habit 3-sphere, all those great circles being equidistant and twisted, and \emph{parallel} in the sense of spherical geometry. This is described in Appendix \ref{appthreesphere} (the Hopf fibration) and studied for its defects in \ref{DefectsoftheS3templateofdouble-twist}. See also \textcite{edv88}.
    \subsubsection{Tetrahedral and icosahedral local orders, disclinations} \label{Tetrahedralandicosahedrallocalorders}
For \emph{amorphous metals} and \emph{Frank and Kasper phases}, the origin of frustration is the tendency towards dense packing of equal or quasi equal spheres, representing atoms.\\

     \noindent \textsl{a. {Frank and Kasper phases.}}   \label{FKphases}
     In complex metallic alloy structures, particularly those of transition metals, it occurs frequently that the structure is entirely determined by the requirements for sphere packing, i.e. that the atoms form tetrahedral clusters, and that the coordination polyhedra are triangulated.

\textcite{frank58b,frank59} thoroughful topological and geometrical study of the \emph{crystalline} structures submitted to such constraints shows $-$ if one admits that the number of neighbors Z$_{S}$ of an atom on the coordination polyhedra to which it belongs is either Z$_{S}$ = 5 or Z$_{S}$ = 6 $-$, that there are only 4 types of coordination polyhedra, with Z = 12, 14, 15 and 16.
Frank and Kasper distinguish the sites Z = 12 as \emph{minor sites}, and the other ones as \emph{major sites}. The edges which join neighbouring major sites form a \emph{skeleton}; sites of Z = 14, 15 or 16 are meeting points of two, three or four \emph{bones}. This skeleton is much simpler to study than the structure as a whole. The description of the Frank and Kasper phases in terms of a skeleton of bones (i.e. a network of line defects) is contemporary to the development of the concept of disclination in liquid crystals (due to \textcite{frank58a} himself), but it was only later that the topological nature of these defects  as true \emph{disclinations} was recognized \cite{nelson83a}.

    The Frank and Kasper networks constitute a remarkable example where the main characteristics of geometrical frustration show up; the frustrated atoms are along lines which structure a sea of unfrustrated atoms Z = 12, and there is a typical distance between lines which here scales with the lattice parameter.  The existence of the skeleton of major sites is not dependent on the existence of a periodic lattice, and the only necessary hypothesis is that the medium be polytetrahedral.\\

\noindent \textsl{b. {Amorphous metals.}}   \label{amorphousmetals}  
Icosahedral order (Z = 12) is met in Frank and Kasper phases, but also amorphous metals, undercooled atomic systems \cite{frank, bernal59,bernal64}, and quasi-crystals \cite{schechtman84}. It is also valid for small clusters less than a few hundreds atoms; for metals and rare gases, cf J. \textcite{friedelprague, friedel77a}.

            Bernal has used a \emph{polyhedral approach} in order to analyze hand-made systems of equal spheres, and has shown that a large majority of the polyhedra (ca. 86 $\%$) are tetrahedra.  Hence the predominance of local icosahedral order.  Furthermore, these tetrahedra arrange frequently into \emph{pseudo-nuclei} that are aggregates of face-sharing tetrahedra, two by two, and tend to build a connected lattice in the whole structure, wrapping themselves so-to-speak around the larger holes (i.e. rather low density polyhedra with V = 8, 9 or 10 vertices). Of course the tetrahedra cannot fill in the whole 3-dimensional space; however, since they can extend freely along one direction, one notices a large number of three-stranded spirals, right- or left-handed, formed by a one dimensional array of regular tetrahedra.  Their local density is large. They are reminiscent of the twisted great circles of S$^{3}$, and therefore fit locally quite well into the curved space crystal representative of the local order. The pseudo-nuclei are the regions of less frustrated order.

            Numerous analyses of atomic packings have followed Bernal's pioneering work; we refer the interested reader to \textcite{zallen79}'s review on dense random packings.\\

\noindent \textsl{c. {The \{3,3,5\} template.}}  \label{The3,3,5template}
    This is an example of a crystalline structure in a curved space. There are no icosahedral Euclidean crystals, but icosahedral order is compatible with a space of constant curvature, namely the 3-sphere S$^{3}$, which can be tiled with regular tetrahedra, 20 of them meeting at a vertex, i.e. generating an icosahedron.  This structure is known after \textcite{coxeter73}, as the \{3,3,5\} polytope:

        elementary facets have 3 edges, hence are equilateral triangles,

    3 facets meet at a vertex, hence elementary cells are regular tetrahedra,

    5 cells share a common edge, hence an edge is a five-fold axis.

    This polytope has ${N_{0}=120}$ vertices, ${N_{1}=720}$ edges, ${N_{2}=1200}$ faces, and ${N_{3}=600}$  cells.  The passage from this template to a disordered system, i.e. the decurving of {\{3,3,5\}} and its infinite extension to a Euclidean space, occurs through the introduction in the perfect {\{3,3,5\}} crystal \cite{kleman79}, \cite{kleman89} of \emph{disclinations} of negative strength $-$ i.e. which introduce extra matter $-$ forming a 3D network in physical space, like the F\&K network.  For a description of quasicrystals in terms of frustration see \textcite{kleman88} and \textcite{kleman90b,kleman89}.

    \subsection{The decurving process}
\label{3,3,5decurved:geometry,topology,anddefects}
\subsubsection{Rolling without glide and disclinations}
 \label{Rollingwithoutglide}
    Let us now focus our attention to the transformation of a curved template into an actual flat medium, under the constraint that the local order of the template is conserved. An isometric mapping continuous all over the medium is not possible, because it would require that the Gaussian curvatures be equal at corresponding points \cite{hilbert64,darboux,singer96}.

     However, isometric mapping can be achieved locally, by \emph{parallel transport} along a line L, according to E. \textcite{cartan}. Letting M (S$^{3}$, say) roll without glide upon E$^{3}$, along any path L $\subset \,\textrm{S}^{3}$, lengths and angles along the path are conserved at corresponding points in E$^{3}$. If L is a geodesic of $\textrm{S}^{3}$, it maps along a straight line L in $\textrm{E}^{3}$: this is the so-called Levi-Civita connection.
    In general, such a mapping transforms a closed line L $\subset$ M into an open line in E$^{3}$. The closure failure can be described as a disclination.

    Consider a material cone: the curvature is concentrated at the apex, a useful feature here.  For it maps isometrically on the plane as a whole, except at the apex, which becomes the vertex of an empty wedge bordered by two generatrices. This is clearly the picture of the Volterra process for a disclination, whose angle, called here the \emph{deficit angle}, is a measure of the concentrated curvature. In order to complete the mapping, it suffices to fill the void with perfect matter. The final object does not carry stresses, if the material cone is amorphous; but it cannot be so if the medium is ordered, since then the disclinations are quantized.

    We show in the next paragraph that the decurving process yields generically two sets of disclinations:

    (i) those resulting from the mapping of  M onto the Euclidean space E$^3$; these disclinations are of negative (resp. positive) strength if M has positive (resp. negative) Gaussian curvature;

    (ii) those resulting from an elastic relaxation (disclinations of a sign opposite to the former ones).

The relationship between disclination lines and curvature has  been emphasized long ago by \textcite{kondo} and \textcite{bilby60}. Their approach is opposite to the present one; Kondo and Bilby start from E$^3$, which they consider as the habit space of the physical crystal, and map it on a space that is curved due to the presence of disclinations in E$^3$.  In contrast, the physics of the disordered system, which lives in E$^3$, is contained in its curved representation.

\subsubsection{The Volterra process in a curved crystal }
\label{VPincurvedcrystal }

It is useful to \emph{approximate} a Riemannian manifold by a piecewise flat manifold. Such a process of  'triangulation' has been proposed by \textcite{regge61} to calculate properties of curved manifolds in general relativity without using coordinates. For instance, in \{3,3,5\}, the edges of the lattice are replaced by straight lines in the embedding euclidean space, and the faces and cells by euclidean faces and cells. The edges, on which all the curvature is now concentrated, form the \emph{skeleton} of the triangulated manifold (the 1-bones), which is articulated at the vertices (the 0-bones).
      \begin{figure}
\includegraphics{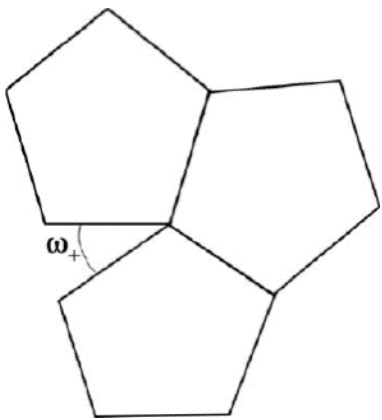}
   \caption{\label{deficit}{Deficit angle at a merging vertex of three regular pentagons, $\omega_{+}= \frac{\pi}{5}$}}
\end{figure}
Consider a D = 2 example, \{5,3\} (three pentagons at each vertex); its Regge image is a dodecahedron with flat pentagonal faces. By mapping the pentagons around a vertex (where the curvature is concentrated) onto the plane, one lets appear a \emph{deficit angle}, :
                         \begin{equation} \label{e100} \omega_{+} = 2\,\pi - 3\times\frac{3\,\pi}{5} = \frac{\pi}{5} \end{equation}
                         Fig. \ref{deficit}
\noindent which is not equal to the angle of a quantized disclination in \{5,3\}, namely $\Omega \,=\,3\pi/5. $ The stress field produced by the Volterra process at such a vertex is that one of a negative disclination of strength $\omega_{-}\,=\, \Omega -\omega_{+}\, =\,-2\pi/5$. This angle also measures the deficit angle of the local \emph{negative} Gaussian curvature generated by the introduction of extra matter; this negative Gaussian curvature would manifest itself as a locally hyperbolic surface element, if the  disclinated \{5,3\} is allowed to relax elastically in 3 dimensions.

It is energetically unfavorable that the only disclinations present in the actual medium be of negative strength. Therefore all the vertices are not the seats of mapping disclinations, and those which are not are flattened by force, which yields stresses characteristic of positive disclinations of strength $\omega_{+}$.

     In 3 dimensions, the curvature is concentrated along the 1-bones; if one moves a vector by parallel transport on a closed circuit L about an edge \emph{i}, it rotates by an angle $\omega _{i}$ (the deficit along \emph{i})
                         \begin{equation}          \label{e103}                                                                            \omega_{i}  = 2\pi -\sum \alpha _{k} \end{equation}
which does not depend on the precise location of L as long as it is traced in euclidean space and does not cross another edge;  $\alpha _{k} $ is the dihedral angle of the local flat polyhedron \emph{k} with edge \emph{i}.  One can perform a Volterra process, now truly reminiscent of the usual Volterra process in flat space, either by gluing the 2 lips across the angle $\omega_{i}$, or by inserting (removing) a lattice unit cell $\Omega$ in the space separating the lips.  According to the case, one introduces a topological disclination of strength $\pm|\omega_{i}|$ or $\mp|\omega_{i} -\Omega|$. These disclinations are \emph{wedge disclinations}, since they lie along a rotation vector $\bm \Omega$. The Volterra process for \emph{twist disclinations} does not add or remove matter; therefore it does not change the curvature.

{The deficit vectors $\bm \omega_{i }$ that are merging at a vertex and are parallel to the corresponding edges obey the Kirchhoff relation  \begin{equation}       \label{e104}                                                                               \sum \bm \omega_{i } =  0,   \end{equation} here written for small $ \omega_{i }$s.
This expression is in fact the Bianchi identity for curvatures when the deficit angles are small \cite{regge61}.}

             To summarize, the actual medium contains two sets of disclinations, call them D$_s$-lines when they carry spherical curvature and are positive strength disclinations, D$_h$-lines otherwise. Notice the duality between D$_s$- and D$_h$- lines; it means that one can start equally from a spherical crystal or from a hyperbolic crystal to construct a disordered system  \cite{kleman83,kleman82a}. The precise location and density $\varrho_s$ and $\varrho_h $ of the D$_s$- and D$_h$- lines is of course subject to a great arbitrariness, and it is the best elastic balance which decides of the final choice. From that point of view an amorphous solid with local icosahedral order is certainly closer to a spherical crystal than an hyperbolic crystal, and one expects $\varrho_s> \varrho_h $.

           \subsection{The concept of a non-Euclidean amorphous medium}
\label{Whynon-Euclideanamorphousmedium}

  The foregoing model of an amorphous medium, based on the existence of a frustrated order, does not forbid the \textit{conceptual} possibility of a homogeneous, isotropic, structureless, medium in $\textrm{S}^{3}$, whose defects we can investigate after the manner of the amorphous medium discussed in \ref{CONTINUOUSDEFECTS}.

  The model we have in mind has no relation whatsoever with any kind of local order.  We denote it \{am/$\textrm{S}^{3}$\}.  Because of the curvature imposed by the \emph{habit space} $\textrm{S}^{3}$, the singularities that break the continuous rotation and (non-commutative) translation symmetries are somehow at variance with disclinations and dislocations in flat Euclidean space.  This is also true for the \{3,3,5\} case, but the continuous case is expected to be easier to grasp. Furthermore, the concept itself of a \emph{curved amorphous medium} is new and worth investigating in its own right.  As we have just observed, a pending question in the {\{3,3,5\}} case is how to decurve such a template in order to get an atomic liquid or amorphous medium with icosahedral local order. The difficulty lies in the fact that the {\{3,3,5\}} disclinations are quantized, and the answer is not unique; stresses remain. The same question for {\{am/$\textrm{S}^{3}$\}} yields a unique answer with no stresses remaining, since defects are continuous in strength and distribution.

Finally there is the question of disclination networks,
analogous in spirit to Frank dislocation networks.  Disclination networks are apparent in polynanocrystals (\ref{Nanocrystals}) and in Frank and Kasper phases (\ref{FKphases}). They might also be important in undercooled liquids, but the true local geometry is that one
of a spherical crystal with icosahedral symmetry (\ref{The3,3,5template}). This situation obviously requires reconsidering the Kirchhoff relations
in a curved habit space, either amorphous (as a generalization of \ref{Generalizedpolygonalloops.Kirchhoff'srelation.}, see \ref{Kirchhoff'srelations}) or icosahedral (i.e. with quantized disclinations, see \ref{The3,3,5defects}).

\section  {Defects in 3-Sphere Templates}
\label{TemplatesintheThree-Sphere}

\subsection {Geometry and topology of a 3-sphere; a reminder}
\label{Geometryandtopologyofa3-sphere}
A point \textbf{M} on the 3-sphere $\textrm{S}^{3}$ of unit radius will be defined by its Cartesian coordinates in $\textrm{E}^{4}$, $\{\textrm{x}_{0},\textrm{x}_{1},\textrm{x}_{2}, \textrm{x}_{3}\}$  or more concisely by the unit quaternion $x$
\begin{equation}
x={\textrm{x}_{0}+\textrm{x}_{1}i+ \textrm{x}_{2}j+ \textrm{x}_{3}k} \\
|x|^{2}=x\tilde{x}=\textrm{x}_{0}^{2}+\textrm{x}_{1}^{2}+ \textrm{x}_{2}^{2}+ \textrm{x}_{3}^{2}=1      \label{(35)}
\end{equation}
where $\tilde{x}$ stands for the (complex) conjugate of $x$.

\subsubsection{The rotation group \textrm{SO(3)} in quaternion notation}  \label{TherotationgroupSO(3)inquaternionnotation}
    The set of unit quaternions forms a group that is related to the group of rotations SO(3) in $\textrm{E}^{3}$, as follows.  The unit quaternion $x$
 \begin{equation}
 x=\cos\vartheta+q\sin\vartheta, \, \, \,  q=\frac{{\textrm{x}_{1}i+ \textrm{x}_{2}j+ \textrm{x}_{3}k}}{\sqrt{x_{1}^{2}+ x_{2}^{2}+x_{3}^{2}}}   \label{(36)}
 \end{equation}

\noindent is representative of a rotation $\vartheta$ along the direction $\{\textrm{x}_{1}, \textrm{x}_{2}, \textrm{x}_{3}\}$; two antipodal points $x,\, -x$ on $\textrm{S}^{3}$ embedded in $\textrm{E}^{4}$ are representative of the same rotation $\vartheta$ and $\vartheta+\pi$ along the same direction ${x_{1}, x_{2}, x_{3}}$; $q$ is called a pure unit quaternion; its real part vanishes, $q^{2}= -1, \,q\tilde{q}=1$.  All SO(3) rotations are therefore represented by a sphere $\textrm{S}^{3}$ with antipodal points identified, namely $\textrm{P}^{3}=\textrm{S}^{3}/Z_{2}$, the projective plane in three dimensions.  Reciprocally, $\mathrm{Q}=\textrm{S}^{3}$, the multiplicative group of unit quaternions,  is the \emph{double covering} of $\textrm{P}^{3}$ and, as a topological group, is isomorphic to SU(2) and homomorphic 2:1 to the group SO(3) of all rotations that leave the origin fixed,
\begin{equation}
\textrm{SO}(3)=\textrm{SU}(2)/\textrm{Z}_{2},      \label{(37)}
\end{equation}
the kernel of the homomorphism being generated by the rotation $2\pi$.

 Because of the validity of Moivre's formula for unit quaternions, $x$ can also be denoted
                            \begin{equation}
                            x=\exp {\tfrac{\vartheta}{2}q}.    \label{(38)}
                            \end{equation}

 \subsubsection{The rotation group SO(4) in quaternion notation}
  \label{Therotation groupSO(4)inquaternionnotation}
    The quaternion notation provides an easy analysis of the basic isometric transformations of $\textrm{E}^{4}$ \cite{coxeter91} that conserve a fixed point (the centre of $\textrm{S}^{3}$). These are:

(i) the {\emph{single rotation}}, with one pointwise fixed 2D plane, the so-called axial plane containing the origin \textbf{O}; this plane is the $\textrm{E}^{4}$-generalization of the rotation axis in $\textrm{E}^{3}$.  For that reason, one shall often call the single rotation in $\textrm{E}^{4}$ as the \emph{rotation}.

(ii)  the \emph{double rotation} (with one fixed point only, the centre of $\textrm{S}^{3}$), which is the commutative product of two rotations about two completely orthogonal axial planes.\\

\noindent \textit{a. {The single rotation.}} \label{Thesinglerotation}
The basic formula for a rotation of angle $\alpha$ about the axial plane $\Pi_{1,w}=(0,1,w)$, defined by three points in $\textrm{E}^{4}$, namely a)  the origin $\{0,0,0,0\}$ $-$denoted $\{0\}$, b) $\{1,0,0,0\}$ $-$denoted $\{1\}$, c) $\{0,\textrm{w}_{1},\textrm{w}_{2},\textrm{w}_{3}\}$ $-$denoted $\{w\}$, is:
                \begin{equation}
                x^{\prime}=e^{-\tfrac{\alpha}{2}w}\,{x}\,e^{\tfrac{\alpha}{2}w} \label{(39)}
                \end{equation}

 The transformation expressed by Eq.~\ref{(39)} leaves invariant any point $x$ of $\Pi_{1,w}$, i.e. $x=\lambda+\mu w$  ($\lambda$, $\mu$ real), and no other points.
 \begin{center}
$\lambda+\mu w=e^{-\tfrac{\alpha}{2}w}({\lambda+\mu w})e^{\tfrac{\alpha}{2}w}$\\
\end{center}

If $x$ is a pure quaternion (not necessarily a unit one), Eq.~\ref{(39)} gives the expression of the rotation of its representative point $\mathbf{x}$  by an angle $\alpha$ about the axis $\mathbf{w}$, on a two-sphere $\textrm{S}^{2}$ of radius $|x|$. Therefore a pair of two conjugate unit quaternions $(e^{-\tfrac{\alpha}{2}w},e^{\tfrac{\alpha}{2}w})$ represents one element of SO(3); the pair $(-e^{-\tfrac{\alpha}{2}w},-e^{\tfrac{\alpha}{2}w})$ represents the same element; the representation is 2:1, as already noted.

$\Pi_{1,w}$ intersects the habit sphere $\textrm{S}^{3}$ of \{3,3,5\} or of \{am/$\textrm{S}^{3}$\}, radius $R$, along a great circle $\textrm{C}_{1,w}$, radius $R$, which is thereby pointwise invariant in the rotation given by Eq.~\ref{(39)}; $\{R\}$ and $\{Rw\}$ belong to this great circle.

\textsf{The vector }$\alpha \mathbf{t}$ \textsf{tangent to} $\textrm{C}_{1,w}$ \textsf{at} \textbf{M} \textsf{is the local rotation vector in} $\textrm{S}^{3}$ \textsf{induced by the single rotation in} $\textrm{E}^{4}$.

With Eq.~\ref{(39)}, we have considered a special axial plane.  A rotation of angle $\alpha$ about a generic axial plane passing through the origin in $\textrm{E}^{4}$ is of the form
                \begin{equation}
                x^{\prime}=e^{-\tfrac{\alpha}{2}p}\,{x}\,e^{\tfrac{\alpha}{2}q} \label{(40)}
                \end{equation}

 \noindent $p\neq \pm q$, with axial plane:
\begin{equation}
\Pi_{p,q}=(0,1-pq,p+q). \label{(41)}
\end{equation}

It is easy to show that $1-pq$  and  $p+q$ are invariant in the transformation given by Eq.~\ref{(40)}.  Notice that $p$ and $q$ are pure unit quaternions.

This is the place to introduce the plane
                    \begin{equation}
                    \Pi_{p,q}^{\bot}=(0,1+pq,p-q) \label{(42)}
                    \end{equation}

 \noindent which is the plane completely orthogonal to $\Pi_{p,q}$; the directions denoted by $1+pq$ and $p-q$ are both orthogonal to  $1-pq$  and to $p+q$.  A rotation of angle $\beta$ about $\Pi_{p,q}^{\bot}$  taken as axial plane reads:
                \begin{equation}
                x^{\prime}=e^{\tfrac{\beta}{2}p}\,{x}\,e^{\tfrac{\beta}{2}q} \label{(43)}
                \end{equation}

Since the four directions  $1-pq$, $p+q$,  $1+pq$  and  $p-q$ are mutually orthogonal, they can be used as the directions of a Cartesian frame of reference in $\textrm{E}^{4}$.\\

\noindent \textit{b. {The double rotation. Right and left helix turns.}} \label{Thedoublerotation.Rightandlefthelixturns.}
The rotations about two completely orthogonal planes are \emph{commutative}. Consider the product of two rotations of the same angle $\alpha$ about $\Pi_{p,q}$  and $\Pi_{p,q}^{\bot}$.  According to Eq. \ref{(40)} and Eq. \ref{(43)}, we have $$x^{\prime}=e^{-\tfrac{\alpha}{2}p}\,(e^{\tfrac{\alpha}{2}p}\,{x}\,e^{\tfrac{\alpha}{2}q})\,e^{\tfrac{\alpha}{2}q},$$ i.e.
                            \begin{equation}
                            x^{\prime}={x}\,e^{{\alpha}q} \label{(44)}
                            \end{equation}

This double rotation conserves $\textrm{S}^{3}$ (and any three-sphere centered at the origin) globally, and leaves only one point invariant, the intersection of the two planes $\Pi_{p,q}$  and $\Pi_{p,q}^{\bot}$, i.e. the centre $\{0\}$ of $\textrm{S}^{3}$. It leaves no point invariant in $\textrm{S}^{3}$; thereby it is akin to a \emph{translation} in Euclidean space; E. Cartan introduced the term \textit{transvection} to connote such an operation in a Riemannian space \cite{cartan}.

 $p$ does not appear any longer in Eq.~\ref{(44)}, hence this transformation can be given a geometric interpretation in any pair of a large set of  completely orthogonal plane pairs.  For instance, it is a transformation that, in the 2-plane $\Pi_{1,q}$ containing the origin \{0\} and the directions \{1,0,0,0\}  and $q=\{0,\textrm{q}_{1},\textrm{q}_{2},\textrm{q}_{3}\}$, is a rotation of angle $\alpha$, and in the completely orthogonal 2-plane $\Pi_{1,q}^{\bot}$ is a rotation of the same angle $\alpha$ \cite{montes}.

 \begin{figure}
\includegraphics{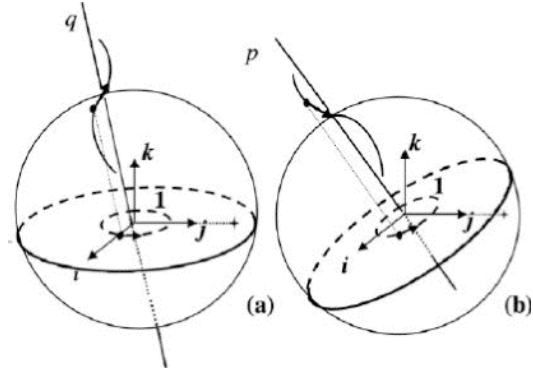}
   \caption{ \label{fig16}{(a) Right helix turn $x^{\prime}={x}\,e^{{\alpha}q}$ and (b) left helix turn $x^{\prime}=e^{{-\alpha}p}\,{x}$ in stereographic projection}}
\end{figure}
It can also be called a \emph{right helix turn}, for the following reason.  Consider the transvection $x^{\prime}={x}\,e^{{\alpha}i}$.  In order to visualize it, let us employ the stereographic projection of $\textrm{S}^{3}$ from its pole $\{-1,0,0,0\}$ onto the hyperplane spanned by the directions $i$, $j$, and $k$ (the 'imaginary' part of the quaternion set); $x^{\prime}={x}\,e^{{\alpha}i}$  turns the \emph{equator} plane spanned by the directions $j$ and $k$ (i.e. the plane $\Pi_{1,i}^{\bot}$ completely orthogonal to $\Pi_{1,i}$) to the \textit{right} by an angle $\alpha$ about the axis $i$, and pushes it \textit{forward} by the same angle, Fig.~\ref{fig16}(a).
The same is true, also in stereographic projection, for the transformation $x^{\prime}={x}\,e^{{\alpha}q}$, where $q$ is any pure unit quaternion chosen as the \emph{axis} of the rotation, since any axis $q$ results from the axis $i$ by a rotation in the 3D $\langle i,j,k \rangle$ hyperplane; all pure unit quaternions are visualizable in stereographic projection, as well as their corresponding equator planes.

The \emph{left helix turn}
 \begin{equation}
x^{\prime}=e^{{-\alpha}p}\,{x} \label{(45)} \end{equation}
\noindent results from the product of two rotations of the same angle and of opposite signs, $x^{\prime}=e^{-\tfrac{\alpha}{2}p}\,(e^{-\tfrac{\alpha}{2}p}\,{x}\,e^{\tfrac{\alpha}{2}q})\,e^{-\tfrac{\alpha}{2}q}$, performed on two completely orthogonal planes; it turns the equator of $p$ also to the \textit{right} by an angle $\alpha$, but pushes its equator plane \textit{backward} by the same angle, Fig.~\ref{fig16}(b).  A left helix turn and a right helix turn are commutative, but two left (or right) helix turns are not.  Therefore these transformations are akin to \textit{non-commutative} translations.

\subsubsection{Group of direct isometries in the habit 3-sphere S$^{3}$}   \label{Groupofdirectisometriesinthehabit3-sphere}
Any product of a right translation by a left translation is a direct isometry in $\textrm{S}^{3}$:
                        \begin{equation}
                        x^{\prime}=e^{{-\beta}p}\,{x}\,e^{{\alpha}q}; \label{(46)}
                        \end{equation}

\noindent it is also the product of two commutative rotations of angles $\alpha+\beta$ and $\alpha-\beta$ about two completely orthogonal axial planes  $\Pi_{p,q}$  and  $\Pi_{p,q}^{\bot}$ ;  a right (or left) helix turn is of the form exhibited by Eq.~\ref{(46)}, with $\beta$ (or $\alpha$) $=0, 2\pi$.  $\{-e^{{\alpha}p},-e^{{-\beta}q}\}$  generates the same isometry as $\{e^{{\alpha}p},e^{{-\beta}q}\}$ .

Any isometry in $\textrm{S}^{3}$ being generated by a combination of two commutative helix turns  $\{e^{{\alpha}p},e^{{-\beta}q}\}$, one right, one left $-$ each belonging to the group SU(2) $-$ the maximal group that leaves $\textrm{S}^{3}$ invariant is
        \begin{equation}\textrm{SO}(4)=\textrm{G}(\textrm{S}^{3}) \\ \sim \textrm{SU}(2) \times \textrm{SU}(2)/ \textrm{Z}_{2}  =\textrm{S}^{3} \times \textrm{SO}(3) \label{(47)}
            \end{equation}

It is a direct product.  Its universal cover is $\bar{\textrm{G}}(\textrm{S}^{3})=\textrm{S}^{3} \times \textrm{S}^{3}$, i.e. the group of unit quaternions, squared.

A reminder of some geometric characteristics of $\textrm{S}^{3}$ can be found in Appendix \ref{appthreesphere}.

\subsection{Disclinations and disvections in {S}$^{3}$}
\label{DisclinationsanddisvectionsinS3}

 The notion of line defect easily generalizes to any ordered medium in {S}$^{3}$. In terms of the Volterra process, we have two types of defects:
 \subsubsection{Disclinations in {S}$^{3}$}
\label{DisclinationsinS3}
Disclinations break a single rotation in {E}$^{4}$, defined by the SO(4) group element $\{e^{{-\tfrac{\Omega}{2}p}},e^{{\tfrac{\Omega}{2}q}}\}$  about some axial plane $\Pi_{p,q}$; it conserves the habit sphere $\textrm{S}^{3}$ globally. The plane $\Pi_{p,q}$ intersects $\textrm{S}^{3}$ along a great circle C. For an observer confined to the habit sphere, this rotation appears as a set of rotation vectors  $\Omega \mathbf{t}$ tangent to C all along, as already indicated.  The points $x$ of the cut surface are displaced by the single rotation $$x'=e^{{-\tfrac{\Omega}{2}p}}\,x\,e^{{\tfrac{\Omega}{2}q}}.$$ Remember that any great circle is a geodesic of $\textrm{S}^{3}$, so that this operation is clearly a generalization of a disclination in $\textrm{E}^{3}$, where the rotation vectors are along straight lines, i.e. Euclidean geodesics.
 \subsubsection{Disvections in {S}$^{3}$}
\label{DisvectionsinS3}
Disvections break transvection symmetries and are thereby the generalizations of dislocations, which break translation symmetries. A right (say) transvection

\begin{center} $x_{i}^{\prime}={x_{i}}\,e^{{\alpha}q},\,\,|x_{i}|=R,$ \end{center} brings $\mathbf{M}_{i}(x_{i})$ to $\mathbf{M'}_{i}(x_{i}^{\prime})$, at a distance $|b|=|x_{i}^{\prime}-x_{i}|=2\,R\,\sin\frac{\alpha}{2}$, along the great circle C$_{i}$ through $\mathbf{M}_{i}$ and $\mathbf{M'}_{i}$.  $\mathbf{M'}_{i}$ in turn is brought by the same distance $|b|$ along the same great circle. Any point $\mathbf{M}_{j}$ outside C$_{i}$ likewise follows another great circle C$_{j}$ which is equidist\emph{}ant to C$_{i}$, with the same $|b|$, i.e. which turns about C$_{i}$ in a double-twisted manner. The points $x$ of the cut surface are displaced by the double rotation $$x'=x\,e^{{\alpha q}}.$$ This operation is clearly a generalization of a dislocation in $\textrm{E}^{3}$, with $\mathbf{b}$ the analog of the Burgers vector.

The term of disvection was introduced to connote line defects that break non-commutative translation symmetries in quasicrystals \cite{kleman92}.

\subsection{Defects of the double-twist S$^{3}$ template}
\label{DefectsoftheS3templateofdouble-twist}

The topological classification of line defects in the double-twist template \{dtw/{S}$^{3}$\} is the same as that one of the uniaxial nematic \cite{seth85}, namely
\begin{equation}  \label{(dtw)}
\Pi_{1}(\mathrm{V}_{dtw})=\mathrm{Z}_{2},
\end{equation}
since $\mathrm{V}_{dtw}= \dfrac{\mathrm{SO}(4)}{\mathrm{SU}(2) \times \mathrm{D}_{2h}}=\mathrm{P}_{2}$, the projective plane. In this expression,

 SO(4)=SO(3)$\times$SU(2) is the group of isometries of S$^{3}$ (Eq.~\ref{(47)}),

 SU(2), in the denominator, is isomorphic to the group of left \emph{or} right transvections $-$ the template has a definite chirality,

  D$_{2h}$ is the rotation point group, as in a N phase. These results can be obtained using the geometric picture of the Hopf fiber bundle (see Fig.~\ref{fig17} in Appendix \ref{appthreesphere}), which is the geometrical representation of \{dtw/{S}$^{3}$\}.

Whereas the topological stability analysis provides only one class of topological defects, namely $|k|=\frac{1}{2}$, the Volterra process provides many more, exactly as for the N phase. Therefore one has to differentiate $k=-\frac{1}{2}$ disclinations, which \emph{add} matter, and $k=\frac{1}{2}$ disclinations, which \emph{remove} matter. In principle, only the first category of disclinations are liable to decurve \{dtw/{S}$^{3}$\} to an Euclidean medium; this is precisely the result which is claimed in the current BP structural models \cite{meiboom83}. However a \emph{caveat} is in order: because viscous relaxation operates at constant density, a negative sign for the decurving disclinations should not be a prerequisite; \textcite{seth85} has indeed proposed a model where there are no disclinations at all, with however the same cubic structure.

\subsection{Continuous defects in a 3D-spherical isotropic uniform medium}
\label{Continuousdefectsina3D-sphericalisotropicmedium}

\subsubsection{The wedge disclination}  \label{ThewedgedisclinationinS3}
First, a few remarks about the great circles of $\textrm{S}^{3}$, which are geodesics of $\textrm{S}^{3}$, i.e. the equivalent of straight lines in flat space.\\

\noindent \textsl {a. Wedge disclinations are along great circles.} \label{Wedgedisclinationsarealonggreatcircles}
\indent Let $\Pi_{p,q}=\{0,1-pq,p+q\}$ be the plane that intersects the habit sphere  along the great circle C.  We have $|1-pq|^{2}=|p+q|^{2}=2-pq-qp,$ where $-pq-qp=2\mathbf{p}\cdot \mathbf{q}=2\,\cos{\phi_{p,q}}$,  ${\phi_{p,q}}$ being the angle between $\mathbf{p}$ and $\mathbf{q}$; $(1-pq)$ and $(p+q)$ are orthogonal.  We then have, for any point $u$ on C,
 \begin{center}
 $u=a(1-pq)+b(p+q) \qquad        |u|=R$
  \end{center}

We choose the origin of the angles in the $\Pi_{p,q}$ plane along the direction $(1-pq)$; one then gets:
\begin{equation}
            u(\vartheta)=R\,\textrm{N}_{p,q}\,e^{{\tfrac{\vartheta}{2}p}}(1-pq)e ^{{\tfrac{\vartheta}{2}q}} \\ =  R\,\textrm{N}_{p,q}\,\left[(1-pq)\,\cos{\vartheta}+(p+q)\,\sin{\vartheta}\right].    \label{(48)}
\end{equation}

\noindent where
$$\textrm{N}_{p,q}=(2-pq-qp)^{-1/2}=(2\,\cos \tfrac{\phi_{p,q}}{2})^{-1}.$$
The first equality expresses the fact that the points of C are obtained by a rotation $\{e^{{\tfrac{\vartheta}{2}p}},e^{{\tfrac{\vartheta}{2}q}}\}$ about $\Pi_{p,q}^{\bot}=\{0,1+pq,p-q\}$  in $\textrm{E}^{4}$.  Observe that the rotation  $\{e^{{\tfrac{\vartheta}{2}p}},e^{{\tfrac{\vartheta}{2}q}}\}$ keeps pointwise invariant the intersection of $\Pi_{p,q}^{\bot}$  with $\textrm{S}^{3}$, which is a great circle $\textrm{C}^{\bot}$  orthogonal to C. The angular distance between any pair of points $\mathbf{M}\in\textrm{C}$, $\mathbf{M}^\bot \in\,{\textrm{C}^\bot}$, is equal to $\pi/2$ (the two vectors $\mathbf{OM}$ and $\mathbf{OM}^\bot$  are orthogonal). C and $\textrm{C}^{\bot}$, which have the same center and the same diameter, have no point in common.\\

\noindent \textsl {b. Volterra elements of a wedge disclination.}  \label{VPelementsofawedgedisclination}
\indent Consider now the rotation $\{e^{{-\tfrac{\Omega}{2}p}},e^{{\tfrac{\Omega}{2}q}}\}$  about $\Pi_{p,q}$; it conserves $\textrm{S}^{3}$ globally. The plane $\Pi_{p,q}$ intersects $\textrm{S}^{3}$ along C. In $\textrm{S}^{3}$, this rotation is a set of rotation vectors  $\Omega \mathbf{t}$ tangent to C all along.  It is therefore a rotation that builds C Volterra-wise as a line of \emph{wedge} character, strength $\Omega$.  Its constitutive dislocations are as follows.  Let $x=x^{\|}+x^{\bot}$ be the quaternion representation of a point $\mathbf{P=x^{\|}+x^{\bot}}$ on the cut surface $\Sigma_{\mathbf{C}}$ of C (it is not useful at this stage to select a special cut surface), split into its two components, one $x^{\|}$ belonging to $\Pi_{p,q}$ $-$ invariant in the rigid Volterra rotation $\{e^{{-\tfrac{\Omega}{2}p}},e^{{\tfrac{\Omega}{2}q}}\}$  that moves apart the two lips of the cut surface $-$, the other one $x^{\bot}$, belonging to $\Pi_{p,q}^{\bot}$, not invariant in the rotation under consideration. Hence

\begin{center}
$x'^{\|}=x^{\|}=e^{-\tfrac{\Omega}{2}p}\,x^{\|}\,e^{\tfrac{\Omega}{2}q},$  \qquad
$x'^{\bot}={e}^{-\tfrac{\Omega}{2}p}\,x^{\bot}\,e^{\tfrac{\Omega}{2}q}.$\end{center}
  Thus, $|x^{\bot}|$ is the distance of $\mathbf{P}$ to the axial plane $\Pi_{p,q}$  of the Volterra process we are considering.

The sum \textit{total} of the Burgers vectors of the constitutive dislocations between the wedge line and $\mathbf{P}$ is, in quaternion notation,
\begin{equation}
b=x^{\prime}-x= {e}^{-\tfrac{\Omega}{2}p}\,x\,e^{\tfrac{\Omega}{2}q}-x \label{(49)},
\end{equation}

\noindent which can also be written
\begin{equation}
b=\sin{\tfrac{\Omega}{2}}\,{e}^{-\tfrac{\Omega}{2}p}\,(xq-px)  =\sin{\tfrac{\Omega}{2}}\,(xq-px)\,e^{\tfrac{\Omega}{2}q} \label{(50)}.
\end{equation}

But, because $x^{\|}$  belongs to $\Pi_{p,q}$ and $x^{\bot}$ to $\Pi_{p,q}^{\bot}$, we have
$$\textit{x}^{\bot}=e^{\tfrac{\vartheta}{2}p}\,\textit{x}^{\bot}\,e^{\tfrac{\vartheta}{2}q},  \qquad
\textit{x}^{\|}=e^{-\tfrac{\vartheta}{2}p}\,\textit{x}^{\|}\,e^{\tfrac{\vartheta}{2}q},$$

\noindent for any $\vartheta$. Hence
\begin{equation}
x^{\bot}q+px^{\bot}=0 \qquad x^{\|}q-px^{\|}=0 \label{(51)}.
\end{equation}

\noindent and eventually
\begin{equation}
b=-2\sin\tfrac{\Omega}{2}\,p\,{x^{\bot}} \,e^{\tfrac{\Omega}{2}q}  \equiv 2\sin\tfrac{\Omega}{2}\,e^{-\tfrac{\Omega}{2}p}\,{x^{\bot}}\,q.  \label{(52)}
\end{equation}

\noindent To make contact with the discussion about Eq.~\ref{e1}, we make use of two remarks

(i) the modulus of $b$, calculated from Eq.~\ref{(52)}, is
                        \begin{equation}
                        |b|=2|x^{\bot}|\sin\tfrac{\Omega}{2} \label{(53)}
                        \end{equation}

(ii) the Burgers vector \textbf{b} is perpendicular to $\Pi_{p,q}$. Notice that it is also orthogonal
to $\textbf{x}^{\bot}$ and to $\textbf{x}'^{\bot}$, which both belong to $\Pi_{p,q}^{\bot}$; hence $\mathbf{b}\cdot \mathbf{OP}=\mathbf{b}\cdot \mathbf{OP'}=0$.

These two properties belong also to the constitutive dislocations of a wedge line in $\{\textrm{am/}\textrm{E}^{3}\}$. Note that Eq.~\ref{(53)} reproduces the integral of Eq.~\ref{e1} performed on the interval between the line and a point \textbf{P} on the cut surface for a wedge disclination, since $|x^{\bot}|$ is in both cases the distance to the invariant manifold (in Eq.~\ref{e1} it is the distance to the rotation axis of the disclination.) The fact that \textbf{b} is orthogonal to the axial plane $\Pi_{p,q}$ compares to the fact that in Eq.~\ref{e1} the Burgers vector is orthogonal to the axis of rotation.

 The Burgers vector $d{b}$ of the constitutive dislocations situated between $x$ and $x+dx$ on the cut surface is
 $$db={e}^{-\tfrac{\Omega}{2}p}\,dx\,e^{\tfrac{\Omega}{2}q}-dx,$$

\noindent which can also be written
\begin{equation}
db=-2\sin\tfrac{\Omega}{2}\,p\,{dx^{\bot}} \,e^{\tfrac{\Omega}{2}q}  \equiv 2\sin\tfrac{\Omega}{2}\,e^{-\tfrac{\Omega}{2}p}\,{dx^{\bot}}\,q.  \label{(54)}
 \end{equation}

To summarize: \textsf{Eq.~\ref{(54)}, valid in} $\{\textrm{am}/\textrm{S}^{3}\},$ \textsf{is the equivalent of Eq.~\ref{e1}, valid in} $\{\textrm{am}/\textrm{E}^{3}\}$. \textsf{It describes the constitutive dislocations of a wedge disclination $\Omega$ located at the intersection of the habit 3-sphere and the 2-plane $\Pi_{p,q}=\{0, 1-pq, p+q\}$.}

It remains to find the manifolds along which $d\mathbf{b}$  is not only constant in magnitude, but also in direction, i.e. the geometry of the constitutive dislocation lines.  The relation $x^{\bot}=\textrm{constant}$ defines a two-plane parallel to $\Pi_{p,q}$, which intersects the habit sphere along a circle $\textrm{C}_{x^{\bot}}$, of radius $|x^{\|}|$, which is not a great circle, except if  $x^{\bot}= 0$, in which case $\textrm{C}_{x^{\bot}}$ is the disclination C itself.  C and $\textrm{C}_{x^{\bot}}$ are parallel, perpendicular to some direction $d_{x_{\bot}}$ belonging to $\Pi_{p,q}^{\bot}$; $b$, which depends only on $x^{\bot}$, is constant on $\textrm{C}_{x^{\bot}}$, which is indeed a constitutive dislocation of the wedge disclination.  The set of constitutive dislocations $\textrm{C}_{x^{\bot}}$ for $0\leq |x^{\|}| \leq {R}$, so chosen that these circles are in parallel planes, all perpendicular to the same direction $d_{x^{\bot}}$ in $\Pi_{p,q}^{\bot}$, describes half a great sphere, which is a geodesic manifold of $\textrm{S}^{3}$, and consequently the equivalent of a half-plane in flat space.  There is a one-parameter family of such half great spheres, depending on the direction $d_{x^{\bot}}$ in $\Pi_{p,q}^{\bot}$, each of them playing the role of a possible cut surface $\Sigma_{\mathbf{C}}$ for the wedge disclination C.

Equation~\ref{(52)} can be given an interpretation in terms of disvections.  We introduce  the quaternion $z_{\bot}=-p\,x^{\bot}=x^{\bot}\, q$; it results geometrically from a rotation by an angle of $\frac{\pi}{2}$  of the vector $\mathbf{x}^{\bot}$ in the  $\Pi_{p,q}^{\bot}$ plane, so that we also have  $z^{\bot}=e^{-\tfrac{\pi}{4}p}\,x^{\bot} \,e^{\tfrac{\pi}{4}q}$;  $z^{\bot}$ is \emph{constant} all along $\textrm{C}_{x^{\bot}}$.  Hence \begin{equation}
            db=2\sin\tfrac{\Omega}{2}\,({dz^{\bot}}\,e^{\tfrac{\Omega}{2}q}) \equiv 2\sin\tfrac{\Omega}{2}\,(e^{-\tfrac{\Omega}{2}p}\,{dz^{\bot}})  \label{(55)}
\end{equation}

\noindent appears either as a left screw or a right screw.

\subsubsection{Defects  attached to a disclination in \{am/S$^3$\}} \label{Defects attachedtoadisclinationinthe3-spherecurvedamorphousmedium}

\noindent \textsl {a. Useful identities and relations.} \label{Usefulidentitiesandrelations}

 We introduce \textit{two} great circles at each point \textbf{M} of a generic disclination line L; one of them $\textrm{C}_{m}$ is tangent to L at \textbf{M} along the direction $\mathbf{m}$, the other one $\textrm{C}_{\mu}$ tangent to the local rotation vector $\mathbf{\Omega}=\Omega \bm \mu$;
 $m\tilde{m}=\mu \tilde{\mu}=1$. Their analogs in $\{\textrm{am}/\textrm{E}^{3}\}$ are the straight lines along \textbf{m}  $-$ the tangent to the disclination line $-$ and $\bm\mu$ $-$ along the rotation vector. The 2-planes to which these circles belong are denoted $\Pi_{p,q}=\{0,1-pq,p+q\}$ for $\textrm{C}_{m}$, $\Pi_{\varrho,\sigma}=\{0,1-\varrho\sigma,\varrho+\sigma\}$ for $\textrm{C}_{\mu}$; $p$, $q$, $\varrho$,  $\sigma$, are defined as in Eq.~\ref{(56a)} and~\ref{(56b)}.
\begin{equation}
p=-\frac{1}{{R}} u \,\tilde{m}=\frac{1}{{R}} m \, \tilde{u},  \, \,  q=\frac{1}{{R}}\tilde{u } \, m=-\frac{1}{{R}}\tilde{m} \, u,
\label{(56a)}
\end{equation}

\begin{equation}
\varrho=-\frac{1}{{R}} u \,\tilde{\mu}=\frac{1}{{R}} \mu \, \tilde{u},  \, \, \sigma=\frac{1}{{R}}\tilde{u } \, \mu=-\frac{1}{{R}}\tilde{\mu} \, u,
\label{(56b)}
\end{equation}

The demonstration is given in Appendix \ref{appthreesphere}. Notice that the relations $m \tilde{u}+u \tilde{m}=0$ and $\tilde{u }m+\tilde{m}u=0$, easily deduced from Eq.~\ref{(56a)}, express the orthogonality of $\bm m$ and \textbf{u}, $\bm m \cdot \textbf{u}=0$. Likewise, $\mu \tilde{u}+u \tilde{\mu}=0$, etc.

Observe that $m$ is invariant in any rotation about  $\Pi_{p,q}$, and that $\mu$  is likewise invariant in any rotation about $\Pi_{\varrho,\sigma}$; this yields the relations
 \begin{center}
 $e^{-\tfrac{\vartheta}{2}p}\,{m}\,e^{\tfrac{\vartheta}{2}q}=m, \qquad e^{-\tfrac{\vartheta}{2}\varrho}\,{\mu}\,e^{\tfrac{\vartheta}{2}\sigma}=\mu,$
 \end{center}
for any $\vartheta$; hence:
                \begin{equation}
                p\,m-m\,q=0, \qquad \varrho\,\mu-\mu\,\sigma=0. \label{(57)}
                \end{equation}

These relations also stem from Eq.~\ref{(56a)} and~\ref{(56b)}, but it is worth retrieving them this way, in order to emphasize the role of invariance by rotation.  Since \textbf{M} belongs to both axial planes  $\Pi_{p,q}$ and $\Pi_{\varrho,\sigma}$ , we have
  \begin{center}
 $e^{-\tfrac{\vartheta}{2}p}\,{u}\,e^{\tfrac{\vartheta}{2}q}=u, \qquad e^{-\tfrac{\vartheta}{2}\varrho}\,{u}\,e^{\tfrac{\vartheta}{2}\sigma}=u,$
 \end{center}
for any $\vartheta$; hence:
                \begin{equation}
                p\,u-u\,q=0, \qquad \varrho\,u-u\,\sigma=0. \label{(58)}
                \end{equation}

\noindent also steming from Eq.~\ref{(56a)} and~\ref{(56b)}.

More generally, the quaternion components $a^{\|}$ and $a^{\bot}$, resp. in $\Pi_{p,q}$ and $\Pi_{p,q}^{\bot}$, of any vector $a=a^{\|}+a^{\bot}$ obey the relations
\begin{equation}
p\,a^{\|}-a^{\|}\,q=0, \qquad  p\,a^{\bot}+a^{\bot}\,q=0 \label{(59a)},
\end{equation}

\noindent   and similar relations for its components $a_{\|}$ and $a_{\bot}$ in $\Pi_{\varrho,\sigma}$ and $\Pi_{\rho,\sigma}^{\bot}$,
\begin{equation}
\varrho\,a_{\|}-a_{\|}\,\sigma=0, \qquad  \varrho\,a_{\bot}+a_{\bot}\,\sigma=0. \label {(59b)}
\end{equation}

By differentiating, one also gets:
\begin{equation*}
e^{\tfrac{\vartheta}{2}p}\,{u\,dq}\,e^{\tfrac{\vartheta}{2}q}=u\,dq \qquad e^{\tfrac{\vartheta}{2}p}\,{dp\,u}\,e^{\tfrac{\vartheta}{2}q}=dp\,u \end{equation*}
  \begin{equation*}e^{\tfrac{\vartheta}{2}\varrho}\,{u\,d\sigma}\,e^{\tfrac{\vartheta}{2}\sigma}=u\,d\sigma \qquad \textrm{e}^{\tfrac{\vartheta}{2}\varrho}\,{d\varrho\,u}\,e^{\tfrac{\vartheta}{2}\sigma}=d\varrho\,u\\
 \,\end{equation*}
   \begin{equation*}e^{\tfrac{\vartheta}{2}p}\,{m\,dq}\,e^{\tfrac{\vartheta}{2}q}=m\,dq \qquad e^{\tfrac{\vartheta}{2}p}\,{dp\,m}\,e^{\tfrac{\vartheta}{2}q}=dp\,m\\
 \, \end{equation*}   \begin{equation}e^{\tfrac{\vartheta}{2}\varrho}\,{\mu\,d\sigma}\,e^{\tfrac{\vartheta}{2}\sigma}=\mu \,d\sigma \qquad e^{\tfrac{\vartheta}{2}\varrho}\,{d\varrho\,\mu}\,e^{\tfrac{\vartheta}{2}\sigma}=d\varrho\,\mu
\end{equation}

 All these relations are easy to establish directly, by employing identities of the type
\begin{equation}
e^{\tfrac{\vartheta}{2}\varrho}\,d\varrho=d\varrho\,e^{-\tfrac{\vartheta}{2}\varrho} \qquad d\sigma\,e^{\tfrac{\vartheta}{2}\sigma}=e^{-\tfrac{\vartheta}{2}\sigma}\,d\sigma \label {(60)}.
\end{equation}

\noindent which express rotational invariance about
 $\Pi_{p,q}^{\bot}$ and $\Pi_{\varrho,\sigma}^{\bot}$:

the quaternions $$u\,dq,\:\:dp\,u,\:\:m\,dq,\:\:dp\,m $$ belong to the plane $\Pi_{p,q}^{\bot}$; they are invariant in any rotation about the axial plane $\Pi_{p,q}^{\bot}.$

the quaternions $$u\,d\sigma,\:\:d\varrho\,u,\:\:\mu\,d\sigma,\:\:d\varrho\,\mu$$ belong to the plane $\Pi_{\varrho,\sigma}^{\bot}$; they are invariant in any rotation about the axial plane $\Pi_{\varrho,\sigma}^{\bot}.$


      More generally, the quaternions $b^{\|}$ and $b^{\bot}$ in $\Pi_{p,q}$ and $\Pi_{p,q}^{\bot}$, and the quaternions $b_{\|}$ and $b_{\bot}$ in $\Pi_{\varrho,\sigma}$ and $\Pi_{\varrho,\sigma}^{\bot}$, $b=b^{\|}+b^{{\bot}}=b_{\|}+b_{{\bot}}$, are such that
\begin{equation*} b^{\|}\,dq,\qquad dp\,b^{\|}, \qquad b^{{\bot}}\,q, \qquad p\,b^{{\bot}},\\
\mathrm{belong\,to\,the\,plane}\:\: \Pi_{p,q}^{\bot},\qquad
\end{equation*} 
\begin{equation}b^{{\bot}}\,dq, \qquad dp\,b^{{\bot}}, \qquad b^{\|}\,q, \qquad p\,b^{\|}, \\ \mathrm{belong\,to\,the\,plane}\:\:\Pi_{p,q} \label {(61a)}.\end{equation} 
   \begin{equation*}b_{\|}\,d\sigma, \qquad d\varrho\,b_{\|}, \qquad b_{{\bot}}\,\sigma, \qquad \varrho\,b_{{\bot}}, \\ \small{\textrm{belong to the plane}}\:\:\Pi_{\varrho,\sigma}^{\bot},\qquad\end{equation*}
   \begin{equation}
   b_{{\bot}}\,d\sigma, \qquad d\varrho\,b_{{\bot}}, \qquad b_{\|}\,\sigma, \qquad \varrho\,b_{\|}, \\ \small{\textrm{belong to the plane}}\:\:\Pi_{\varrho,\sigma} \label{(61b)}.
   \end{equation}

   Other useful relations obtain by considering small rotations $d\vartheta$ about an axial plane. For instance we have
    $$u+du=e^{\tfrac{d\vartheta}{2}p}\,u\,e^{\tfrac{d\vartheta}{2}q},$$
    which yields:
   \begin{equation}
   du=\tfrac{1}{2}(p\,u+u\,q)d\vartheta \label{(62)}.
   \end{equation}

   Since $p\,u-u\,q=0$ (Eq.~\ref{(58)}), Eq.~\ref{(62)} yields
   \begin{equation}
   du=p\,u\,d\vartheta=u\,q\,d\vartheta=R\,m\,d\vartheta \label{(63)},
   \end{equation}

   \noindent the last equality also originating in the expressions for $p$ and $q$, Eq.~\ref{(56a)}.\\

\noindent \textsl {b. A general expression for the attached defect density.} \label{Ageneralexpressionfortheattacheddefectdensity.}

 We reproduce the extended Volterra process approach we have used for Euclidean crystals.  Because \{am/$\textrm{S}^{3}$\} is amorphous, any disclination L, whatever its strength, can be constructed as a sum of infinitesimal constitutive defects, attached to L if L has twist character.  The wedge disclination case has already been discussed; the constitutive defects are disvections (one set of disvections, whose 'chirality' is ambiguous, left or right, Eq.~\ref{(55)}). Disvections are the analogs in $\{\textrm{S}^{3}\}$ of dislocations in $\{\textrm{E}^{3}\}$, so that one can consider this result as the  $\{\textrm{S}^{3}\}$ analog of the result in  $\{\textrm{E}^{3}\}$. Contrariwise, as we show, in the twist or mixed line case, the attached defects are generically disclinations, which however can be defined as the sum of two sets of disvections of opposite 'chiralities'.

Now, assume that L is any loop in the habit 3-sphere, and consider two neighboring points \textbf{M} and $\mathbf{M}+d\mathbf{M}$ on this line, $u$ and $u+du$  in quaternion notations, with $u\tilde{u}=(u+du)(\tilde{u}+d\tilde{u})={R}^{2}$, i.e. $d\tilde{u}\,u+\tilde{u}\,du=0$.  In order to compare the Volterra processes at two neighboring points \textbf{M} and $\mathbf{M}+d\mathbf{M}$, we proceed as follows.

Let   $\mathbf{\Omega}=\Omega \bm{\mu}$ be the rotation vector of the disclination at \textbf{M}.  Being a rotation vector, $\bm{\mu}$ belongs to the axial plane $\Pi_{\varrho,\sigma}=\{0,1-\varrho\sigma,\varrho+\sigma\}$  that intersects the habit sphere along the great circle $\textrm{C}_{\mu}$
 running through \textbf{M}, and to which $\bm{\mu}$  is tangent.  This is akin to the situation investigated in the case of the wedge disclination, where $\textrm{C}_{\mu}$ is in fact the wedge line itself and where $\bm{\mu}$ is tangent to the wedge line (observe that $\bm{\mu}$ does not belong to the habit 3-sphere but to the 3D flat tangent space to the sphere at \textbf{M}).  We therefore have  $\bm{\mu}\cdot \mathbf{OM}=0$.

Consider now two close points \textbf{M} and $\mathbf{M}+d\mathbf{M}$ on the disclination L, with rotation vectors along $\mu$ and $\mu+d\mu$, and a point \textbf{P} ($x$ in quaternion notation) on the cut surface $\Sigma_{\mathbf{L}}$.  The variation in displacement observed from \textbf{M} to  $\mathbf{M}+d\mathbf{M}$ at the same point \textbf{P} is:
 \begin{equation}
                db_{\mathbf{M}}=
    e^{-\tfrac{\Omega}{2}\,(\varrho+d\varrho)}\,x\,e^{\tfrac{\Omega}{2}\,(\sigma+d\sigma)}
-e^{-\tfrac{\Omega}{2}\varrho}\,x\,e^{\tfrac{\Omega}{2}\sigma}
    \\  =\sin{\tfrac{\Omega}{2}}\,(e^{-\tfrac{\Omega}{2}\varrho}\,x\,d\sigma
-d\varrho\,x\,e^{\tfrac{\Omega}{2}\sigma})
\label{(64)}.
\end{equation}

If the line L is a great circle and the local rotation vector is along the tangent of this circle, which means that one can choose $\varrho$, $\sigma$ constant (independent of the point on C), then $d\varrho=d\sigma=0$ ; the line is of wedge character, as expected, and $db_{\mathbf{M}}=0$ .  There are no \textit{attached} defects. We retrieve the results of \ref{VPelementsofawedgedisclination}.

\textsf{Equation~\ref{(64)} is the fundamental equation related to attached defect densities in} $\{\textrm{am}/\textrm{S}^{3}\}$.

We show now that it cannot be interpreted the same way as Eq.~\ref{e4} in $\{\textrm{am}/\textrm{E}^{3}\}$, although the extended VP from which it results  is similar. \\

\noindent \textsl{c. Attached disclination densities.} \label{Attached disclination densities.}

As a matter of fact, Eq.~\ref{(64)} appears as the difference between two disclinations of axial planes $\Pi_{\varrho,\sigma}$ and $\Pi_{\varrho+d\varrho,\sigma+d\sigma}$; this difference can be expressed as a \emph{unique disclination carrying an infinitesimally small angle of rotation}:

The quaternions $\varrho+d\varrho\, \textrm{and}\, \sigma+d\sigma$ are pure unit quaternions, if  second order terms in $d\sigma^{2}, \,d\varrho^{2}$, etc. are neglected; $d\varrho \, \textrm{and}\,d\sigma$ are pure quaternions, of equal moduli $|d\varrho| =|d\sigma|=d\lambda,\,\textrm{real, \,positive}$. We introduce the two pure unit quaternions $t_{\varrho},\,t_{\sigma}$ such that $d\lambda\,t_{\varrho}=d\varrho,\,d\lambda\,t_{\sigma}=d\sigma$.

Let us define
\begin{equation}
d_\mathbf{M}=e^{-\tfrac{d \Omega_{\mathrm{F}}}{2}\,t_{\varrho}}\,x\,e^{\tfrac{d \Omega_{\mathrm{F}}}{2}\,t_{\sigma}}-x
\label{(65)}.
\end{equation}

\noindent Equation \ref{(65)} is the expression of the displacement $d_\mathbf{M}(x)$ at \textbf{P}{(x)} due to an infinitesimal disclination of angle $d \Omega_{\mathrm{F}}$ about the axial plane $\Pi_{t_{\varrho},t_{\sigma}}$.  Notice that $d_{\mathbf{M}}(u)=0$, which confirms that  the just defined infinitesimal disclination is attached at {\textbf{M}} to L. We now show that Eq.~\ref{(64)} can be given the same form as Eq.~\ref{(65)}, with the following choice of  $d \Omega_{\mathrm{F}}$:
\begin{equation}
d \Omega_{\mathrm{F}}
=2\,\sin{\tfrac{\Omega}{2}} \cos{\tfrac{\Omega}{2}}\,d\lambda,
\label{(66)}
\end{equation}

\noindent justified below. Equation~\ref{(66)} can also be written ${d \Omega_{\mathrm{F}}}=2\,d( \sin{\tfrac{\Omega}{2}})$, ${d \Omega_{\mathrm{F}}}$ thus appearing as the differential of the modulus of the Frank vector introduced in \ref{Generalizedpolygonalloops.Kirchhoff'srelation.}, if one assumes that the variation $d \Omega$ of the argument equals  ${2}\,d\lambda\,\sin{\tfrac{\Omega}{2}}$; we see later on why it should be so.

We now give hints how to identify Eq.~\ref{(64)} and Eq.~\ref{(65)}, and find \textit{{passim}} an interesting simplification of the expressions of $db_{\mathbf{M}}$, $d_\mathbf{M}$.

Split $x=x_{\|}+x_{\bot}$ into its components $x_{\|}\,\in\,\Pi_{\varrho,\sigma}$ and $x_{\bot}\,\in\,\Pi_{\varrho,\sigma}^{\bot}$.  According to Eq.~\ref{(61b)}, $x_{\|}\,d\sigma $ {and}  $d\varrho\,x_{\|}$ belong to $\Pi_{\varrho,\sigma}^{\bot}$. Hence $e^{-\tfrac{\Omega}{2}\varrho}\,x_{\|}\,d\sigma=x_{\|}\,d\sigma\,e^{\tfrac{\Omega}{2}\sigma}$, and:

\begin{equation} db_{\mathbf{M}}=\sin{\tfrac{\Omega}{2}}\,
(\{ x_{\|}\,d\sigma-d\varrho\,x_{\|}\}\,e^{\tfrac{\Omega}{2}\sigma}
\\ +
\{ e^{-\tfrac{\Omega}{2}\varrho}\,x_{\bot}\,d\sigma-d\varrho\,x_{\bot}\,e^{\tfrac{\Omega}{2}\sigma} \})
\end{equation}

Because $x_{\|}\in \Pi_{\varrho,\sigma}$, we have $$x_{\|}\,\sigma-\varrho\,x_{\|}=0.$$

Likewise, because $x_{\bot}\in \Pi_{\varrho,\sigma}^{\bot}$, we have $$x_{\bot}\,\sigma+\varrho\,x_{\bot}=0.$$

We differentiate these relations, keeping $x$ constant, and get:
$$x_{\|}\,d\sigma-d\varrho\,x_{\|}=0 \qquad x_{\bot}\,d\sigma+d\varrho\,x_{\bot}=0 .$$

Hence
\begin{equation}
db_{\mathbf{M}}=\sin{\tfrac{\Omega}{2}}\,
\{
e^{-\tfrac{\Omega}{2}\varrho}\,x_{\bot}\,d\sigma-d\varrho\,x_{\bot}\,e^{\tfrac{\Omega}{2}\sigma}
\}  \label{(67)}.
\end{equation}

The terms depending on $x_{\|}$ have disappeared. The remaining terms  can be transformed in many ways, using the equalities just derived and the rotation invariances of Eq.~\ref{(61a)} and Eq.~\ref{(61b)}. We get for $db_{\mathbf{M}}$,
\begin{equation}
db_{\mathbf{M}}
=-2\,\sin{\tfrac{\Omega}{2}}\,\cos{\tfrac{\Omega}{2}}\,d\lambda\,t_{\varrho}\,x_{\bot} \\ =2\,\sin{\tfrac{\Omega}{2}}\,\cos{\tfrac{\Omega}{2}}\,d\lambda\,x_{\bot}\,t_{\sigma}
 \label {(68a)},
 \end{equation}
and using similar transformations for $d_{\mathbf{M}}$,
\begin{equation}
d_{\mathbf{M}}
=-d \Omega_{\mathbf{F}}\,t_{\varrho}\,x_{\bot}
=d\Omega_{\mathbf{F}}\,x_{\bot}\,t_{\sigma}
\label{(68b)}.
\end{equation}

Hence, to summarize: \textsf{$db_{\mathbf{M}}$ and $d_{\mathbf{M}}$ describe the same infinitesimal disclination attached to }\textrm{L} at \textbf{M}, \textsf{of angle} $d \Omega_{\mathbf{F}}$, \textsf{of axial plane} $\Pi_{t_{\varrho},t_{\sigma}}$.

Notice that the definition of $\delta \Omega={2}\,d\lambda \,\sin{\tfrac{\Omega}{2}}$ is not mere chance; it originates in expressions of the type $e^{\tfrac{\Omega}{2}\,(\sigma+d\sigma)}$, which can also be written
$$e^{\tfrac{\Omega}{2}\,(\sigma+d\sigma)}
=e^{\tfrac{\Omega}{2}\,\sigma}+\sin{\tfrac{ \Omega}{2}}\,d\lambda\,\,t_{\sigma}
=e^{\tfrac{\Omega}{2}\,\sigma}+\tfrac{d \Omega}{2}\,t_{\sigma}.$$

{\footnotesize For the sake of completeness, we write some other expressions for $db_{\mathbf{M}}$
\begin{equation}
db_{\mathbf{M}}=\sin \tfrac{\Omega}{2}\,\{-e^{-\tfrac{\Omega}{2}\varrho}\,d\varrho\,x_{\bot}+x_{\bot}\,d\sigma\,e^{\tfrac{\Omega}{2}\sigma}\} \label{(69a)},
\end{equation}
$$
\quad \quad =\sin \tfrac{\Omega}{2}\,\{-d\varrho\,e^{\tfrac{\Omega}{2}\varrho}\,x_{\bot}+x_{\bot}\,e^{-\tfrac{\Omega}{2}\sigma}\,d\sigma\} \label{(69b)}.
$$
 We shall make use of Eq.~\ref{(69a)}.\\

 Also,  because $x_{\bot}\in \Pi_{\varrho,\sigma}^{\bot},$
$$v_{\bot}\equiv e^{\tfrac{\Omega}{2}\varrho}\,x_{\bot}
=x_{\bot}\,e^{-\tfrac{\Omega}{2}\sigma},$$

\noindent we can write
$$db_{\mathbf{M}}= \sin \tfrac{\Omega}{2}\,(-d\varrho\,v_{\bot}+v_{\bot}\,d\sigma).$$

 \noindent   where $v_{\bot} \in \Pi_{\varrho,\sigma}^{\bot},$
{ because}
$$e^{-\tfrac{\Omega}{2}\varrho}\,(e^{\tfrac{\Omega}{2}\varrho}\,x_{\bot})\,e^{-\tfrac{\Omega}{2}\sigma}
\equiv x_{\bot}\,e^{-\tfrac{\Omega}{2}\sigma}
= e^{\tfrac{\Omega}{2}\varrho}\,x_{\bot}$$
(the first member of this equation is indeed a rotation of $v_{\bot}$ with axial plane
$\Pi_{\varrho,\sigma}^{\bot})$.}\\

\noindent \textsl {d. Infinitesimal Burgers vectors and disclination lines.} \label{InfinitesimalBurgersvectorsanddisclinationlines.}

Starting from Eq.~\ref{(64)} or Eq.~\ref{(69a)}, we calculate
$|db_{\mathbf{M}}|$,
 using the relation
 $|db_{\mathbf{M}}|^{2}=db_{\mathbf{M}}\,d\tilde {b}_\mathbf{M}$.
 The calculation, not reproduced here, uses some of the equalities established above. One finds:
\begin{equation}
|db_{\mathbf{M}}|=2\sin^{2} \tfrac{\Omega}{2}\, |x_{\bot}| \,d\lambda \label{(70)}
\end{equation}

 This equation compares to Eq.~\ref{(55)} (wedge disclination) by the common presence of the $|x_{\bot}|$ term $-$ again, the relevant distance is the distance to the axial plane that carries the rotation vector of the disclination. The presence of the $\sin^{2} \tfrac{\Omega}{2}$ term is interpreted as follows. We can split $db_{\mathbf{M}}$ into two Burgers vectors, appearing in Eq.~\ref{(69a)}, related to two disvections, one right, one left,
 \begin{equation}
 db_{\mathbf{M},\sigma}=\sin{\tfrac{\Omega}{2}}\,x_{\bot}\,d\sigma\, e^{\tfrac{\Omega}{2}\sigma},
\\
db_{\mathbf{M},\varrho}=-\sin{\tfrac{\Omega}{2}}\,e^{-\tfrac{\Omega}{2}\varrho}\,d\varrho\,x_{\bot} \label {(71)}.
\end{equation}

$db_{\mathbf{M},\varrho}$ and $db_{\mathbf{M},\sigma}$ have the same Burgers vector modulus, namely $|db_{\mathbf{M},\varrho}|=|db_{\mathbf{M},\sigma}|=\sin \tfrac{\Omega}{2}\,|x_{\bot}| \, d \lambda$. The extra $\sin \tfrac{\Omega}{2}$ factor in $|db_{\mathbf{M}}|$ means that the two vectors $d\mathbf{b}_{\mathbf{M},\varrho}$ and $d\mathbf{b}_{\mathbf{M},\sigma}$ make a constant angle, equal to $\Omega$. One can check directly that $d\mathbf{b}_{\mathbf{M},\rho}\cdot d\mathbf{b}_{\mathbf{M},\sigma}=\tfrac{1}{2}\,(d{b}_{\mathbf{M},\varrho}{d \tilde{b}}_{\mathbf{M},\sigma}+{d \tilde{b}}_{\mathbf{M},\varrho}d{b}_{\mathbf{M},\sigma})=\cos \Omega$.

The description of the infinitesimal defects related to the disclination L in terms of disvections is equivalent to the description in terms of attached disclinations. Notice that the disvection lines, which are those lines along which $db_{\mathbf{M},\varrho} \, \textrm{and} \, db_{\mathbf{M},\sigma}$ are constant, are \emph{not} attached to L $-$ which would require that $x_{\bot}=0$, since the points of L are characterized by the quaternion coordinate $u$ which obeys $u_{\bot}=0$. The line L itself is a particular disvection line, of vanishing Burgers vector. Henceforth the infinitesimal disvection lines form \emph{two} sets of lines surrounding L, each comparable to the set of disvections surrounding a wedge disclination line.
\subsubsection{Twist disclination along a great circle} \label{Twistdisclinationofconstantrotationvectoralongagreatcircle}

By constant rotation vector we mean that the local rotation axis of the disclination L, namely $\Omega \mu$ in quaternion notation, with $\mu\tilde{\mu}=1$, is parallel-transported along a line with the natural connection of the sphere $\textrm{S}^{3}$.  If such a line is a great circle, i.e. a geodesic, the rotation axis makes a constant angle with this circle, $\bm{\mu} \cdot \mathbf{m} =\cos \varphi$.  This configuration is obviously equivalent to a straight disclination line in $\{\textrm{am}/\textrm{E}^{3}\}$, with constant rotation vector.  A point $\mathbf{M}$ on L depends on the variable $u$ or the arc angle $\vartheta$. Let $\textrm{C}^{\vartheta}_{\mu}$ be the great circle tangent to $\mu(\vartheta)$ at $\mathbf{M}(\vartheta)\in \textrm{L}$; it belongs to the plane $\Pi_{\varrho,\sigma}=\{0,~1~-~\varrho~\sigma,\varrho+\sigma\}$;  $\varrho= \varrho(\vartheta)$ and $\sigma=\sigma(\vartheta)$ are given by Eq.~\ref{(56b)}.  $\Pi_{p,q}=\{0,~1~-~p~q,p+q\}$ is the two-plane that contains the great circle L; $p \, \textrm{and} \,q$ are constant.

The parallel transport of $\mu$ along L, a geodesic, is also a rotation about the axial plane $\Pi_{p,q}^{\bot}=\{0,~1+pq,p-q\}$, completely orthogonal to $\Pi_{p,q}$. Such a rotation can be written:
$$\mu+d\mu=e^{\tfrac{d \vartheta}{2}p}\,\mu\,e^{\tfrac{d \vartheta}{2}q} \label{(72)},$$

\noindent which yields
\begin{equation}
d\mu=\tfrac{1}{2}\,(p\,\mu+\mu \,q)\,d\vartheta \label{(73)}.
\end{equation}

Using  Eq.~\ref{(73)} and Eq.~\ref{(56b)}, the following expressions for $d\varrho$ and $d\sigma$ obtain:

\begin{equation}
d\varrho=\tfrac{1}{2}\,(p\,\varrho-\varrho \, p)\,d\vartheta, \,\,
d\sigma=-\tfrac{1}{2}\,(q\,\sigma-\sigma \, q)\,d\vartheta
 \label{(74)}.
 \end{equation}

 We also have
 $$d\lambda=\tfrac{1}{2}\,d\vartheta\,|p\,\varrho-\varrho \, p|=\tfrac{1}{2}\,d\vartheta\,|q\,\sigma-\sigma \, q|$$

 We have $\bm{\mu} \cdot \mathbf{m}=\cos \varphi=\tfrac{1}{2}\,(m\,\tilde{\mu}+\mu\,\tilde{m})$, a constant. We can show, again using Eq.~\ref{(56a)} and Eq.~\ref{(56b)}, that
 \begin{equation*}(p\,\varrho-\varrho \, p)^{2}=(q\,\sigma-\sigma \, q)^{2}
 =4\,\sin^{2} \varphi,\end{equation*}
 which yields
 \begin{equation}
 d\lambda=|\sin \varphi|\, d\vartheta \label{(75)}.
 \end{equation}

As indicated by the presence of the  coefficient $|\sin \varphi|$ in the expression of $ d\lambda$, the only component of $\bm\mu$ which is relevant in the twist properties of the line is its component orthogonal to $\mathbf{m}$. Hence we split the rotation vector as follows
                        \begin{equation}
                        \mu(\vartheta)=\mu^{\bot}+\mu^{\|} \label{(76)}.
                        \end{equation}
                        where $\mu^{\bot}$ belongs to the plane $\Pi_{p,q}^{\bot}$ and is therefore invariant in any rotation about this plane; $\mu^{\|}=m\,\cos \varphi$ belongs to the plane $\Pi_{p,q}$. It is a question of simple algebra to show that the $\mu_{\|}$ component does not contribute to $(\sigma\,q-q\,\sigma)$  and to $(p\,\varrho-\varrho\, p)$ .  The only component of the rotation vector that contributes to the  dislocation densities attached to C is  $\mu_{\bot}$, (which is a constant all along the disclination), through the $| \sin \varphi |$ factor.  But notice that the \emph{rotation vector} is still $\Omega \,\bm \mu$.

The same discussion as in  \ref{ThewedgedisclinationinS3} and \ref{Defects attachedtoadisclinationinthe3-spherecurvedamorphousmedium}
is valid, with the simplifications of Eq.~\ref{(74)} and Eq.~\ref{(75)}.

\subsection{Kirchhoff relations}
\label{Kirchhoff'srelations}
Any isometry of S$^{3}$ has of the form $x'=e^{-\beta q}\,x\,e^{\alpha p}$, Eq.~\ref{(46)}, and can be split into the product of two commutative helix turns, one right, one left.  Therefore a product of isometries can be written
\begin{equation}
 x'=  {l}_{{\mathrm{Q}}}^{(1)} \cdots{l}_{{\mathrm{Q}}}^{(i)} \,{l}_{{\mathrm{Q}}}^{(i+1)}\cdots\,x\,\cdots{r}_{{\mathrm{Q}}}^{(i+1)} \,{r}_{{\mathrm{Q}}}^{(i)} \cdots{r}_{{\mathrm{Q}}}^{(1)},
 \label{(isom)}
  \end{equation}

\noindent where ${l}_{{\mathrm{Q}}}^{(i)},{r}_{{\mathrm{Q}}}^{(i)}\in{\mathrm{Q}}=$ SU(2).  $n$ disclinations meeting at a node can be split into two sets of disvections meeting at this node, $n$ left disvections, $n$ right disvections.  Thus, we first investigate the Kirchhoff relations for (left or right) disvections.

\subsubsection{Three disvections meeting at a node} \label{Kirchhoff'srelationsfordisvections}

Let ${h}_{{\mathrm{Q}}}^{(1)}=e^{\alpha p}$, ${h}_{{\mathrm{Q}}}^{(2)}=e^{\beta q}$, ${h}_{{\mathrm{Q}}}^{(3)}=e^{\gamma r}$ be the elements of symmetry of three right (say) disvections meeting at a node.  We then have
  \begin{equation}
  {h}_{{\mathrm{Q}}}^{(1)} \,{h}_{{\mathrm{Q}}}^{(2)} \,{h}_{{\mathrm{Q}}}^{(3)}=\{\pm 1\} \label{(81)}
  \end{equation}

  Remember that the group of all unit quaternions is 2:1 homomorphic to the group of all the \textit{rotations} of a 2-sphere that leave the origin fixed. In that sense, ${h}_{\mathrm{Q}}$ and $-{h}_{\mathrm{Q}}$ represent the same rotation in SO(3), but they do not represent the same transvection in SU(2). $\{-1\}$ transforms $x$ into $-x$ by an helix turn of angle $\pi$ about \emph{any} axis $\mathbf{w}$:  $\{-1\}=e^{\pi w}$. It however will appear as a fundamental necessity to introduce the quaternion \{-1\} in Eq.\ref{(81)}, as we shall see.

Denote by $a$, $b$, $c$ the angles between the directions defined by the three pure unit quaternions $p$, $q$, $r$, namely $a = \angle (\mathbf{q,r})$, $b = \angle (\mathbf{r,p})$, and $ c = \angle (\mathbf{p,q})$.  Equation~\ref{(81)}, which also reads, e.g., ${h}_{\mathrm{Q}}^{(1)} \,{h}_{\mathrm{Q}}^{(2)}=\pm \tilde{{h}}_{\mathrm{Q}}^{(3)}$, yields:

\noindent - three relations coming from the real part of this relation, of the type
                \begin{equation}
                \cos \alpha \,\cos \beta-\cos c \,\sin \alpha \,\sin \beta=\pm \cos \gamma \label {(82)}.
                \end{equation}

\noindent where $\cos c =\mathbf{p}\cdot\mathbf{q}=p_{1} \,q_{1}+p_{2} \,q_{2}+p_{3} \,q_{3}$,

\noindent - nine relations coming from the pure quaternion part, of the type
            \begin{equation}
            q_{1} \,\sin \beta \, \cos \alpha+p_{1}\sin \alpha \,\cos \beta \\ +s_{c,1} \,\sin c\,\sin \alpha\,\sin \beta  =\mp r_{1}\sin \gamma \label{(83)}
            \end{equation}

\noindent where the unit vector $\mathbf{s}_{c}$ (i.e. the pure unit quaternion $s_{c}$), defined by the cross product $\mathbf{s}_{c}\,\sin c=(\mathbf{p}\times\mathbf{q})$ has been introduced; similarly $\mathbf{s}_{a}\,\sin a=(\mathbf{q}\times\mathbf{r})$ and $\mathbf{s}_{b}\,\sin b=(\mathbf{r}\times\mathbf{p})$.

Multiplying  Eq.~\ref{(83)} by $s_{c,1}$ and summing over the three Eq.~\ref{(83)} containing $s_{c,1},s_{c,2},s_{c,3}$, one gets
                     \begin{equation}
                        \frac{\sin^{2}a}{\sin^{2}\alpha}=\frac{\sin^{2}b}{\sin^{2}\beta}=\frac{\sin^{2}c}{\sin^{2}\gamma}  =\mp\frac{V_{p,q,r}}{\sin \alpha\,\sin \beta \,\sin \gamma} \label{(84)}
                        \end{equation}
                        where $V_{p,q,r}=\textbf{r}\cdot(\textbf{p}\times\textbf{q})$ is the scalar triple product; $V_{p,q,r}=\mathbf{p}\cdot\mathbf{s}_{a}\,\sin a=\mathbf{q}\cdot\mathbf{s}_{b}\,\sin b=\mathbf{r}\cdot\mathbf{s}_{c}\,\sin c$.

One recognizes in Eq.~\ref{(82)} and \ref{(84)} expressions much akin to those met in 2D spherical trigonometry.  The geometric interpretations of Eq.~\ref{(81)} in terms of spherical triangles are different, whether one has a $\{-1\}$ or a $\{1\}$ in the right-hand member.\\

\noindent \textsl {a. ${h}_{{\mathrm{Q}}}^{(1)} \,{h}_{{\mathrm{Q}}}^{(2)} \,{h}_{{\mathrm{Q}}}^{(3)}=\{ -1\}.$}

\begin{figure}
\includegraphics{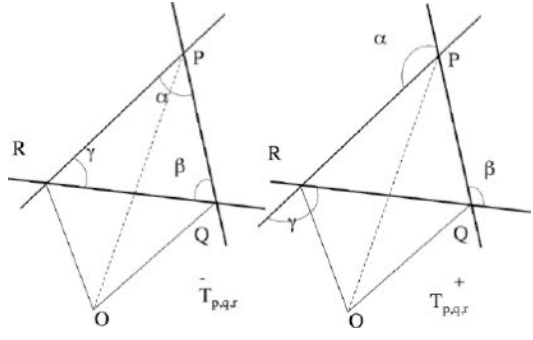}
    \caption{
    \label{figT}{Spherical 2-D representation of Kirchhoff relations for disvections. \textbf{P}, \textbf{Q}, and \textbf{R} are on the sphere centered at \textbf{O}. See text}
     }
 \end{figure}

One finds that there are \textit{two} \textit{conjugate} spherical triangles, hereunder denoted ${T}_{p,q,r}^{-}$ and ${T}_{s_{a},s_{b},s_{c}}^{-}$, whose angles and angular arcs yield  $\{ -1\}$ in the right-hand member:

(i)- the vertices of ${T}_{p,q,r}^{-}$ are the extremities $\mathbf{P, Q, R,}$ of the 3-vectors $\mathbf{p}$, $\mathbf{q}$, $\mathbf{r}$; the angles of the triangles are $\alpha,\beta,\gamma$, the angular arcs are $a, b, c$. See Fig.~\ref{figT}. Standard results in spherical trigonometry yield \cite{weisstein}:
\begin{equation} -\cos \gamma=\cos \alpha \,\cos \beta-\cos c \,\sin \alpha \,\sin \beta \label{(82i)} \end{equation}
\begin{equation} \frac{\sin \alpha}{\sin a}=\frac{\sin \beta}{\sin b}=\frac{\sin \gamma}{\sin c}= \frac{V_{p,q,r}}{\sin a\,\sin b \,\sin c}. \label{(84i)}  \end{equation}

\noindent Equation~\ref{(82i)} obtains by applying the 'cosine rules for the angles', specialized to the angle $\gamma$.

The construction of a triangle of angles $\alpha,\beta,\gamma$, of angular arcs $a, b, c$, requires that be obeyed the following inequalities
\begin{equation}\alpha+\beta+\gamma\geqslant \pi,\,\,\,\,a+b+c\leqslant 2\pi \label{(ineq1)} \end{equation}

(ii) the vertices of ${T}_{s_{a},s_{b},s_{c}}^{-}$ are the extremities $\mathbf{A, B, C,}$ of the 3-vectors $\mathbf{s_{a},s_{b},s_{c}}$; the angles of the triangles are $\pi - a,\pi - b,\pi - c$, the angular arcs are $\pi - \alpha,\pi -  \beta,\pi - \gamma$. One checks that \hbox{$
\mathbf{s}_{a}\cdot\mathbf{s}_{b}=(\mathbf{q}\times \mathbf{r})\cdot(\mathbf{r}\times \mathbf{p})=\tfrac{\cos a\,\cos b -\cos c}{\sin a\,\sin b} $ etc.,} i.e. $\mathbf{s}_{a}\cdot\mathbf{s}_{b}=-\cos\gamma$, etc., by applying the 'cosine rules for the angles', specialized to the angle $\pi - c$. Equation~\ref{(82i)} is also satisfied (it obtains by applying the 'cosine rules for the edges', specialized to the edge of arc $\pi -\gamma$), and Eq.~\ref{(84i)} is replaced by
\begin{equation} \frac{\sin a}{\sin \alpha}=\frac{\sin b}{\sin \beta}=\frac{\sin c}{\sin\gamma}=\frac{V_{s_{a},s_{b},s_{c}}}{\sin \alpha \,\sin \beta \,\sin \gamma}. \label{(84ii)}  \end{equation}

{\footnotesize Eq.~\ref{(84i)} and~\ref{(84ii)} are identical and both identical to Eq.~\ref{(84)}, as can be shown by employing the standard identities
\begin{center} $V_{p,q,r} \equiv \sin a\,\sin b \,\sin \gamma \equiv \sin b\,\sin c \,\sin \alpha \equiv \cdots,$ \\
$V_{s_{a},s_{b},s_{c}} \equiv \sin \alpha \,\sin \beta \,\sin c \equiv \cdots,$\end{center}
\noindent which yield
\begin{equation*}V_{p,q,r}\,V_{s_{a},s_{b},s_{c}} =\sin \alpha \,\sin \beta \,\sin \gamma \,\sin a\,\sin b \,\sin c \end{equation*}
\begin{center}$V_{p,q,r}^{2}=\sin a\,\sin b \,\sin c \,V_{s_{a},s_{b},s_{c}},$ \end{center}
 \begin{center}$V_{s_{a},s_{b},s_{c}}^{2}=\sin \alpha \,\sin \beta \,\sin \gamma \,V_{p,q,r}.$ \end{center}}

The construction of a triangle of angular arcs $\pi-\alpha,\pi-\beta,\pi-\gamma$, of angles $\pi-a, \pi-b, \pi-c$, requires that be obeyed the same inequalities as for ${T}_{p,q,r}^{-}$.

 \emph{Remark.} The two tetrahedra $\mathbf{OABC}$ and $\mathbf{OPQR}$ are in conjugate positions on the 2-sphere; the edge $\mathbf{s_{a}}$ of $\mathbf{OABC}$ is perpendicular to the facet $(\mathbf{OQR})$ of
$\mathbf{OPQR}$, and the edge $\mathbf{p}$ of $\mathbf{OPQR}$ is perpendicular to the facet $\mathbf{(OBC)}$ of
$\mathbf{OABC}$, etc. Hence the appearance of the arc angles $\pi -\alpha$, conjugate to the angles $\alpha$, etc., of the angles $\pi -a$, conjugate to the arc angles $a$, etc.

Both geometrical representations $\mathbf{OABC}$ and $\mathbf{OPQR}$ of the Kirchhoff relation for three disvections,  are equivalent; we retain the first one. They express that the set of disvections terminate on a disvection of angle $\pi$, whose direction $w$ is not given a fixed value. In that sense, the terminal node can be interpreted as a singular point.

Notice also that the same $\{ - 1\}$ disvection can be assigned to the spherical triangle $\mathbf{(PQR)^{*}}$ \emph{outside} the \underline{smaller} triangle $\mathbf{(PQR)}$. This shows that, in the present representation of transvections, \textsl{the full 2-sphere }S$^{2}$ \textsl{has to be assigned the identity disvection }$\{+1\}$, \textsl{with angle} $2\pi$.\\

\noindent \textsl {b. h$_{\mathrm Q}^{(1)} \,h_{\mathrm Q}^{(2)} \,h_{\mathrm Q}^{(3)}=\{1\}.$}

Notice that
$e^{-\pi p}=e^{-\pi q}=e^{-\pi r}=\{-1\}.$ Hence the $\{+1\}$ Kirchhoff relation of Eq. \ref{(81)}, for the angles $\alpha,\beta,\gamma$ and the angular arcs $a, b, c$, can be transformed to the folllowing one
\begin{equation}
 e^{(\pi -\gamma) r}\,e^{(\pi -\beta) q}\,e^{(\pi -\alpha) p}=\{-1\}
 \label{(Kir+1)}   \end{equation}

\noindent which can be discussed as the previous one for the angles $\pi -\alpha,\, \pi -\beta,\, \pi -\gamma$ and the angular arcs $\pi -a, \, \pi -b, \, \pi -c$. The inequalities of Eq.~\ref{(ineq1)} are now replaced by
\begin{equation}a+b+c\geqslant \pi,\,\,\,\,\alpha+\beta+\gamma\leqslant 2\pi \label{(ineq2)} \end{equation}
i.e. the complementary ones to the case $\{-1\}.$

 For the spherical representation of Eq. \ref{(Kir+1)}, see Fig. \ref{figT}. Now the relevant angles are the external angles. There is no $\{-1\}$ disvection at the point where the three disvections merge.

\subsubsection{Orientation \textit{vs}. handedness of a disvection} \label{Orientationvs.handedness} The \emph{handedness} of a disvection is a topological concept; its \emph{orientation} is related to the Volterra process, it does not make sense topologically.

We haven't yet taken care of the orientations (of the disvection lines, of the edges of the triangles T$_{{p,q,r}}^{\pm}$) and of the signs of the angles $\alpha, \,{\beta},\,\mathrm{and}\, \gamma$. In the usual dislocation theory, the orientation of a dislocation line is fixed arbitrarily, from which orientation the sign of the related Burgers vector is deduced, but still depending on a convention, generally the so-called FS/RH convention \cite{nabarro}. For a given orientation, the change of Burgers vector has a topological meaning, because $\mathbf{b}$ and $-\mathbf{b}$ are different translation symmetries. We have here a somewhat more subtle situation.\\

\noindent \textsl {a. Topological considerations only.\\}  \label {Topologicalconsiderationsonly}
(i) The change of sign of a transvection, viz. $x\,\to \,x\,e^{\alpha \,p}$ changed to $x\,\to \,-\,x\,e^{\alpha \,p}$, does not connote the same disvection. For in such a case the triple node $e^{\alpha \,p}\,e^{\beta \,q}\,e^{\gamma \,r}=\{\pm1\}$ changes sign: $e^{\alpha \,p}\,e^{\beta \,q}\,e^{\gamma \,r}=\{\mp 1\}$ is a different disvection.  Hence a change of sign does not correspond to an arbitrariness in the orientation.

(ii) The change of sign of the angle of a transvection, viz. $x\,\to \,x\,e^{\alpha \,p}$ changed to $x\,\to \,x\,e^{-\alpha \,p}$, does not connote the same disvection. The triple nodes $e^{\alpha \,p}\,e^{\beta \,q}\,e^{\gamma \,r}=\{\pm1\}$ and $e^{-\alpha \,p}\,e^{-\beta \,q}\,e^{-\gamma \,r} \neq \{\pm1\}$ are different disvections, because of the non-commutativity of the transvections. A change of sign of the angle does not correspond to an arbitrariness in the orientation.
   \begin{figure}
\includegraphics{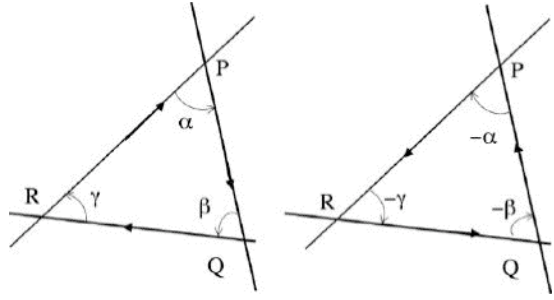}
    \caption{
    \label{figort}{The two sketches represent conjugate disvections. See text}}
 \end{figure}

(iii) Two conjugate disvections, viz. $x\,\to \,x\,e^{\alpha \,p}$ and $x\,\to \,e^{-\alpha \,p}\,x$ are \emph{equivalent} disvections. They are not \emph{equal}, because they differ by the disclination $e^{-\alpha \,p}\,x\,e^{-\alpha \,p}$, which breaks a proper rotation, but a proper rotation does not modify chirality in the usual sense. This emphasizes the concept of conjugacy (change of handedness). The two equations $e^{\alpha \,p}\,e^{\beta \,q}\,e^{\gamma \,r
}=\{\pm1\}$ and $e^{-\gamma \,r}\,e^{-\beta \,q}\,e^{-\alpha \,p}=\{\pm1\}$ are conjugate. The two oriented triangles of Fig.~\ref{figort} (illustrating the case $ \{-1\}$) are conjugate triangles. \\

\noindent \textsl {b. Volterra process.\\}  \label {Volterraprocess4}
\indent The topological considerations above indicate that the lips of the cut surface of a disvection are displaced at $x$ by the quantity $b(x)\,=\,x\,e^{\alpha \,p}\,-x$ (expressed as a quaternion), but there is no hint whether this process adds or removes matter. The choice being made, it can be related to an arbitrary orientation and a convention on the sign $\pm|\mathbf{b}|$, in the manner of dislocations. The situation is then much alike, and is independent of the handedness. \\
\indent The situation is different with disclinations, because the opposite rotations $e^{-\alpha \,p}\,x\,e^{\alpha \,p}$ and $e^{\alpha \,p}\,x\,e^{-\alpha \,p}$ are both allowed topologically.

  \subsubsection{More than three disvections meeting at a node} \label{Morethanthreerightdisvectionsmeetingatanode}
 Let \begin{equation}
 e^{\alpha \,p}\,e^{\beta \,q}\,e^{\gamma \,r}\,e^{\delta \,s}=\{1\}
 \label{(Kir89)}   \end{equation}
 \begin{figure}
\includegraphics{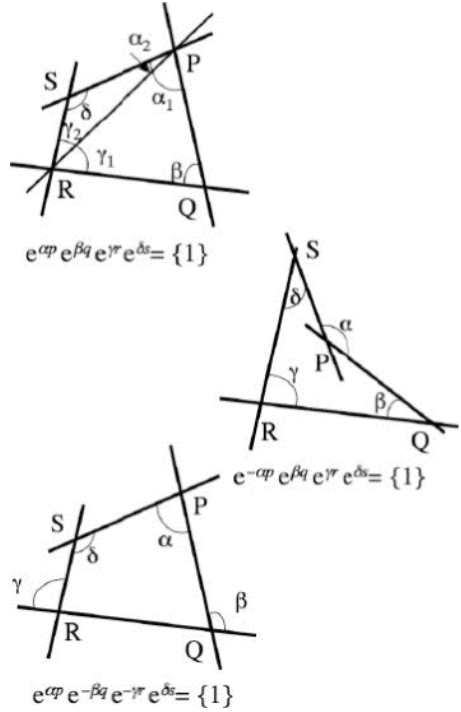}
   \caption{\label{figkirfour}{Four disvections meeting at a node; representation on the unit 2-sphere}}
\end{figure}
 \noindent be four disvections meeting at a node, \textbf{P, Q, R, S} the terminations of the directions $\mathbf{p,\,q,\,r,\,s}$ on the unit 2-sphere, $\alpha =\angle (\mathbf{SPQ}),\,\beta =\angle (\mathbf{PQR}),\,\gamma =\angle (\mathbf{QRS}),\,\delta =\angle (\mathbf{RSP})$.  They are represented Fig.~\ref{figkirfour}, top, as a spherical quadrangle that can be split into two triangles ${T}_{p,q,r}^{-}$ and ${T}_{r,s,p}^{-}$.  Equation \ref{(Kir89)} also holds if the triangles are of types ${T}_{p,q,r}^{+}$ and ${T}_{r,s,p}^{+}$, i.e., if the angles are all complementary to the inside angles of the quadrangle (not represented Fig.~\ref{figkirfour}). The two other sketches of Fig.~\ref{figkirfour} correspond to other possibilities mixing inside and outside angles.

More generally, a $n$-gon is the graphical representation of the meeting at a node of $n$ disvections; the $n$-gon can be split into $n$ triangles (in many ways), each of them of type ${T}_{p_{i},p_{j},p_{k}}^{+}$ or ${T}_{p_{i},p_{j},p_{k}}^{-}$.

\subsubsection{Kirchhoff relations for disclinations}  \label{Kirchhoff'srelationsfordisclinations}
It suffices to consider one set of right disvections and one set of left disvections, separately. To each set the results of the previous section apply. \\

\noindent  {\textsl{a. Three disclinations meeting at a node.}} \label {Threedisclinationsmeetingatanode}

Assume first that the right and left transvections in Eq. \ref{(isom)} are \emph{complex conjugate}, ${l}_{{\mathrm{Q}}}^{(i)}={\tilde{r}}_{{\mathrm{Q}}}^{(i)}$, and that the angles $\alpha_{i}$ and the directions $p_{i}$ are such that they form a $\{-1\}$ triangle. The same triangle represents as well the left and the right transvections, Fig.~\ref{figort}. Hence, because $\{-1\}^{2}=\{1\}$, the left and the right transvections all together amount to an identity isometry.  The representative triangle, which is at the same time 'left' and 'right', has then to be thought of as a $\{1\}$ triangle, i.e. a disclination of  angle $2\pi$. Taking into account the complementary triangle on the 2-sphere, \textsl{the full  }S$^{2}$ \textsl{has to be assigned the identity isometry}$\{+1\}^{2}$, \textsl{with angle} $4\pi$.

Also, the rotation ${l}_{{\mathrm{Q}}}^{(i)}\,x\,{{r}}_{{\mathrm{Q}}}^{(i)}$ (with ${l}_{{\mathrm{Q}}}^{(i)}={\tilde{r}}_{{\mathrm{Q}}}^{(i)}$) is of angle $2\alpha_{i}$, each disvection taking part by an angle $\alpha_{i}$ . Let us consider the Euclidean limit in the $\mathbf{OABC}$ representation. Keeping in mind that the rotation ${l}_{{\mathrm{Q}}}^{(i)}\,x\,{{r}}_{{\mathrm{Q}}}^{(i)}$ (${l}_{{\mathrm{Q}}}^{(i)}={\tilde{r}}_{{\mathrm{Q}}}^{(i)}$) is of angle $2\alpha_{i}$, ($\alpha_{i}= \displaystyle \frac{\omega_{\alpha_{i}}}{2}$, half the angle of the related disclination), and noticing that, in this limit, the quantities $\sin \alpha_{i}$, etc, are proportional to the edge lengths of a flat triangle, it appears that in this limit the vectors $\mathbf{f}_{\alpha_{i}}=2 \,\sin \displaystyle \frac{\omega_{\alpha_{i}}}{2}\,\mathbf{p_{i}}$, etc. obey a Kirchhoff relation, as already observed for disclinations in $\{am/\textrm{E}^{3}\}$.\\

\noindent  \textsl{b. Extension to the generic case, when ${l}_{{\mathrm{Q}}}^{(i)}\,\textrm{and}\,{{r}}_{{\mathrm{Q}}}^{(i)}$ are not complex conjugate}. \label {Extensiontothegenericcase}

It is possible, as noted in \ref{Groupofdirectisometriesinthehabit3-sphere}, to split a rotation ${l}_{{\mathrm{Q}}}^{(i)}\,x\,{{r}}_{{\mathrm{Q}}}^{(i)}$ into two rotations with complex conjugate transvections. Such a splitting having been performed, we are back to the previous case.

\subsubsection{Mixed case} \label{Mixedcase}
  \begin{figure}
\includegraphics{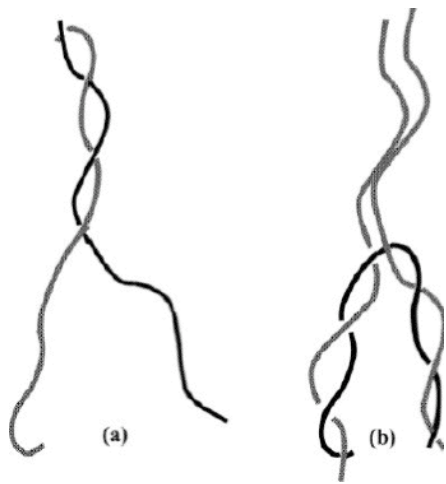}
   \caption{\label{fig27}{Node relations for disvections and disclinations, mixed. A disvection is either left (black lines) or right (shaded lines). (a)- disclination split into two disvections of opposite hands; (b)- disvection split into two disclinations}}
\end{figure}
Figure \ref{fig27} indicates on two simple examples how disclinations and disvections can merge at nodes, in the particular case when the involved disvections all have the same strength, up to chirality (complex conjugacy).  Notice that three disclinations of the same strength cannot meet at the same node, a seemingly obvious result, easily demonstrated by using the splitting of each disclination into two disvections of opposite hands.

These results are entirely topological, and do not take into account the orientations and signs of the disvections that have been previously discussed (\ref {Volterraprocess4}). For example, in Fig.~\ref{fig27}(b), one can assume that the two bundled right disvections are of opposite signs and same 'Burgers vectors'. Therefore the bundle annihilates and the two remaining disclinations are oriented the same way, making them together a unique disclination. Hence Fig.~\ref{fig27}(b) is not at variance with the result of \ref{Attached disclination densities.}, established in the extended Volterra process fashion, which states that a twist disclination carries continuous disclinations, not disvections.

\subsection{The \{3,3,5\} defects}
\label{The3,3,5defects}
A short reminder of the \{3,5\} and \{3,3,5\} groups is given in Appendix \ref{appfive}.

We distinguish between disclinations, disvections, and defects that combine both types.  In this section, we restrict to defects in the \textit{non-decurved} \{3,3,5\} polytope.
\subsubsection{Disclinations} \label{Disclinations5D}
There are 60 Volterra disclinations, when identifying 'antipodal' elements in $\overline{\textrm{Y}}$.  This number goes up to 120 in the TS classification, for which $\Pi_{1}(\textrm{V} _{\{3,3,5\}})\sim \overline{\textrm{Y}}\times \overline{\textrm{Y}}$.  Each element of rotation symmetry leaves invariant a great circle $\textrm{C}_{\omega}$.  For instance, for $|\omega| = 2\pi/5$, $\textrm{C}_{\omega}$ follows a sequence of 10 spherical edges of the \{3,3,5\} polytope; in the locally Euclidean version of \{3,3,5\}, in which each spherical facet is approximated by a planar facet, each spherical edge by a straight segment in $\textrm{E}^{4}$, $\textrm{C}_{\omega}$ is a polygon with 10 edges.  Observe that a \underline{wedge} disclination is a disclination loop along such a $\textrm{C}_{\omega}$.  The local rotation vector $\omega$\textbf{t} is along the tangent to $\textrm{C}_{\omega}$; no attached dislocations (disvections) are necessary to curve the disclination line. See \textcite{nicolis} for a quantitative description of a disclinated \{3,3,5\} crystal in its habit 3-sphere.
\subsubsection{The disvection 'Burgers vectors'}  \label{ThedisvectionBurgersvectors}
There are 60 left disvections $h_{\overline{\mathbf{Y}}}\,x\,\{1\} \equiv h_{\overline{\mathbf{Y}}}\,x$, and 60 right disvections $\{1\}x\,h_{\overline{\mathbf{Y}}} \equiv x\,h_{\overline{\mathbf{Y}}}$, when identifying 'antipodal' elements in $\overline{\textrm{Y}}$.  These numbers go up to 120 right disvections and 120 left disvections in the TS classification.  Disvections leave no fixed point in \{3,3,5\}.

Consider the right screw $x \,e^{\alpha q}$, $x=\textrm{x}_{0}+\textrm{x}_{1} \,i+\textrm{x}_{2} \,j+\textrm{x}_{3} \,k$ being some \{3,3,5\} vertex \textbf{M} on the habit sphere $\textrm{S}^{3}$ of radius $R$.  The displacement of \textbf{M} is $\bigtriangleup x=x\,e^{\alpha q}-x$, whose absolute value $|\bigtriangleup x|$ is $|x| \,|e^{a q}-1|=2R\sin \tfrac{\alpha}{2}$.  The left screw $e^{-\alpha q}\,x$ has the same Burgers modulus $|\bigtriangleup x|$ as the right screw; it is a constant independent of $x$.  However, the direction of $\bigtriangleup \mathbf{x}$ varies with \textbf{M} in \{3,3,5\}: it is tangent to the Clifford parallels belonging to the Hopf fibration related to $q$, Fig.~\ref{fig17}, Appendix \ref{appthreesphere}.

 There are two such Burgers vectors at each point, one 'right', one 'left'.  Thereby, $\bigtriangleup \mathbf{x}$ is not a Burgers vector in the ordinary sense, but it would tend to a constant vector if the radius $R$ of \{3,3,5\} is allowed to increase without limit; however this process has no physical meaning, and we shall see later on another mean to transform $\bigtriangleup \mathbf{x}$ into a genuine Burgers vector.

Taking $\alpha_{5}=\tfrac{\pi}{5}$, we find $|\bigtriangleup \mathbf{x}|_{5}=R\tau^{-1}$, which is precisely the length of an edge of the polytope \{3,3,5\}. This 'Burgers vector' is the smallest possible.  It displaces \textbf{x} to one of its twelve nearest neighbors, which are at the vertices of an \textit{icosahedron}.  The orbit of \textbf{x} under the repeated action of the right screw is a geodesic circle $\textrm{C}_{5,right}$ that is approximated by a ten-edge polygon.  The coordinate $\textrm{x}_{0}$ measures along the 0-axis the distance to the geodesic circle $\textrm{C}_{p}$.  The angle $\beta_{5}$ made by the orbit of \textbf{x} with $\textrm{C}_{p}$ is $\tfrac{\textrm{arc}(\textrm{x}_{0})}{2R}$, \cite{sommer}.  The same considerations apply to a left screw, yielding the same absolute value for the Burgers vector, but directed now along another great circle $\textrm{C}_{5,left}$ at \textbf{x}, with angle $ \tfrac{-\textrm{arc}(\textrm{x}_{0})}{2R}$.  Hence the spherical angle between the two circles $\textrm{C}_{5,right}$ and $\textrm{C}_{5,left}$ at \textbf{x} is $ \tfrac{\textrm{arc}(\textrm{x}_{0})}{R}$, which has to be equal to $2\pi/5$ (five ten-edge great circles emanating from any vertex and belonging to a geodesic 2-sphere).  Observe that the notion of right or left is relative to $\textrm{C}_{p}$; a $\textrm{C}_{5}$ has an intrinsic meaning, there are 72 different $\textrm{C}_{5}$'s, since there are 120 vertices and since 6 $\textrm{C}_{5}$'s are incident at each vertex (6$\times$120/10 = 72).  There are 48 right (48 left) screws whose angle is a multiple of $\pi/5$ (±72 degrees of arc, ±288 degrees of arc, ±144 degrees of arc, ±216 degrees of arc) along six possible directions from each vertex.  The Burgers vectors have respective lengths $R\tau^{-1}$, $R(3-\tau)^{1/2}$, $R\tau$ and $R(2+\tau)^{1/2}$ .  They all join \textbf{x} to the vertices of icosahedra of respective edge lengths $R\tau^{-1}$ for the first and $R$ for the three others.  The Burgers vectors of length $R(3-\tau)^{1/2}$ join \textbf{x} to its twelve 3rd nearest neighbors, which also form an icosahedron \cite{coxeter73}.

We also have:

\noindent $|\bigtriangleup x|_{3}=R$ for $\alpha_{3}=\tfrac{\pi}{3}$ $-$this is the distance from the centre x of a \textit{dodecahedron} (edge length $R\tau^{-1}$) to its twenty vertices, which are the next nearest neighbors of \textbf{x}; these are not the vertices of the \{3,3,5\} polytope, but the centers of the \{3,3,5\} triangular facets;

\noindent $|\bigtriangleup x|_{2}=\sqrt{2}R$ for $\alpha_{2}=\tfrac{\pi}{2}$ $-$this is the distance from the centre \textbf{x} of an \textit{icosidodecahedron} (edge length $\sqrt{2}R$) to its thirty vertices, which are the 4rth nearest neighbors of \textbf{x}; observe that the edge length can be larger than the radius of the habit sphere \cite{coxeter73}. These vertices are the midpoints of the \{3,3,5\} edges.

\subsubsection{Screw disvections} \label{Screwdisvections}
In analogy with the Euclidean case, we define a \emph{screw disvection} as a disvection line parallel to the Burgers vector, which in our case points out a great circle $\textrm{C}_{n}$, $n$ = 2, 3 or 5.  We have emphasized above the $n$ = 5 case.  The cut surface can be any surface bound by $\textrm{C}_{5}$, for example half a great 2-sphere.  Now, again in analogy with the classical Euclidean case, we are interested in an \textit{ideal} cut surface $\Sigma_{s}$ that slips along itself in the Volterra process.  An ideal cut surface of a screw disvection (i.e., a \textit{loop}) has to be parallel to the Burgers vectors, which are supported by a set of Clifford parallels.  Several possibilities exist: \\
\indent (i) $\textrm{C}_{p}$ being the axis about which $\textrm{C}_{5}$ is twisted, drop a segment of geodesic $\textrm{C}_{\bot}$ (an arc of great circle) perpendicular to $\textrm{C}_{p}$ between any point of $\textrm{C}_{5}$ and $\textrm{C}_{p}$; this operation determines a unique point on $\textrm{C}_{p}$ if $\textrm{C}_{5}$ and $\textrm{C}_{p}$ are not conjugate, any point on $\textrm{C}_{p}$ if $\textrm{C}_{5}$ =  $\tilde{\textrm{C}}_{p}$.\\
  \indent (ii) the surface generated by the great circles Clifford-parallel to $\textrm{C}_{5}$ and lying on $\textrm{C}_{\bot}$. In effect, consider two conjugate great circles $\textrm{C}_{5}$ and $\tilde{\textrm{C}}_{5}$, generated by a five fold axis; these conjugate geodesics are indeed ten-edge great circles, as shown in \textcite{coxeter91}.  Assume that the screw disvection line is along $\textrm{C}_{5}$.  The Burgers vector varies in direction along $\textrm{C}_{5}$ and the cut surface, but has a constant modulus, as shown above.  It is along the tangent to $\textrm{C}_{5}$.  The cut surface is made of Clifford parallels to $\textrm{C}_{5}$ and $\tilde{\textrm{C}}_{5}$ which stretch on the skew surface $\Sigma_{s}$ between $\textrm{C}_{5}$ and $\tilde{\textrm{C}}_{5}$.  The Burgers vector attached to each point of $\Sigma_{s}$ is by construction along the local Clifford parallel, and the Volterra process consists in a movement of $\Sigma_{s}$ in its surface.
\subsubsection{Edge disvections}  \label{Edge disvections}
An ideal cut surface $\Sigma_{e}$ of an \emph{edge disvection loop} has to be orthogonal to the Burgers vectors, which are supported by a set of Clifford parallels.  Therefore, according to the previous discussion, $\Sigma_{e}$ is a cap of a great sphere, bound by a loop of any shape.

\section{Discussion}
\label{Discussion}

{The foregoing pages present a general view of
disclinations and dislocations. The emphasis is put on

- their interplay with \textit{continuous} defects (themselves characterized as continuous dislocations, disclinations, or dispirations) in the constitution of defect textures in all media with continuous and frustrated symmetries,

- their various relaxation processes as they can be approached by the consideration of defects, in particular \textit{continuous} ones.

These questions are not entirely new, as can be seen from the citations, but they have been largely ignored in the ill-ordered media community. The purpose of the present essay is therefore  to stress the changes of perspective which occur in the theory of defects in mesomorphic phases (liquid crystals) and other 'ill-ordered' media, when the notions of disclination and continuous defects and the correlated use of the Volterra process, are fully used, in concurrence with the topological stability theory of defects.
}
Central to that analysis is the concept of \textit{extended} Volterra process, essential to the understanding of disclinations: this considers, in the last stage of the process, a viscous or plastic relaxation of the elastic stresses developed.

The other approaches to the defects in ill-ordered media are of two sorts. Only the second one is directly comparable to the Volterra process approach. \\

\noindent \textsl {a.} In liquid crystals with quantized translation symmetries, like smectics or columnar phases, a large part of the observations can be described in pure geometrical terms as \textit{isometric singular textures}, in a way largely inspired by the pioneering work of Georges Friedel in the first decades of the twentieth century \cite{friedel22,friedel10}; focal conic domains and developable domains, which obey rather restrictive geometric properties, are of the sort. But the geometrical approach to defects is limited to situations where the role of continuous defects can be taken aside. For a review, see \textcite{kln}.  It is a rather remarkable fact that some mesomorphic \emph{liquid} phases adopt, on rather huge sizes compared to molecular sizes, configurations that obey very precise geometrical rules and that these configurations are directly related to the structural symmetries of the medium. These configurations are worth being investigated for themselves, as they continue to be to-day. But these investigations did not inspire considerations leading to the present  theory of defects. \\

\noindent \textsl {b.} In contrast to the emphasis on the geometrical point of view, and since the late seventies, when the theory appeared, the defect classification basic concepts in mesomorphic phases and in frustrated media are those of the topological theory.  The topological theory relates the stability of defects to the topological properties of the order parameter; it only concerns discrete, quantized, defects.  That topological stability leads to energetic metastability is in many cases a reasonable assumption, which however cannot distinguish between different topologically equivalent configurations.

    \subsection{The extended Volterra process}  \label{TheextendedVolterraprocess}
\subsubsection{Pure {Volterra process} in the absence of plastic relaxation. Constitutive defect densities}  \label{PureVPintheabsenceofplasticrelaxation} The pure Volterra process has been thoroughly employed to the study of material deformation \cite{frank50b,friedeldisloc}; it applies to the construction of a Volterra \textit{disclination} in a solid, but only to a limited extent. One distinguishes two cases.\\

\noindent \textsl {a. Perfect disclinations.} The only case is that one of a straight, wedge, disclination line in a Volterra continuous elastic body or in a crystalline solid, with rotation vector $\bm{\Omega}$ parallel to the line, $\bm{\Omega}$ being a symmetry operation of the crystal. \\

\noindent \textsl {b. Imperfect disclinations.} This includes

(i) straight wedge Frank grain boundaries, that are analyzable: in an \textit{amorphous medium}, in terms of
parallel infinitesimal dislocations, uniformly distributed, and in a \textit{solid crystal}, in terms of finite dislocations, at least for small rotations,

(ii) straight twist lines or lines of a more general shape, which absorb/emit \emph{constitutive} dislocations attached to the line or in its vicinity. In a solid medium with \emph{no relaxation}, all these disclinations require very large energies to be created and also to move: their creation involves large stress concentrations extending over large regions; their motion would, whatever their nature, leave behind a stream of dislocations that could only be dispersed by slip and/or climb. These properties are related to the facts that, in such a solid without relaxation, the long range elastic energy is very large as soon as ${\Omega}$ is finite and the solid is of macroscopic size; the core energy is also very large as soon as the position of the lines deviates from the axis of rotation. Except for border cases involving disclinations of very small rotations, connected with weakly polygonized grain boundaries in crystals, one has not observed disclinations in solids as isolated objects, but only regrouped in close parallel strands with compensating strengths: parallel wedge pairs of opposite rotations in single crystals (described e.g. in \textcite{friedeldisloc}) or triplets with rotations following a Kirchhoff relation, at the edges of the grains in relaxed polycrystals, as analyzed above in \ref{crystallinesolidsandnanocrystals}. In this last case, the twist components of the involved disclinations exchange their constitutive dislocations. In the absence of possible relaxation by slip or climb, all these are \emph{sessile} defects.

 \emph{Remark}  The pure Volterra process does apply only if ${\Omega}>-2\pi$, since below this limit there is no matter left. But any value in the range so defined is allowed, even if the configuration is topologically unstable. Notice also that there is no limitation on the choice of the cut surface $\Sigma$ with respect to crystalline symmetries, but this yields different dislocation configurations with different energies, e.g.: a \textit{tilt} grain boundary if the rotation vector $\bm{\Omega}$ (Eq.~\ref{e1}) is in the plane of $\Sigma$, with a set of parallel edge dislocations; a \textit{twist} grain boundary if the rotation vector is perpendicular to the plane of $\Sigma$, with two crossing sets of parallel screw dislocations.

\subsubsection{Extended {Volterra process}. Relaxation defect densities} \label{Disclinationsandrelaxationdefects} Some plasticity of the medium can allow dislocations to relax. Disclinations are relaxed either by Nye's dislocations, or/and by absorbed/emitted attached defects.\\

\noindent \textsl {a. Nye dislocations} compensate as best as possible the dislocations of the grain boundaries associated to the disclination. They  widen the singular core of topologically unstable disclinations (as they do for a $k=\mathrm{n}$ line in a nematic). They can be included in the extended Volterra process, in a last step of the process that takes into account the plastic relaxations allowed by the boundary conditions and the symmetries.

Notice that the extended Volterra process allows ${\Omega}\leq-2\pi$. \\

\noindent \textsl {b. Emitted/absorbed dislocations} play a prevailing role when the disclination moves or changes its shape, in interplay with Nye's dislocations. And, as Nye's dislocations, they can be included in the extended Volterra process, also taking into account the allowed plastic relaxations. They can be induced: in a \textit{solid}, either by high temperature diffusion, or by plastic deformation at low temperature (or twinning, or crack formation), in a \textit{liquid crystal} by the anisotropic flow of matter.

\subsubsection{Mostly liquid crystals} \label{Mostlyliquidcrystalsandfrustratedmedia}

It has been known for long, well before the emergence of the topological theory, that the geometrical variability, the energetic (as opposed to topological) stability, and the elastic relaxation of defects in cholesterics \cite{friedel69} is largely controlled by relaxation defect densities.  This occurs through mechanisms, not yet enough investigated, that involve the continuous symmetries of the medium. A \underline{continuous defect}, just like a quantized defect in the Volterra classification, is related to a symmetry element of the liquid crystal, a continuous symmetry precisely. In terms of the topological theory, it always belongs to the identity element $\{1\}$ of the first homotopy group $\Pi_1(\textrm{V})$; in that sense, a continuous defect is never topologically stable. Continuous defects are often attached to quantized disclinations, whose flexibility and relaxation they control.  Similar results apply to all mesomorphic phases, as they all display continuous symmetries. The extended Volterra process was present in \textcite{friedel69}, but has not been developed since.  It is interesting to note that even the Georges Friedel's 'rigid' defect geometries get a different light when discussed in terms of continuous dislocations.

\textit{Remark} Some non-continuous Volterra disclinations also belong to
$\{1\}$. For instance, in uniaxial nematics N the Volterra process provides two
types of \textit{quantized} disclinations, those whose rotation
angle is an odd multiple of $\pi$, and those whose rotation angle is a
multiple of $2\pi$. Only the defects of the first type are
recognized as being topologically stable (\textit{i.e.} differing from $\{1\}$), the
defects of the second type belong to $\{1\}$. This is so because
the liquid crystal symmetries contain a \textit{continuous}
rotation subgroup (not present in solid crystals) whose presence
drives the first homotopy group finite, whereas the group of Volterra defects is
infinite.  In practice quantized defects that are not allowed by
the topological theory are also of interest, as it may happen, for instance with suitable boundary conditions, that the
\emph{energetic stability} of the defect prevails over its
topological stability. In this sense these quantized disclinations {which carry an angle of rotation}
multiple of $2\pi$ are true disclinations.\\
\indent A deformed mesomorphic phase being a liquid, there are no strains at rest; but obviously there are large curvature deformations that can be analyzed in terms of Nye's densities \cite{kleman82b}, of continuous dislocations or continuous dispirations.  Since continuous defects are topologically related to the continuous symmetries of the mesomorphic phase; as already stressed, this puts limitations to the possible curvature deformations of the medium in the vicinity of quantized disclinations, and to the shape modifications and mobility of those defects.\\
\indent Indeed a continuous transition between Hooke strain elasticity for solids and Frank-Oseen curvature elasticity for mesomorphic phases is insured by a continuously growing density of dislocations whose Burgers vectors decrease and vanish, in such a way that the total Burgers vector keeps constant.

\subsubsection{Extended {Volterra process} \textit{vs}. {topological stability}} \label{VPvsTS}

\noindent \textsl {a. The topological stability theory} only considers defects that cannot be suppressed by plastic relaxation. It provides an \emph{a posteriori} description that results from a mapping of the distorted medium onto the order parameter space V = E$^3$/H. Quantized dislocation and disclination
invariants $\{a_i\}$ belong to the first homotopy group of the
order parameter space $\Pi_1(\textrm{V})$, which is usually non-abelian.

The topological approach
has the advantage to be extendible to defects of any
dimensionality (point, line and wall defects, and configurations). It is a more condensed process but  a poorer one than the Volterra process, as it neglects the boundary conditions and makes equivalent all the configurations that can be deduced one from the other by continuous deformations. For such deformations to occur, it needs a high degree of plasticity which is not met in solids at low temperatures, whether crystals or glasses and other amorphous systems, but which is met in liquid crystals and ferromagnetism. It does not always allow to predict the stabler configuration in ordered media, still less to foresee the finer details of any configuration. By neglecting the topologically unstable situations, TS makes the bet that the corresponding singularities do reduce their energy by dissipating in some way. This bet is most often justified; however it might fail in the presence of special boundary conditions, as the $k=1$ nematic line in a capillary tube \cite{cladis72,meyer73} and more generally in thin samples, where their broad core was well-known by G. \textcite{friedel22}. Also, in the presence of strong material constant anisotropies, a $k=1$ disclination core may be singular $-$ this latter situation is met in nematic main-chain polymers \cite{mazelet86}.\\

\noindent \textsl {b. The extended Volterra process} is an \textit{a priori} description of defects, as it gives a process of
\textit{creation} of line defects (only)
whose characteristic invariants are classified by the elements of
the symmetry group H of the medium) in a medium free of stresses.

It yields the same conclusions as the topological stability theory, but at a finer level, by properly handling all the plastic relaxations, including those related to line-attached defects. This approach can be particularly useful when investigating dynamical aspects, when the viscosity is large as in solids, in most smectics, and in polymeric liquid crystals \cite{friedel79, kleman1984}. It is also useful when dealing with nanostructures. This is at the price of an often much more complex analysis, as already stated.
\subsubsection{Flashback on a posteriori Volterra description of a defect in an amorphous medium}
\label{aposterioriVolterra}
A Volterra constructed disclination is qualified by the line direction
with respect to the rotation vector: it is a wedge, or a twist, or a
mixed disclination line. Similarly a dislocation is a screw, edge,
or mixed dislocation line.  Such a specificity does not belong to
a line defined by its topological invariant.  The question therefore arises
whether it is possible to qualify any given line of the deformed medium in the same terms.
This requires an \textit{a posteriori} description of the defect,
{i.e.} some kind of mapping that brings back the deformed
medium to a stress-free, reference, medium, from which the line
has been supposedly constructed in the Volterra mode.

This program seems feasible when the reference medium
is endowed with a 3D \textit{crystal} lattice L whose equivalent
directions and equivalent points can be recognized in the
deformed medium. Directions can be followed along circuits
surrounding disclinations, and a rotation or Frank vector
obtained, which is unique in the reference lattice. This is how the Kirchhoff relations are established\footnote{Notice that the Burgers vector of a dislocation surrounding a
disclination $\mathbf{\Omega}$ is not uniquely determined this way; it is defined
but to a rotation $\mathbf{\Omega}$, as first noticed by
\textcite{sleeswyk66}, see also \textcite{dewit71,harris70, harris71}. This is
a small (and interesting) complication stressed in \ref{Frank'sgrainboundaryandFriedel'sdisclinationcompared}, whose effect on the
Kirchhoff relations has not been investigated.}.

These
considerations, namely the possibility of fully characterizing
an isolated dislocation or disclination, extend to the
continuous theory of defects of Bilby, Kr\"{o}ner, etc, which
always assumes explicitly the presence of a lattice, whose
repeat distances are most certainly taken infinitesimally small,
but this does not invalidate our reasoning. The continuous
theory, in a way, contemplates \textit{continuous sets} of
\emph{isolated} defects, whose Burgers vector is
nonetheless infinitesimally small.  For this purpose it employs
the methods of differential geometry on manifolds.  It is
precisely the existence of a lattice that justifies the use of
Cartan-Levi-Civita's \textit{distant parallelism} method in the
continuous theory of defects. For, {if} equivalent points
and directions{did not }exist in the deformed medium, the comparison of
vectors at two distant points would make no sense. It then
follows that there are quantized disclinations in an amorphous medium, but their characteristics depend on the gedanken lattice drawn on it; therefore they do not present physical interest. We have considered only continuous disclinations in amorphous media, in this essay.

\subsection{Volterra processes in various media, compared} \label{Vprocess}

The Volterra process applies to amorphous and frustrated media as well as to liquid, solid crystals, and templates of frustrated media. There are however some differences worth stressing.\\

\noindent \textsl {a. In an amorphous medium,} the translation vector and the
rotation vector that define the relative displacement of the cut
surface are elements of the full Euclidean group $\textrm{E}^3$. The
related defects are not topologically stable.
Dislocations and finite disclinations are \textit{non-independent}
defects.  A finite
disclination carries a field comprising constitutive and
relaxation dislocations. It also possibly carries infinitesimal
disclinations. The constitutive and relaxation dislocations relate
to the strength of the line and to a part of its curvature
(through the kink mechanism); the infinitesimal disclinations
relate to another curvature component.  These infinitesimal
disclinations correspond to infinitesimal
rotations belonging to the rotational part of the Euclidean
group, whereas the strength of a finite disclination is
characterized by the invariants associated to the constitutive and relaxation dislocations, i.e. belonging to the translational part of
the Euclidean group only; these translations generate the Frank vector
associated with the finite disclination.\\

\noindent \textsl {b. In a solid crystal,} the characteristic Volterra invariants
of the defects are the translation and the rotation elements of
its Euclidean symmetry group H, whose elements classify
dislocations and disclinations (necessarily finite) as, now,
\textit{independent} defects.  However, the concepts of
constitutive and relaxation dislocations still make sense.
Because there are no
infinitesimally small rotations in H, a line can be curved only by
the kink mechanism, which makes use of finite relaxation dislocations. On the other hand, a straight disclination line is necessarily of wedge character, as there are no infinitesimal dislocations to give it a twist character.

Notice that this description applies better to coarse-grained crystals with a quasi perfectly polygonized, annealed, structure. Most grain boundaries in a polycrystal are strongly misorientated, like polynanocrystalline materials below. \\

\noindent \textsl {c. In a polynanocrystal,} the nanograins are separated by generally large misorientation grain boundaries which thereby are not analyzable in terms of quantized dislocations (those of the crystal) but rather in terms of continuous dislocations. In an ideal picture, three grains meet along a segment of (continuous) disclination, and these segments meet at quadruple nodes, forming then a 3D disclination network. The plastic deformation of polynanocrystals is at variance with that one of usual polygonized crystals; it is governed by the disclinations, which yields considerable stresses. Disclination-governed plastic deformation has long been a favorite subject of the Saint Petersburg russian school (see \textcite{rom92} for a review), without attracting much attention elsewhere.\\

\noindent \textsl {d. In a liquid crystal,} the characteristic invariants of
the defects are as above the translation and the rotation elements
of a Euclidean group H, whose elements classify dislocations and
disclinations as, again, \textit{independent} defects. But,
according to the case, the corresponding translation and rotation
symmetries are finite or/and continuous, including infinitesimally
small elements.  In a nematic (N), translation symmetries are
continuous, including infinitesimally small elements, and rotation
symmetries are of both types. In a SmA, there are
quantized and continuous translation and rotation symmetries.
Line defect curvatures are therefore related to both the kink
mechanism through (quantized or continuous) dislocations and the
presence of infinitesimally small strength disclinations.\\

\noindent \textsl {e. In the curved habit spaces of frustrated media,} like $\{\textrm{am}/\textrm{S}^{3}\}$, $\{3,3,5/\textrm{S}^{3}\}$, and $\{dtw/\textrm{S}^{3}\}$, dislocations (which we call there disvections) are not commutative, and disclinations can still be defined, with the same differences as above between amorphous media and crystalline media. Generically, disvections do not attach to disclinations, but disclinations do. Three-D disclination networks are therefore important features of the habit curved spaces, if one draws one attention to the plastic deformation of such spaces at constant Gaussian curvature.

Disclinations also form 3D networks in actual, Euclidean, amorphous systems, liquid crystals such as cholesteric blue phases, Frank and Kasper phases, and probably in undercooled liquids or quasicrystals. In such 3D networks, the disclinations have to be somewhat flexible, which is possible, whether these disclinations are quantized or not, only if other defects, dislocations (defined in the Euclidean decurved medium) or disclinations, continuous or not, attach to them.

\section*{Acknowledgments}

The authors thank Yves Br\'echet, Jean-Luc Martin and Helena Van Swygenhoven for discussions on nanocrystals. They are also grateful to Claire Meyer, Yuriy Nastishin and Oleg Lavrentovich for permission to reproduce some of their artworks. MK would like to thank Claire Meyer and Yuriy Nastishin for discussions on continuous defects in mesomorphic phases.

\appendix

\section{Continuous dislocations in solids and Nye's dislocation densities} \label{appnye}
This is a very simplified presentation of a topic that has been much inspiring for the theory of continuous dislocations.

We consider a deformed body described in terms of deformations $\beta_{{ij}}$ or of strains $e_{ij}=\tfrac{1}{2}(\beta_{ij}+\beta_{ji})$; we assume that these quantities are so small that it is licit to ignore second order terms.  As it is well known, it suffices to know the strains  $e_{ij}$ in order to derive the stresses and the elastic energy, so that the deformations  $\beta_{ij}$ are generally ignored.  In the absence of elastic singularities (dislocations, disclinations), the $\beta_{ij}$s derive from a displacement:  $\beta_{ij}=u_{j,i}$ (we use the notation $u_{i,j}=\dfrac{\partial{u}_{i}}{\partial{x}_{j}}$).  The conditions (due to Beltrami, 1888) to which the strains have to obey in order to insure the existence of a displacement function are
    \begin{equation}(inc\textbf{e})_{ij} \equiv {\varepsilon_{ijk}}{\varepsilon_{lmn}}e_{{jm,kn}}=0 \label{(A1)} \end{equation}
 \noindent Here, we have introduced the incompatibility tensor $(inc\textbf{e})_{{ij}}\equiv{\varepsilon_{ijk}}{\varepsilon_{lmn}}e_{jm,kn}$ ; it does not vanish in the presence of elastic singularities.

The condition for the existence of a displacement function takes a much simpler form in terms of the $\beta_{ij}$s
    \begin{equation}
    \alpha_{ij} \equiv -{\varepsilon_{ikl}}{\beta_{lj,k}}=0 \label{(A2)}
    \end{equation}

The theory of continuous dislocation densities emphasizes the role of the tensor $\alpha_{ij}$; when there is no displacement function, it does not vanish, and it can be interpreted in terms of dislocation densities; we refer for details to \textcite{kroner}, see also \textcite{39} and \textcite{nabarro}, chapt.~1 and~8.

The integral $\triangle{u_{i}}=-\int_{\partial{\sigma}}{\beta_{ij}}{dx_{i}}=-\int\!\!\!\int_{\sigma}\alpha_{kj}{dS_{k}}$ on a loop $\partial{\sigma}$ that bounds the surface element $\sigma$ vanishes if the strain derives from a displacement function ($\beta_{ij}=u_{j,i}$); if this is not the case, the 1-form $\beta_{ij} dx_{i}$  is not integrable and the integral measures a displacement vector  $\triangle{u_{i}}$.  $\triangle\textbf{u}$  is interpreted as the Burgers vector of the dislocation densities that pierce the surface bounded by $\partial{\sigma}$; ${\alpha_{ij}}$  is the density of dislocations along the $x_{i}$-axis; it measures the total Burgers vector component along the $x_{j}$-axis of the set of dislocations through the unit area perpendicular to the $x_{i}$-axis.  For a dislocation L of finite Burgers vector $\mathbf{b}$ along a direction $\mathbf{t}$, one gets  $\alpha_{ij}=-t_{i} b_{j} \delta(\textrm{L})$.

Now, again after Kr\"oner, we write $\alpha_{ij}$  as follows:
     \begin{equation}
     \alpha_{ij} = -\varepsilon_{ikl} {e_{lj,k}}+\omega_{i,j}-\omega_{k,k}\delta_{ij}  \label{(A3)}
    \end{equation}

 \noindent by splitting $\beta_{ij}=e_{ij}+\omega_{ij}$  into a symmetric part $e_{ij}$  and an antisymmetric part $\omega_{ij}$
 We have introduced $\omega_{i}=\tfrac{1}{2} \varepsilon_{ikl} \,\omega_{kl}$, (reciprocally $\omega_{ij}=\varepsilon_{ijk} \,\omega_{k}$), which yields  $\omega_{i}=\tfrac{1}{2} \varepsilon_{ikl} \, \beta_{kl}$.  Hence the relation~\ref{(A3)}.  If $e_{ij}$  is 'compatible', i.e.  $\alpha_{ij}=0$, we have  $\omega_{i}=\tfrac{1}{2} \varepsilon_{ikl} \, u_{l,k}$.  In the generic incompatible case, the  $\beta_{ij}$s and the  $\omega_{i}$s have to be given new interpretations, which we detail below.

\begin{figure}
\includegraphics{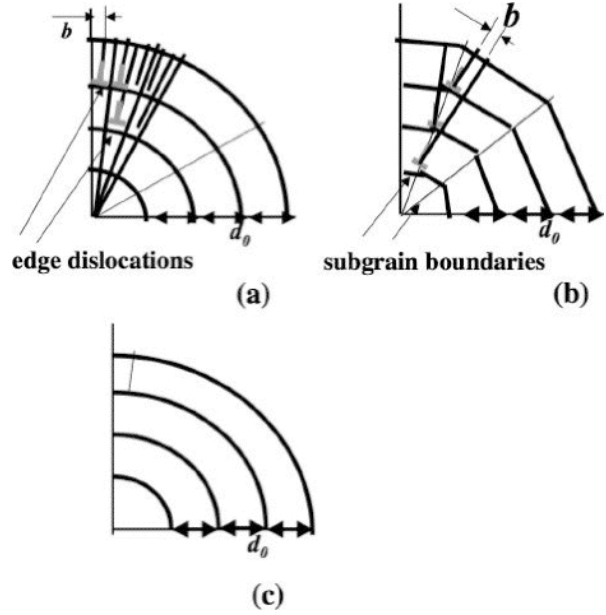}
   \caption{ \label{figA1}{Genealogy of Nye's dislocations. These figures represent the gradual passage from a situation where (a) a random density of finite dislocations imposes curvature \textit{and} strain to a situation where (c) the Burgers vectors becoming infinitesimal, the strain vanishes but the curvature subsists. In-between (b), the dislocations of (a) have organized into subgrain boundaries}}
\end{figure}

For the time being, notice that  $\omega_{i}$ measures the rotation of an element of volume with respect to the lattice directions ($\omega_{3}$ is a rotation about the $x_{3}$-axis with respect to the $x_{1}$-axis and the $x_{2}$-axis); according to Eq.~\ref{(A3)} it contains \textit{two} contributions to the rotations, one coming from the strains and the other from the dislocation densities $\alpha_{ij}$.  This splitting into two contributions makes sense, because it is possible to conceive an ordered medium with no strains but a density of dislocations, the \emph{Nye's dislocation densities} \cite{nye53}, at the origin of the rotation distortions.  An example is given in Fig.~\ref{figA1},
directly inspired by Nye's analysis.  In a strainless medium, these densities read:
     \begin{equation}
     \alpha_{ij}=\omega_{i,j} -\omega_{k,k} \delta_{i,j} \label{A4a}
     \end{equation}

 \noindent which can also be written
     \begin{equation}
     \omega_{{i,j}}={\alpha_{ij}}-\tfrac{1}{2}{\alpha_{kk}\delta_{i,j}} \label {A4b}
     \end{equation}

In the general case, the gradient tensor of the rotations is
    \begin{equation}
    \omega_{{i,j}}=\varepsilon_{ikl}{e_{lj,k}}+{\alpha_{ij}}-\tfrac{1}{2}{\alpha_{kk}\delta_{i,j}} \label {A5}
    \end{equation}

 \noindent where the two contributions (from the strains, from the dislocation densities) are separately non integrable.

Of course, this model is valid as long as the rotations $\omega_{i}$s  are small.  In the smectic case discussed in the text, the $\omega_{i}$s  measure the rotation of the orthonormal Darboux-Ribaucour trihedron from its equilibrium position along the principal axes of the shell \cite{darboux}.

There is no equivalence between a description in terms of Nye's dislocation densities and a description in terms of strain dislocation densities; in other words the solutions in $\beta_{ij}$  of the equation  $\alpha_{ij}\equiv-\varepsilon_{ikl} \, \beta_{lj,k}=\omega_{i,j} -\omega_{k,k} \, \delta_{i,j}$ are such that $\beta_{ij}+\beta_{ji}=0$ .  Reciprocally the solutions in $\beta_{ij}$ of the equation $\alpha_{ij}\equiv-\varepsilon_ {ikl} \, \beta_{lj,k}=-\varepsilon_{ikl} \, e_{lj,k}$  are such that  $\beta_{ij}-\beta_{ji}=0$.  Descriptions in terms of Nye's dislocations and of strain dislocations are exclusive one from the other.

The gradual passage depicted in Fig.~\ref{figA1} can also be thought of as a gradual transformation of a solid (with quantized dislocations carrying strains) into a SmA phase (with continuous Nye's dislocations only).  In this transformation, the distorted solid crystal elasticity gradually becomes the elasticity of the distorted smectic phase.  More generally, the prevalence in mesomorphic phases of Nye's dislocations over quantized dislocations is the sign of their liquid character, and the origin of the Frank$-$Oseen elasticity \cite{frank58a} replacing the Hookean elasticity.

Coming back to the physical meaning of the  $\beta_{ij}$s and of the  $\omega_{{i}}$s in the generic incompatible case:

  \indent $\beta_{21}$ describes a \emph{plastic} distortion that displaces an element of volume with a finite section $\triangle{x_{1}}\triangle{x_{2}}$  along the ${x_{1}}$-axis, by a quantity $\beta_{21}\triangle{x_{2}}$.  This distortion can be managed by the $\underline{glide}$ along the same axis of a set of edge dislocations parallel to the ${x_{3}}$-axis, with Burgers vectors along the ${x_{1}}$-axis.  There is no change of volume density.

    \indent $\beta_{22}$ describes a \emph{plastic} distortion that elongates (or shortens, according to the sign) an element of volume with a finite section   $\triangle{x_{1}}\triangle{x_{2}}$ along the ${x_{2}}$-axis, by a quantity $\beta_{22}\triangle{x_{2}}$.  This distortion can be managed by the \emph{climb}, along the same axis, of a set of edge dislocations parallel to the ${x_{3}}$-axis, with Burgers vectors along the ${x_{2}}$-axis. There is a relative change of volume density equal to $\beta_{22}$ .

     $\omega_{3}$, discussed above, can also be thought of as the result of two glide operations, $\beta_{21}$ and $-\beta_{12}$.

  In the present theory, no account is taken of $\textit{disclination}$ densities, since the rotation $d{\omega_{{i}}}=\omega_{{i,j}}dx_{j}$ is of course integrable: the density of disclinations vanishes.

  In fact this theory computes the \emph{minimum} density of dislocations necessary to produce a given distortion of the medium. It assumes that this results from dislocation creation and annihilation, as well as from motion by glide and climb. Such hypotheses are well fulfilled in magnets or liquids or (partly) in liquid crystals. They are not generally fulfilled in solids, except partly at very high temperatures. As in TS, it assumes in fact perfect plastic relaxations but for given boundary conditions.

 \section{Topological stability and Volterra process compared.  Conjugacy classes and
homology} \label{appconjug}

To start with, notice that $\textrm{L} \mapsto (0, a)$ and
$\textrm{L"} \mapsto (2, a)$ are in the same conjugacy class of
the first homotopy group; in fact, all the elements $(2p, a)$, $p
\in Z$, belong to the same conjugacy class and represent the same
core-type of disclination; the other type of core corresponds to
the conjugacy class $(2q+1, a)$, $q \in Z$.  Each core-type of
disclination, i.e. each conjugacy class can be
identified in the Volterra classification with \textit{two} types
of disclinations, $k =\pm  \tfrac{1}{2}$, differing by
the sign: the topological classification does not distinguish $k
= \tfrac{1}{2}$ and $k =- \tfrac{1}{2}$.
There is an infinite number of other conjugacy classes,
corresponding to the dislocations, each defined by a positive
integer $r$ and containing two elements $\{(r, e), (-r, e)\}$,
i.e. two Burgers vectors equal in modulus, opposite in
signs. All together
one has $\infty+2$ conjugacy classes:

$$\begin{array} {c|c|c}
C_r: \{(r, e), (-r, e)\} \qquad r \in Z^+ \bigcup \{0\}\\
C_1: \{(2p+1, a)\}   \qquad p \in Z \qquad \\
C_2: \{(2q, a)\} \qquad \quad \, \, q \in Z
\end{array} \eqno{(\textrm{B}1)} $$

Each conjugacy class therefore corresponds to an element of the
Volterra classification, although it does not specify whether the
line is twist, wedge,... This assertion can be made more accurate
as follows.  The classes $C_{2r}$ - they represent dislocations
whose Burgers vector is even -, are commutators of
$\Pi_1(V_{SmA})$; it is indeed not difficult to prove that any
element of the form $uvu^{-1}v^{-1}$, where $u, v = (n, a)$,
belong to a $C_{2r}$. The commutators generate a normal subgroup
of $\Pi_1(V_{SmA})$  $-$ the so-called derived group \textrm{D} $-$, such that $\Pi_1(\emph{V}_{SmA})$ can be partitioned into cosets $c_i$\\

\indent
  $$ \Pi_1(\textrm{V}_{SmA})= \sum_{i} c_{i} = \sum_{i}u_i \textrm{D} \, \,=
\textrm{D}+u_{1} \textrm{D}+u_{2} \textrm{D}+u_{3} \textrm{D}, \quad  \eqno{(\mathrm{B}2)} $$
\noindent with $c_0 = \textrm{D}$; $c_1 = (2p+1, a) \textrm{D}$; $c_2 = (2q, a) \textrm{D}$;
$c_3 = (2r+1, e) \textrm{D}$. It is easy to check that the content of each
coset is independent of the values of the integers $p$, $q$, and $r$.
Furthermore, all the elements of $C_1$ belong to $c_1$, those of
$C_2$ belong to $c_2$, and those of all the $C_{2r+1}$ belong to
$c_3$. Finally, $\Pi_1(\textrm{V}_{SmA})/\textrm{D}= \{c_0, c_1, c_2, c_3\}$ is an
abelian group whose identity element is $c_0$. It is the dihedral
group $\textrm{D}_2$; its multiplication rules $c_1 c_3 = c_2$, $c_2 c_3 =
c_1$ reproduce the effects described above of the
emission/absorption of a dislocation with an odd Burgers vector at
the core of a disclination.  Two disclinations of different core
types give an odd dislocation, when merging: $c_1 c_2 = c_3$.
\textit{All the odd dislocations appear as one and the same
element $c_3$; the even dislocations appear in the identity
element $c_0$}. More generally, $\Pi_1(\textrm{V})/\textrm{D}$
is an abelian group, called the first homology group, noted
$\textrm{H}_1(\textrm{V},Z)$.

Let $\Omega_1$ be some rotation symmetry axis in a SmA liquid
crystal; an axis $\Omega_2$ obtained by any rotation of
$\Omega_1$ about the normal to the layers is also a rotation
symmetry axis. However the two wedge disclinations $\textrm{L}_1$ and
$\textrm{L}_2$ are not distinguished by the topological classification, which
assigns indifferently the element $(0,a)$ to both. The latter
result is another weakness of the topological classification, which
originates in the presence of a continuous rotation symmetry
about the layer normal in a SmA.  The same difficulty does not
arise with a SmC phase, whose first homology group is also
$D_2$, whereas the conjugacy classes are twice as numerous as in
a SmA phase; $\Pi_1(\textrm{V}_{SmC})= Z \Box Z_4$, see
\textcite{bouligand5,bouligand7}.  Observe that the
symmetry point group of a SmC is $\textrm{H}_{SmC}=Z \Box Z_2$
(reflections excluded), and that $D_2$ is precisely the quotient
of the point group by its derived group. $Z \Box Z_2$ is the Volterra
group of defects (it is easy to check that it contains the
\textit{two} kinds of rotation axes, those in the layer planes,
and those in the mid-layer planes, all perpendicular to the
mirror plane), and $D_2$ can be understood as the Volterra group
restricted to its abelian properties. The SmC point group is a
quantized group, and the relation just indicated between the Volterra
group and the homology group can be extended to any medium with
a quantized point group (reflexions excluded).  The same
relation does not hold when the point group is continuous, as it
is for SmA
($\textrm{H}_{SmA}= Z \Box D_{\infty}$), see \textcite{kleman82b}, chapter 10.

But in both cases the even dislocations are lumped in the identity
element of $\textrm{H}_1(\textrm{V},Z)$. In both cases, eventually in all cases of
ordered media, the Volterra classification has a stronger kinship with
the set of conjugacy classes of $\Pi_1(\textrm{V})$ than with $\textrm{H}_1(\textrm{V},Z)$.
The lumping of even dislocations in the identity element of the
first homology group is of course a weakness of the homology
classification; a tentative explanation is given in
\textcite{kleman82b}, chapter 10, where can be found a more general discussion of
the conjugacy classes and of the homology
classification in any ordered medium(see
also \textcite{kleman77}).

\section{The ellipse in a FCD as a disclination} \label{appell} 
\begin{figure}
\includegraphics{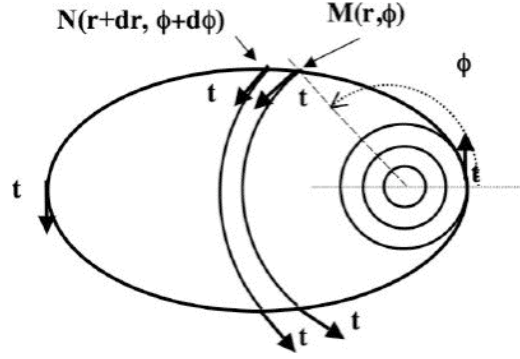}
   \caption{ \label{fig15}{Distribution of the rotation vectors
$\mathbf{\Omega} = \pi \mathbf{t}$ on an ellipse
in a focal domain. The layers intersect the plane of the ellipse
along circles centered in one of the foci}}
\end{figure}

In a generic FCD (Fig.~\ref{fig14}), the hyperbola H is completely embedded in the domain and is a wedge disclination of strength $k = 1$; indeed any
direction along it is a $2\pi$ symmetry axis. On the other hand the ellipse E, which is usually on the boundary of the domain, is a disclination of strength $k = \frac{1}{2}$ for the full layer geometry, which contains the FCD and the two sets of outside layers that meet on the plane of E. But E is not in general of wedge
character, because the tangent to the ellipse is not a folding
axis for the local smectic layers.  An obvious
solution is given Fig.~\ref{fig15}, where the local rotation vector $\Omega\,
\mathbf{t}$, with $\Omega = \pi$, is tangent to the
layers, which are folded symmetrically with respect to the plane
of the ellipse \cite{pml06}.
Other solutions are possible, with
$\mathbf{\Omega}$ off this plane, (but then necessarily
non-symmetric with respect to the ellipse plane), which would
correspond to a different set of attached
dislocations, most probably of higher energy.

We apply Eq.~\ref{e1}, rewritten
\begin{equation}
d \mathbf{b}_{tr}=2 \mathbf{t} \times \mathbf{MN}, \label{(32)}
\end{equation}

\noindent to the ellipse, Fig.~\ref{fig15}; $d \mathbf{b}_{tr}$ is
perpendicular to the plane of the ellipse, and one easily gets
$d\mathbf{b}_{tr}=2\,dr$; therefore the total
Burgers vector attached to one side of the ellipse  ($0
< \phi < \pi$) is \cite{kleman00}\\
\begin{equation}
\mathbf{b}_{tr} =\displaystyle \int_{\phi=0}^{\phi=\pi} d\mathbf{b}_{tr}= 4c. \label{(33)}
\end{equation}

Taking $dr=d_0$, $-$ an approximation which makes sense (up to
second order), since $d_0$ is so small compared to the size $a$ of
the ellipse $-$ it is visible that the points M, of polar
coordinates $\{r, \phi\}$, and N, of polar coordinates $\{r + dr,
\phi + d\phi\}$, are on two parallel smectic layers at a distance
$d_0$, and the total Burgers vector attached to a layer is equal
to $2d_0$. There are no dislocations attached to the singular
circle of a toric FCD, as the eccentricity $e$ vanishes, and
$\delta r=0$. The Burgers vector attached to a layer that is
transverse to the perimeter of the ellipse is a constant, $2d_0$.
Consider one such dislocation of Burgers vector $2d_0$; it spreads
outside the ellipse in the shape of a quantized dislocation, on
both sides of the ellipse, and goes across it in the shape of
curved layers, in a manner akin to the $k =
\tfrac{1}{2}$ case (Fig.~\ref{fig10}). The full balance of Burgers vectors
has to take into account the Nye's dislocations, including those
related to the
infinitesimal disclinations,
\begin{equation}
-d\mathbf{f}=d\mathbf{t} = - \{ \cos \phi, \sin \phi\}\,d\phi,  \label{(34)}
\end{equation}

\noindent $d\mathbf{t}$ is a vector along the normal of the
layer, as required by symmetry (the normal to the layer is an axis
of continuous rotational symmetry).  But there is no \textit{a
priori} reason that the total Burgers vector carried by the Nye's
dislocations balances topologically the quantized attached
dislocations; it is the attachment that provides a balance
equation.

 If the dislocation segments outside the ellipse belong
to a plane, they build a grain boundary limited by
the ellipse.  It is a simple matter to check that
the misorientation angle is $\omega=2\,\sin^{-1} e$. This grain boundary is a pure tilt
boundary if it occupies the plane of the ellipse. Notice that $\bm\omega$ (the tilt vector, along the direction of the ellipse minor axis) and $\Omega\,\mathbf{t}$ (variable along the ellipse) define both the same tilt grain boundary; we are in the case described in \ref{Frank'sgrainboundaryandFriedel'sdisclinationcompared}, Eq. \ref{e10}.

We know from topological stability theory that the defect geometry of the ellipse
as a disclination does not break the constancy of its conjugacy
class in $\Pi_1(\textrm{V}_{SmA})$ along E. The variation of the
representative homotopy class (inside the same class of conjugacy)
for a circuit embracing the line depends on the quantized
dislocations (not the infinitesimal dislocations which, as already
stated, belong to the identity class of the homotopy group)
attached to it, which it also embraces.

 \section{A few geometrical characteristics of the 3-sphere} \label{appthreesphere}

 Details on the topics that follow can be found in \textcite{sommer}, \textcite{montes}, \textcite{coxeter98}.

\subsection {Clifford parallels and Hopf fibration}
\label{Clifford parallels and Hopf fibration}
\begin{figure}
\includegraphics{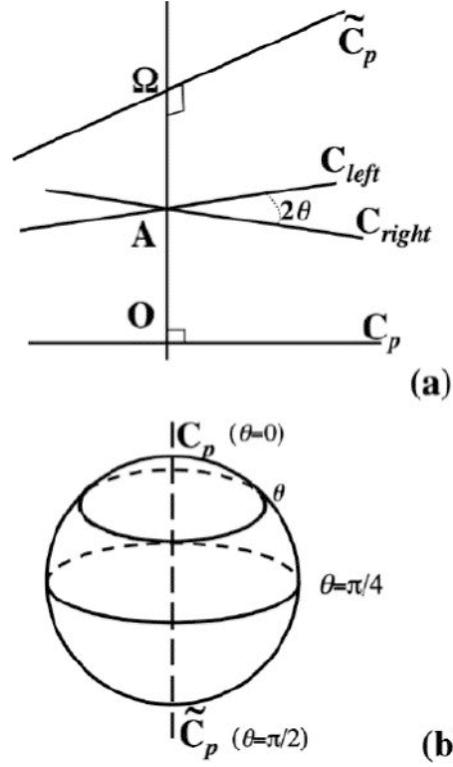}
   \caption{\label{fig17}{(a)- $\textrm{C}_{p}$ and $\tilde{\textrm{C}}_{p}$ are conjugate great circles (represented as straight lines in reason of their geodesic character); $\textrm{C}_{right}$ and $\textrm{C}_{left}$ are equidistant to $\textrm{C}_{p}$ and $\tilde{\textrm{C}}_{p}$ (OA= $\vartheta$, A$\bm \Omega=\pi/2 -\vartheta$).  All right and left Clifford parallels at the same distance $\vartheta$ generate a ruled Clifford surface with two axes of revolution $C_{p}$ and $\tilde{\textrm{C}}_{p}$; (b)- representation on the basis $\textrm{S}^{2}$ of a Hopf fibration of $\textrm{S}^{3}$, with $\textrm{C}_{p}$ and $\tilde{\textrm{C}}_{p}$ as particular fibers. Each fiber is represented by a point on the basis; Clifford surfaces (the small circles on $\textrm{S}^{2}$) $\vartheta=constant$; spherical torus $\vartheta=\pi/4$}}
\end{figure}

The trajectory of a point M $\in\textrm{S}^{3}$  under the action of a right screw, say, $x^{\prime}={x}\,e^{{\alpha}q}$, $0<\alpha\leq2\pi$, is a great circle $\textrm{C}_{right}$ of $\textrm{S}^{3}$.  Two right (resp. left) great circles that rotate helically about the same $q$ are equidistant all along their length and mutually twisted with a pitch $\varpi=2\pi$ (resp. $-2\pi$): they form a congruence of so-called right (resp. left) \emph{Clifford parallels} that fill $\textrm{S}^{3}$ uniformly.  All the great circles of this congruence are equivalent.  Because any of those great circles can be considered as the axial line of cylindrical double-twist geometry for the other parallel great circles, the geometry of Clifford parallels has inspired a non-frustrated model of the Blue Phase \cite{seth1}.  The great sphere $\textrm{S}^{2}$ defined by the intersection of the hyperplane $q_{1}x_{1}+q_{2}x_{2}+q_{3}x_{3}=0$ and $\textrm{S}^{3}$ intersects the great circles of the congruence orthogonally; this geometry defines $\textrm{S}^{3}$ as a fiber bundle of great circles $\textrm{S}^{1}$ over a great sphere $\textrm{S}^{2}$, the \emph{Hopf fibration}, see Fig.~\ref{fig17}.
Notice that great circles are the geodesic lines of $\textrm{S}^{3}$ and great spheres the geodesic surfaces of $\textrm{S}^{3}$.

\subsection {Spherical torus} \label{Sphericaltorus}
The axis $p$ and the circle of unit radius in its equator plane (Fig.~\ref{fig16}) just play identical roles in the double rotation: $p$ is the stereographic projection of a great circle of $\textrm{S}^{3}$, here denoted $\textrm{C}_{p}$, which is the intersection of the  sphere of unit quaternions $x_{0}^{2}+x_{1}^{2}+x_{2}^{2}+x_{3}^{2}=1$  with the 2-plane ${\dfrac{x_{1}}{p_{1}}=\dfrac{x_{2}}{p_{2}}=\dfrac{x_{3}}{p_{3}}}$ ; the equator circle, here denoted $\tilde{\textrm{C}}_{p}$, is a great circle intersection of the unit quaternion sphere with the 2-plane $x_{0}=0$, $p_{1}x_{1}+p_{2}x_{2}+p_{3}x_{3}=0$  .  These two great circles are conjugate; they have this remarkable property that the arc of any great circle $\textrm{C}_{g}$ joining any point on $\textrm{C}_{p}$ to any point on $\tilde{\textrm{C}}_{p}$  is perpendicular to both and has arc length $\pi$/2.  Select any $\textrm{C}_{g}$ and transform it under the action of the right screw $x^{\prime}={x}e^{{\alpha}p}$, $0<\alpha\leq2\pi$, say; this yields a set $<\textrm{C}_{g}>$ of great circles, all stretching between $\textrm{C}_{p}$ and $\tilde{\textrm{C}}_{p}$. The surface so obtained is called the spherical torus; it is generated by two families of great circles: $<\textrm{C}_{g}>$ just defined, and $<\textrm{C}_{p}>$ made of the Clifford parallels that are parallel to $\textrm{C}_{p}$ (and to $\tilde{\textrm{C}}_{p}$). These two families form a rectangular network on the spherical torus.
\subsection {Clifford surfaces} \label{Cliffordsurfaces}
 Consider now the action of a left screw $y'=e^{{\beta}p}\,{x}'$  about the same axis $p$, i.e.  $y'=e^{{\beta}p}\,{x}\,e^{\alpha}p$ on $\textrm{C}_{right}$ when \textbf{M}($\alpha$) traverses the geodesic line $\textrm{C}_{right}$, and $\beta$ varies in the range $0<\beta\leq 2\pi$: each point on $\textrm{C}_{right}$ develops into a full geodesic trajectory $\textrm{C}_{left}$ (a great circle) whose entire set forms a \emph{Clifford surface}, which is a closed surface with two conjugate axes of revolution; these axes are precisely the conjugate geodesic circles of the double rotation $p$.  The geometric properties of the Clifford surface depend on the angular distance $\vartheta$ of $\textrm{C}_{right}$ to $\textrm{C}_{p}$. This is illustrated Fig.~\ref{fig17}.

The particular Clifford surface $\vartheta=\pi/4$ is a spherical torus, whose axes of revolution are $\textrm{C}_{p}$ and $\tilde{\textrm{C}}_{p}$ .

 \section{Geometrical elements related to a great circle in $\textrm{S}^3$} \label{appcircle}

\subsection{Great circle defined by two points}
\label {Greatcircledefinedbytwopoints} Consider two points \textbf{M} and \textbf{M}'  on the 3-sphere $\textrm{S}^{3}$, in quaternion notation $u$ and $u'$, in vector notation $\mathbf{u}$ and $
\mathbf{u}'$.  We have
\begin{equation}
\textbf{u}\cdot \textbf{u}'={R}^{2} \cos\vartheta \qquad u\tilde{u}'+u'\tilde{u}=2{R}^{2} \cos\vartheta.  \label{C1}
\end{equation}

$\vartheta$ is the angle between the directions $\textbf{u}$ and $\textbf{u}'$.

We wish to define the geometric elements related to the great
circle C (centered at the origin $\{0\}$) going through $u$ and $u'$.

We introduce the two quaternions $u+u'$ and $u-u'$; they
correspond to orthogonal vectors. We now show that there are two
pure unit quaternions $\varrho$ and $\sigma$ such that
$$
u+u'\propto1-\varrho \sigma,$$ \vspace*{-1cm}
 \begin{equation}
 u'-u\propto \varrho+ \sigma, \label{C2}
\end{equation}

\noindent with
the same coefficient of proportionality, a real number. In effect, multiply the second equation by $\varrho$ (on the left) in order to get an expression for $\varrho$, and by $\sigma$ (on the right) in order to get an expression for $\sigma$, substitute then in the first equation, it comes:
$$
\varrho\propto (u'-u)(\tilde{u}'+\tilde{u})=u'\tilde{u}-u\tilde{u}'$$
\begin{equation}
\sigma\propto (\tilde{u}'+\tilde{u})(u'-u)=\tilde{u}u'-\tilde{u}'u \label{C3}
\end{equation}

\noindent where we have used $u\tilde{u}=u'\tilde{u}' \enskip (={R}^{2}).$ It is easy to check that $\varrho$ and $\sigma$ are pure quaternions. It remains to renormalize, in order to get pure unit quaternions.  Using
\begin{equation*} |u\tilde{u}'-u'\tilde{u}|^{2}=(u\tilde{u}'-u'\tilde{u})(u'\tilde{u}-u\tilde{u}') =4{R}^{4}\sin^{2}\vartheta \end{equation*}
 and the same expression for $|\tilde{u}'u-\tilde{u}u'|^{2},$  $\varrho$ and $\sigma$ eventually read
\begin{equation}
\varrho= \dfrac{u'\tilde{u}-u\tilde{u}'}{2{R}^{2}|\sin \vartheta|}\qquad \sigma=\dfrac{\tilde{u}u'-\tilde{u}'u}{2{R}^{2}|\sin \vartheta|} \label{C4}
\end{equation}

 The 2-plane in which lies the great circle C, defined by $u$, $u'$, and the origin of the coordinates (which is the center of C), contains the directions $1-\varrho\sigma$ and $\varrho+\sigma$. It is denoted $\Pi_{\varrho,\sigma}=\{0,1-\varrho\sigma,\varrho+\sigma\}$. Any rotation of angle $\vartheta$ that leaves invariant this plane leaves invariant the great circle C that contains $u$ and $u'$; it transforms any point $y\in \textrm{E}^{4}$ into  $y'\in \textrm{E}^{4}$ according to the relation $$y'={e}^{-\tfrac{\vartheta}{2}\varrho}y{e}^{\tfrac{\vartheta}{2}\sigma}.$$
  $\Pi_{\varrho,\sigma}$ is the axial plane of the rotation. The plane $\Pi^{\bot}_{\varrho,\sigma}\{0,1+\varrho\sigma,\varrho-\sigma\}$
is completely orthogonal to $\Pi_{\varrho,\sigma}$.

\subsection {Great circle defined by the tangent at a point}  \label {Greatcircledefinedbythetangentatapoint} Let us look for the great circle going through two very close points $u$ and $u+du$, when $du$ vanishes continuously; one gets:
$$\varrho=\dfrac{du\thinspace \tilde{u}-u\thinspace d\tilde{u}}{2{R}^{2}d\vartheta}=\dfrac{1}{2{R}}(\dfrac{du}{ds} \thinspace \tilde{u}-u\thinspace \tfrac{d\tilde{u}}{ds})$$ \vspace*{-0.5cm} $$\sigma=\dfrac{\tilde{u}\thinspace du-d\tilde{u}\thinspace u}{2{R}^{2}d\vartheta}=\dfrac{1}{2{R}}(\tilde{u}\thinspace \dfrac{du}{ds}-\dfrac{d\tilde{u}}{ds}\thinspace u)$$
where $ds={R}d\vartheta$ is the arc element on C.

Let $\mu$ be the unit tangent at $u$ to C, $|\mu|=1$, we have
\begin{equation}
\varrho=\dfrac{1}{2R}(\mu\tilde{u}-u\tilde{\mu}), \, \, \sigma=\dfrac{1}{2{R}}(\tilde{u}\mu-\tilde{\mu}u). \label{C5}
\end{equation}

Because of the relation of orthogonality $\vec{\mu}\cdot\vec{u}=0$, we can also write:
\begin{equation}
\varrho=\dfrac{1}{R}\mu\tilde{u}=-\dfrac{1}{R}u\tilde{\mu}, \, \, \sigma=\dfrac{1}{R}\tilde{u}\mu=-\dfrac{1}{R}\tilde{\mu}u. \label{C6}
\end{equation}

One easily checks that $\varrho$ and $\sigma$ are two pure unit quaternions; $\varrho^{2}=\sigma^{2}=-1$.

\section{The \{3,5\} and \{3,3,5\} symmetry groups} \label{appfive}

    The symmetry group of the \{3,3,5\} curved crystal is related to the symmetry group Y of the icosahedron, since any vertex is the centre of an icosahedron.

\subsection {The group of the icosahedron \{3,5\}}  \label{The group of the icosahedron 3,5}  Y is a finite group with 60 elements, represented by rotations about the six 5-fold axes, the ten 3-fold axes, and the fifteen 2-fold axes of the icosahedron.  The order parameter space of the icosahedron \{3,5\} is $\textrm{V} _{\{3,5\}}=\textrm{SO}(3)/\textrm{Y}$, whose first homotopy group $\Pi_{1}(\textrm{V} _{\{3,5\}})\sim \overline{\textrm{Y}}$ is the lift of Y in the covering group of SO(3), i.e. in SU(2). $\overline{\textrm{Y}}$  is also known as the binary group $<5,3,2>$ \cite{coxeter73}; it has twice as many elements as Y.

The topological theory defect classes of the icosahedron have been investigated by \textcite{nelson}. $\overline{\textrm{Y}}$ is \emph{perfect} (i.e. the commutator subgroup $\textrm{D}[\overline{\textrm{Y}}]=\overline{\textrm{Y}}$), so that in principle all defects can mutually annihilate \cite{trebin84}.

The icosahedron vertices themselves can be expressed in terms of quaternions, so that we have all the analytical tools to represent the symmetry actions on \{3,5\}.  A caveat: since \textit{all} the symmetry elements $h_{\mathbf{Y}}\in\textrm{Y}$ are rotations about axes having the centre of the icosahedron as a fixed point, going through the two-sphere $\textrm{S}^{2}$ on which \{3,5\} lives, any action on a point $x$ of the icosahedron has the form $h_{\tilde{\mathbf{Y}}}\,x\,h_{\mathbf{Y}}$ and connotes a \textrm{disclination}, in the Volterra process terms.  Therefore \textsl{there are no disvections generated by the point group} Y \textsl{in} \{3,5\}.

\subsection {The group of the \{3,3,5\} polytope}  \label{Thegroupofthe3,3,5polytope}  The 120 elements of $\overline{\textrm{Y}}$, in quaternion representation, occupy on $\textrm{S}^{3}$ precisely the locations of the 120 vertices of a \{3,3,5\}.  Thus the binary icosahedral group constitutes a group isomorphic to the group built by these vertices \cite{coxeter73}. The \{3,3,5\} symmetry point group, (\textit{indirect} isometries excluded), is
\begin{equation}
\textrm{H}=\textrm{Y} \times \overline{\textrm{Y}} \label{(78)}
\end{equation}

\noindent whose lift in  $\bar{\textrm{G}}(\textrm{S}^{3}) = \textrm{S}^{3}\times \textrm{S}^{3}$ is
\begin{equation}
\overline{\textrm{H}}=\overline{\textrm{Y}} \times \overline{\textrm{Y}} \label{(79)}.
\end{equation}

H has 7200 elements.  The displacement of a point $x$ of \{3,3,5\} under the action of the symmetry group is
        \begin{equation}
        x'=h_{\mathbf{Y}}\,x\,h_{\overline{\mathbf{Y}}} \qquad h_{\overline{\mathbf{Y}}} \in \overline{\textrm{Y}}, \quad h_{\mathbf{Y}} \in \textrm{Y} \label{(80)}
        \end{equation}

\noindent (or, taking the conjugate, a point $y'=\tilde{h}_{\overline{\mathbf{Y}}}\,y\,\tilde{h}_{\mathbf{Y}}$),
where $h_{\overline{\mathbf{Y}}}$ and $\tilde{h}_{\mathbf{Y}}$ are not necessarily conjugate, i.e. Eq.~\ref{(80)}
composes any right action and any left action.

\bibliography{RMP230207}
\end{document}